# Evolution of Geometric Phase of light since 1956: A Catalog Review

A. Srinivasa Rao[1,2,3,4]

[1]Graduate School of Engineering, Chiba University, Chiba 263-8522, Japan
[2]Molecular Chirality Research Centre, Chiba University, Chiba 263-8522, Japan
[3]Institute for Advanced Academic Research, Chiba University, Chiba, 263-8522, Japan
[4]Quantlight and High Harmonics Lab Pvt. Ltd., Door No. 3-7-400/PGRC/304, Survey No. 30/P, Street No. 16, Nalanda Nagar, Hyderguda, Hyderabad 500048, India
asvrao@chiba-u.jp, sri.jsp7@gmail.com

**Abstract:**
The geometric phase of light is a fascinating phenomenon in optics and arises whenever there is a change in the polarization state of light. It is a fundamentally well-established concept and has recently found extensive applications, particularly in the development of geometric phase elements that enable efficient manipulation of light. In this tutorial review, we discuss the evolution of the geometric phase of polarization on the Poincaré sphere, from its inception by Shivaramakrishnan Pancharatnam in 1956 to its recent advances and applications. This review article aims to focus on core papers related to the geometric phase of polarization rather than providing an exhaustive literature survey. In this review, first, we introduced the basic parameters and corresponding parameter spheres involved in the geometric phase of light. Then, we provide an in-depth analysis of geometric phase in polarization modes, spatial modes, vector modes, and electromagnetic fields. A brief discussion of applications of the geometric phase is also provided. The intriguing explanation given in this review can awaken new ideas related to the geometric phase of light and can open new directions in fundamental and applied optics. Finally, the tutorial is structured as a comprehensive catalog of the geometric phase of light.

## 1. Introduction

Phase is one of the fundamental properties of a light wave; however, it has no physical significance when considering a single wave. It becomes a crucial parameter when multiple waves are involved, playing a pivotal role in both fundamental and applied optics. Light beams exhibit three types of phases, viz., dynamical phase, Gouy phase, and geometric phase. The dynamical phase arises from the optical path length traversed by light and can be measured relative to a reference beam. This phase is commonly encountered in a wide range of optical experiments. The Gouy phase [1] comes as a result of focusing the light beam, which is predominantly observed when we interfere with a collimated light beam. The major application of the Gouy phase is optical bottle/bubble generation [2,3]. The third type, the geometric phase, is a topological phase of light. This intriguing phase has generated significant interest within the optical community due to the new physics it reveals [4].

The geometric phase in light beams has been experimentally created in various ways by producing a cycle of changes in one of the properties of light in different physical phenomena. For instance, Cyclic changes in the state of polarization [4], periodic change in the direction of propagation vector [5], the transformation of transverse spatial mode [6], change in the state of vector beam [7], squeezed states of light [8], reflection of the light at optical multilayers [9], by light transmitted via smoothly inhomogeneous isotropic medium [10], surface plasmon polaritons travel along the boundary of a metal and a dielectric medium [11], electro-magnetic field [12], and nonlinear frequency conversion [13]. In this review, we primarily focus on the geometric phase of light generated through cyclic changes in the propagation vector, polarization, transverse spatial modes, vector beams, and the electromagnetic field.

In the initial decades following the inception of the geometric phase, it was extensively studied in the context of light polarization. With the subsequent development of scalar and vector laser beams, the geometric phase was also observed in the transverse phase and in the combined effects of transverse phase and polarization. The geometric phase is conceptually different from other fundamental properties of light since its creation completely depends on the spatial/temporal evolution of fundamental properties like polarization, propagation vector, phase, and electromagnetic field. Understanding the geometric phase in generalized Gaussian modes is essential when polarization and phase are employed as probing parameters in both fundamental and applied optics. It is worth noting that in the literature, the geometric phase is often referred to by various names, including the Pancharatnam phase, Berry phase, and Pancharatnam-Berry (PB) phase. In this article, we use these terms interchangeably, as they all refer to the same phenomenon—the geometric phase of light.

To date, only a few review articles on this topic have been published. In the initial period, R. Bhandari wrote a full-fledged review article on geometric phase in classical and quantum systems, and it is published in 1997. He covered most of the experimental techniques used for geometric phase measurement [4]. P. Hariharan Briefly discussed the experimental techniques used till the year 2005 for the detection of the geometric phase of light [14]. Y Ben-Aryeh wrote a short review on theoretical calculations of the Pancharatnam-Berry phase of atomic and optical systems [15]. E. Cohen et al. given a brief review on Geometric phase from Aharonov–Bohm to Pancharatnam–Berry by theoretical methods based on quantum wave functions [16]. C. Pannian Jisha et al. provided a brief discussion on the wavefront manipulation and wave-guiding of light through geometric phase [17]. Detailed discussion on PB phase optical elements, such as lens, grating, and detector provided in the review article of [18]. C. Cisowski et al. wrote a full-fledged review but only on the fiber bundle theory of geometric phase [19]. K. Y. Bliokh et al. briefly discussed geometric, dynamical, and total phases calculated along a closed spatial contour in a multi-component complex field, with particular emphasis on 2D (paraxial) and 3D (non-paraxial) optical fields [20].

In this review, we provide in depth analysis of the geometric phase of light from its inception to the most recent developments. The analysis is entirely grounded in the fundamental concepts of light, making it accessible even to readers who are not experts in fundamental optics. The geometric phase is discussed in three types of beams. The first one is the geometric phase in a scalar fundamental Gaussian beam of uniform polarization. In this, first, we start with fundamental properties of light and their representation on the Poincaré Sphere (PS). Next, we give a brief discussion on the polarization optical gadgets which can be used for the state transformation of polarization on PS. The theoretical geometric phase analysis was carried out on the PS with the optical gadgets, and the corresponding experiments are briefly discussed. Subsequently, we gave a short discussion on geometric phase due to a change in the direction of propagation vector of the light beam with fixed state of polarization. The second one is the geometric phase in scalar structured (spatial) modes. Here, we used a similar treatment of polarization. First, we discussed the orbital/modal PS and related modal optical gadgets, and then we compared spatial mode analysis with previously discussed polarization analysis. We analysed the geometric phase in spatial modes of uniform polarization on modal PS. Further, we discuss the geometric phase in vector beams. In this case, the PS is formed by the combination of polarization PS and modal PS, and desired optical gadgets were developed by using a suitable combination of polarization and modal optical gadgets. Finally, the geometric phase in the electromagnetic field, a concept introduced only recently, is briefly discussed to complete our analysis of the geometric phase of light. One of the most fascinating and prominent phenomena of light is interference, which can be generated using simple and cost-effective techniques and is widely employed to measure the phase of light. Most experimental methods utilize the interference phenomenon as a probing tool to investigate the geometric phase of light. In this review, we discuss all types of interferometers employed in the analysis of the geometric phase. Finally, we provide a brief overview of the applications of the geometric phase and conclude the review with a summary.

"*This review is a small token of gratitude to scientists Shivaramakrishnan Pancharatnam and Sir Michael Victor Berry for their extensive contribution towards inception and development of the geometric phase of light*"

## 2. Geometric phase in polarization

### *2.1. Representation of polarization states on the parametric unit 2-sphere (Poincaré sphere)*

The electric and magnetic fields of an electromagnetic wave oscillate in directions transverse to the wave's propagation. The magnitude of the electric field is much larger than that of the magnetic field ($E = cB$), and the electric field is also much easier to detect. Therefore, the oscillation of the electric field is generally considered as defining the polarization of electromagnetic waves. In typical light sources, such as light emitted spontaneously from molecules or atoms, the waves exhibit random polarization. In such randomly polarized light, the electric field can oscillate in any direction within the transverse plane, spanning a full $2\pi$ angle. In contrast, in laser beams, the process of stimulated emission allows control over the oscillation of all waves, resulting in completely polarized light. As shown in Fig. 1, the polarized electric field can oscillate in a to-and-fro motion around its centre called Linear Polarization (LP) or can rotate around its centre in right-/left-handed direction called Right-/Left-Handed Polarization (R/LHP). The detailed information on polarization of light can be found in [21]. In this sense, generalized polarization is considered to be an elliptical polarization. Here, the electric field vector over one cycle of oscillation traces out an ellipse. The equation for generalized polarization distribution can be expressed in terms of two orthogonally polarized fields who collinearly propagating, called the equation of an ellipse.

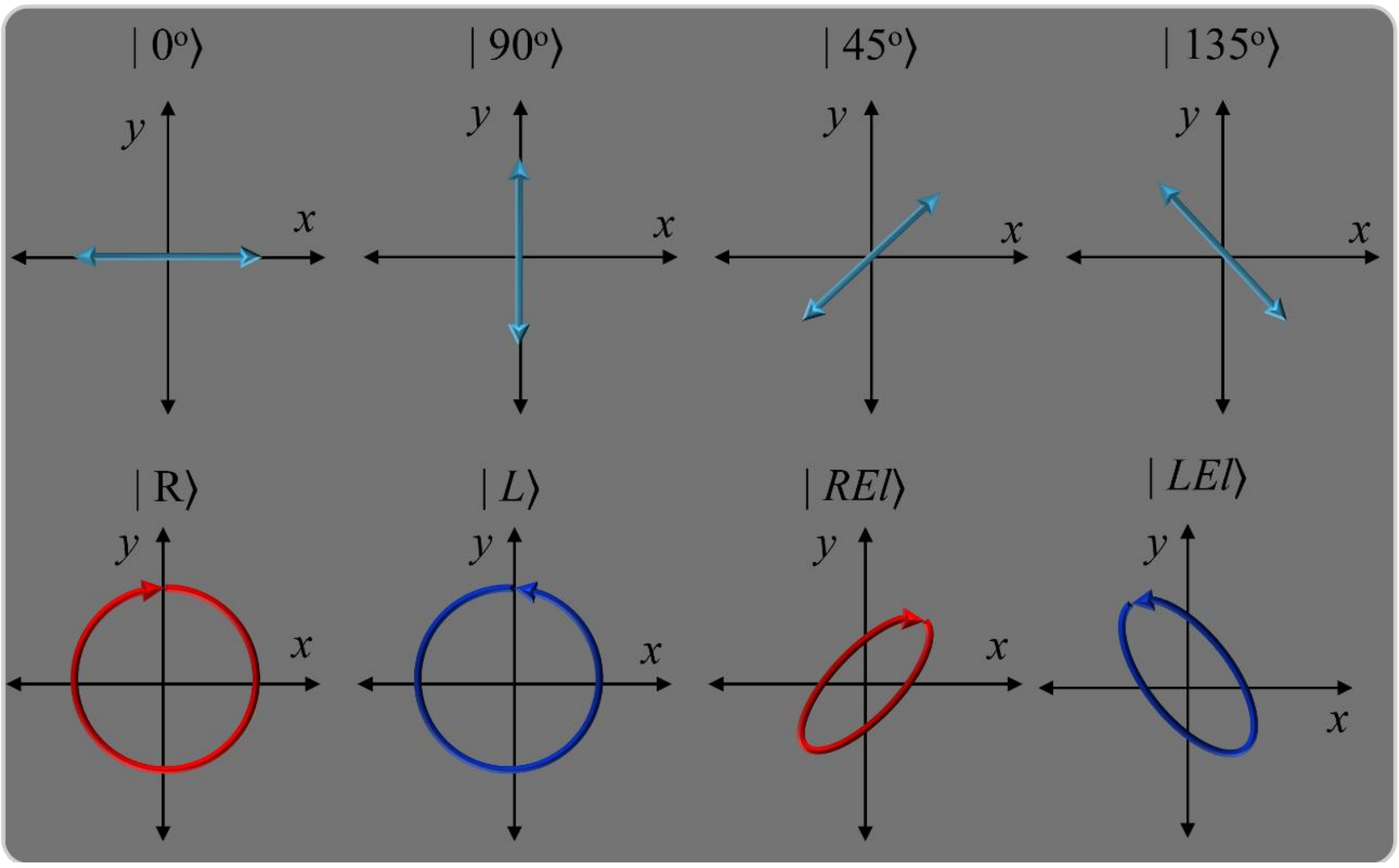

Fig.1. Various kinds of polarization states of light. The top row corresponds to linear polarization states making different angles with respect to the horizontal. Rotational polarization states are given in the bottom row. *R* and *L* represent right and left circular polarization states, respectively. *REl* is right-elliptical polarization state, and *LEl* is left-elliptical polarization state.

For a mathematical expression, consider two orthogonally polarized light waves oscillating along *x*- and y-directions of respective amplitudes $E_x$ and $E_y$, and are collinearly propagating along the *z*-direction

$$E_x(z,t) = E_{0x}e^{i\,\delta_x}\,e^{i\,(kz-\omega t)}, \qquad (1a)$$

and

$$E_y(z,t) = E_{0y}e^{i\,\delta_y}\,e^{i\,(kz-\omega t)}. \qquad (1b)$$

Here, $\delta_x$ is the phase of the *x*-component of the electric field, and $\delta_y$ is the phase of the *y*-component of the electric field. The phase difference between the two polarization modes is $\delta = \delta_x - \delta_y$. The phase difference between the two waves can be varied within the wave periodicity and ($0 \le \delta \le 2\pi$). The propagation vector, $k = 2\pi/\lambda$, and $\omega$ and $\lambda$ are the angular frequency and wavelength of light waves, respectively. The relative amplitudes and phase factors can be written in a complex number form as

$$\frac{E_x(z,t)}{E_y(z,t)} = \frac{E_{0x}}{E_{0y}} e^{i\,\delta} = u + iv \tag{2}$$

with $u = E_{0x}\cos\delta$ and $v = E_{0y}\sin\delta$. The resultant electric field of a single light beam generated by the superposition of the above two orthogonally polarized and collinearly propagating light beams can be obtained by considering the interaction of two beams at a given time in the presence of interactive matter. The resultant equation is the equation of an ellipse

$$\left(\frac{E_x(z,t)}{E_{0x}}\right)^2 + \left(\frac{E_y(z,t)}{E_{0y}}\right)^2 - 2\frac{E_x(z,t)E_y(z,t)}{E_{0x}E_{0y}}\cos\delta = \sin^2\delta. \tag{3}$$

The state of polarization can be quantitatively parameterized in terms of geometrical parameters of the ellipse. The inclination angle ($\psi$) of the major axis of the ellipse ($E_b$) with respect to the horizontal is described as the orientation of the ellipse, and it is also called the angle of azimuth, and it can be expressed in terms of the parameters of the polarization ellipse as

$$\psi = \frac{1}{2}\tan^{-1}\left[\frac{2E_{0x}E_{0y}}{E_{0x}^2 - E_{0y}^2}\cos\delta\right]. \tag{4}$$

The ellipticity of the polarization is given by the ratio of the ellipses' major axis to minor axis, $e = E_b / E_a$. It can also be parameterized in terms of angle as

$$\chi = \tan^{-1}\left[\frac{\pm E_b}{E_a}\right] \tag{5a}$$

or

$$\chi = \frac{1}{2}\sin^{-1}\left[\frac{2E_{0x}E_{0y}}{E_{0x}^2 + E_{0y}^2}\sin\delta\right]. \tag{5b}$$

These angles are constrained by limits: $0 \leq \psi \leq \pi$, $-\pi/4 \leq \chi \leq \pi/4$. The plus and minus sign in the elliptical angle represents respective left (counter-clockwise) and right-hand (clockwise) rotations. Another angular parameter of the ellipse is the auxiliary angle $\zeta$ ($0 \leq \zeta \leq \pi/2$) and is given by

$$\zeta = \tan^{-1}\left[\frac{E_{0y}}{E_{0x}}\right]. \tag{6}$$

Further, the angular coordinates can be written in terms of the auxiliary angle as

$$\psi = \frac{1}{2}\tan^{-1}[\tan(2\zeta)\cos\delta], \tag{7a}$$

and

$$\chi = \frac{1}{2}\sin^{-1}[\sin(2\zeta)\sin\delta]. \tag{7b}$$

The coordinates of the state vector and the phase are related by

$$\cos(2\chi) = \cos(2\zeta)\cos(2\psi) + \sin(2\zeta)\sin(2\psi)\cos(\delta). \tag{8}$$

When $E_a = E_b$, the polarization is circular, and $E_a >> E_b$ or $E_a << E_b$ corresponds to LP. The polarization states of electromagnetic waves form a complete two-dimensional space. The basis of this space can be created by two orthogonal polarization states in Hilbert space. The generalized polarization state can be represented as a point on Argand plane formed by the complex number $u + iv$. Here, $u = 0$ and $v = \pm 1$ correspond to Right/Left Circular Polarization (R/LCP), and $v = 0$ and $u = -\cot\psi$ or $\tan\psi$ lead to the LP states. The rest of the points are elliptically polarized states. All points on this Argand plane can be stereographically projected onto a 2-sphere called the PS [22] in terms of its spherical angles [Fig. 2] [21]. When we connect the coordinates in this projection, the angles on the planar surface become double on the spherical surface. The detailed analysis can be found in [23-25]. In this sphere, north and south poles are in Left-Circular Polarization (LCP) and Right-Circular Polarization (RCP) states, and the polarization states present at the equator are in a LP state. The rest of the points are in an elliptical polarization state. The latitude and the longitude coordinates on the sphere are given by 2χ and 2ψ, respectively. The polar angle 2χ starts from the *xy*-plane, and the azimuthal angle 2ψ is considered from the *xz*-plane. Moreover, the *x*, *y*, and *z* coordinates in terms of 2χ and 2ψ are given by

$$x = \cos 2\psi \cos 2\chi, \tag{9a}$$

$$y = \sin 2\psi \cos 2\chi, \tag{9b}$$

and

$$z = \sin 2\chi. \tag{9c}$$

The *x*-axis polarization states are horizontal polarization $|H\rangle = [1\ 0]^T$ and vertical polarization $|V\rangle = [0\ 1]^T$. The *y*-axis polarization states represent diagonal polarization $|D\rangle = {}^1\!/_{\sqrt{2}}[1\ 1]^T$ and anti-diagonal polarization $|A\rangle = {}^1\!/_{\sqrt{2}}[1\ -1]^T$. The *z*-axis corresponds to circular polarization states: RCP $|R\rangle = {}^1\!/_{\sqrt{2}}[1\ -i]^T$ and LCP $|V\rangle = {}^1\!/_{\sqrt{2}}[1\ +i]^T$. The rotation of the electric field in the circular polarization states creates angular momentum in the electromagnetic waves called Spin Angular Momentum (SAM) with the quantitative value of $\sigma\hbar$. The value of σ for $|R\rangle$ state is +1 and for $|L\rangle$ state is -1. The SAM of LP is zero. The SAM has a range of values 0 < σ < ±1 for the elliptical polarization state $|El\rangle$. The advantage of representing polarization states on a sphere rather than a plane is that it is independent of the choice of reference axes, allowing the dynamics of polarization to be easily monitored, such as in birefringent crystals. Another major advantage and a ground breaking insight provided by the PS is the ability to investigate and understand the geometric phase associated with polarization states. Changes in the polarization state as a light beam passes through optically active media can be easily traced, including the path taken and the final position on the sphere, either by rotation of the sphere itself or by following the polarization trajectory around the appropriate axis defined by antipodal points. The transformation of a polarization state can be achieved by introducing an appropriate phase, or equivalently, by transferring the polarization state so that the desired phase is accumulated. This process is analogous to the transition of a particle from one state to another through the absorption or release of energy. A simple way to understand this is as follows: any polarization state can be decomposed into two orthogonal components, and if a specific phase difference is applied between them, the resulting state transforms into a new state that reflects this phase difference relative to the initial state. The phase accumulated during this process is known as the geometric phase. This phenomenon can be quantitatively analysed within the parameter space. This phase was first independently observed in the polarization of classical light by S. Pancharatnam in 1956 [26] and in molecular physics by H. C. Longuet-Higgins in 1958 [27]. It was later investigated and formalized in quantum systems by M. V. Berry in 1984 [28].

In polarization analysis, experimental results are typically obtained on a planar surface corresponding to the transverse cross-section of the light beam, and these results can be readily interpreted by projecting them onto the PS. The evolution of polarization of light when it passes through a medium that exhibits birefringence, optical activity, or simultaneously both, can be easily understood through PS.

In the above discussion, the PS was constructed with the instantaneous electric field oscillation of elliptical polarization, and it could be difficult to measure the polarization states of the PS. However, we can overcome this difficulty by obtaining the intensity of each polarization by the time average of the electric field. The time-averaged polarization components are called Stokes parameters [29,30] and are given by

$$\begin{pmatrix} S_0 \\ S_1 \\ S_2 \\ S_3 \end{pmatrix} = \begin{pmatrix} E_{0x}^2 + E_{0y}^2 \\ E_{0x}^2 - E_{0y}^2 \\ 2E_{0x}E_{0y}\cos\delta \\ 2E_{0x}E_{0y}\sin\delta \end{pmatrix} = \begin{pmatrix} I_0 \\ I_x - I_y \\ I_D - I_A \\ I_R - I_L \end{pmatrix}. \tag{10}$$

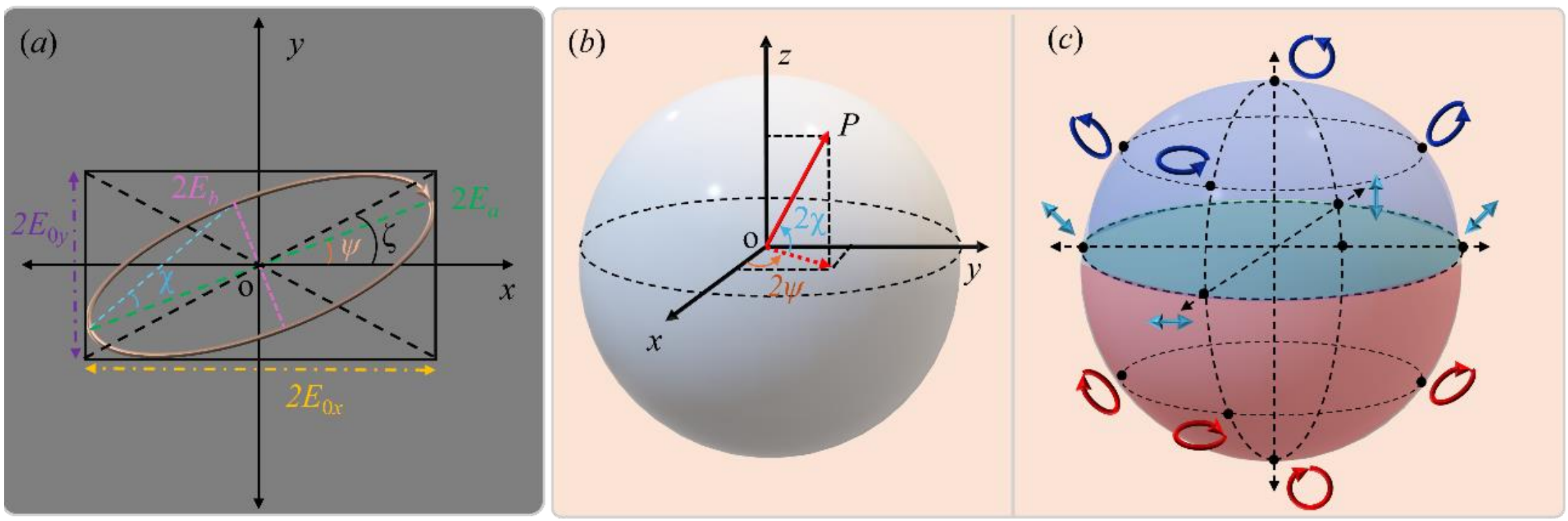

Fig. 2. (*a*) Illustration of properties of elliptical polarization of state $P$ in the light beam transverse cross-section. (*b*) The state $P$ on the Poincaré sphere. (*c*) Polarization states on the Poincaré sphere. In the polarization states, blue colour (upper hemisphere) corresponds to left-handed polarization and red colour (lower-hemi-sphere) represents right-hand polarization. The linear polarization states at the equator are given in cyan colour.

We can construct the PS whose axes are Stokes parameters $S_1$, $S_2$, and $S_3$ [Fig. 3(*a*)] [31]. It is also noted that Eq. 10 provides the generalized state of polarization on the PS. The intensity ($S_0$) of light at an arbitrary position on the PS is projected on the three coordinates, which can be written in terms of 2χ and 2ψ as

$$\begin{pmatrix} S_0 \\ S_1 \\ S_2 \\ S_3 \end{pmatrix} = \begin{pmatrix} I_0 \\ I_0 \cos 2\psi \cos 2\chi \\ I_0 \sin 2\psi \cos 2\chi \\ I_0 \sin 2\chi \end{pmatrix}. \tag{11}$$

Here, $I_0$ is the intensity of the light beam given by the first Stokes parameter. The second Stokes parameter is in terms of $I_x$ (horizontally polarized intensity and state vector $|H\rangle = [1\ 1\ 0\ 0]^T$) and $I_y$ (vertically polarized intensity and state vector $|V\rangle = [1\ -1\ 0\ 0]^T$). The third Stokes parameter is formed by the $I_D$ (45° polarized intensity and state vector $|D\rangle = [1\ 0\ 1\ 0]^T$) and $I_A$ (135°/-45° polarized intensity and state vector $|A\rangle = [1\ 0\ -1\ 0]^T$). The final Stokes parameter is given by $I_R$ (RCP intensity and state vector $|R\rangle = [1\ 0\ 0\ 1]^T$) and $I_L$ (LCP intensity and state vector $|L\rangle = [1\ 0\ 0\ -1]^T$). The spherical coordinates of PS can be written in terms of Stokes parameters as

$$\psi = \frac{1}{2}\tan^{-1}\left[\frac{S_2}{S_1}\right], \tag{12}$$

and

$$\chi = \frac{1}{2}\sin^{-1}\left[\frac{S_3}{S_0}\right]. \tag{13}$$

The Stokes parameters can be easily estimated by intensity measurements of polarization components by wave plates [one polarizer and one Quarter-Wave Plate (QWP)] [32-34].

In the above two representations, we consider the polarization distribution on the PS. We can represent the same states on the same PS by considering the SAM of the polarization state in the form of spinor notation, and the spherical coordinates are polar angle, $\theta$, and azimuthal angle $\phi$. Here, the polar angle reference axis is the axis of the north pole instead of the equator plane. The position and direction of polar angle $\theta$ (= 2χ) and the azimuthal angle $\phi$ (= 2$\psi$) on the PS are visually presented in Fig. 3(*b*). Here, the angles are constrained by limits: $0 \le \theta \le \pi$, $0 \le \phi \le 2\pi$.

For clear visualization, let's consider the superposition of two laser modes in respective RCP state $|R\rangle$ and LCP state $|L\rangle$ with their weight factor provided by polar angle $\theta$, and their relative phase determined by azimuthal angle $\phi$. We can write this superposition state in the mathematical form as

$$|P(\theta,\phi)\rangle = \cos\left(\frac{\theta}{2}\right)e^{-\frac{i\phi}{2}}|R\rangle + \sin\left(\frac{\theta}{2}\right)e^{\frac{i\phi}{2}}|L\rangle \tag{14a}$$

or

$$|P(\theta,\phi)\rangle = \cos\left(\frac{\theta}{2}\right)|R\rangle + \sin\left(\frac{\theta}{2}\right)e^{i\phi}|L\rangle. \tag{14b}$$

The antipodal points on the PS are orthogonal and form a basis for a 2D Hilbert space. By considering arbitrary antipodal points of $|1\rangle$ and $|2\rangle$ on PS, we can also construct the full PS. Therefore, the generalized PS equation is given by

$$|P(\theta,\phi)\rangle = \cos\left(\frac{\theta}{2}\right)e^{-\frac{i\phi}{2}}|1\rangle + \sin\left(\frac{\theta}{2}\right)e^{\frac{i\phi}{2}}|2\rangle. \tag{15}$$

The spinor (it is the same as the third component of the Stokes vector $S_3$ of the SAM direction) in the upward direction (vertically upward) at the north pole transforms to an orthogonal direction (horizontally oriented) at the equator, and then the direction becomes anti-parallel at the south pole (vertically downward) [Fig. 3($c$)]. The upper hemisphere corresponds to Left-Handed Polarization (LHP), and the spinor can have all angular orientations between up and horizontal. Similarly, the lower hemisphere has the Right-Handed Polarization (RHP) states, and the spinor can have all angular directions between horizontal and down. This process is equivalent to the electron spin on the Bloch sphere. The direction of spinor can be controlled by the weight factor in Eq. 15, and it can be quantitatively understood with reference to the propagation direction of the corresponding light wave [Fig. 3($d$)]. The arbitrary direction of spinor created in the laser beams can be projected onto the three Cartesian coordinates with the angular coordinates ($\theta$, $\phi$). We can systematically control the direction of spinor on the beam cross-section through a suitable combination of spatial amplitude distribution and phase distribution in the superposition. This way, we can produce a non-uniformly distributed spinor, and the resulting mode is called a Poincaré beam [35]. Recently, it was shown that some of the Poincaré beams have 2D and 3D optical quasiparticle nature in the spinor [36]. If we consider $\vec{d} = (d_x, d_y)$ as normalized complex unit vector of a polarized light wave, then it has a spinor in the form of

$$|P_s\rangle = \frac{1}{\sqrt{2}}\begin{pmatrix} d_x + id_y \\ d_x - id_y \end{pmatrix}. \tag{16}$$

The unit vector of the spinor $|P_s\rangle$ on the PS, is given by [37]

$$\vec{r} \equiv (\sin\theta\cos\phi, \sin\theta\sin\phi, \cos\theta) \equiv \langle P_s|\vec{\sigma}|P_s\rangle \tag{17}$$

with

$$\vec{\sigma} \equiv \left\{\begin{bmatrix} 0 & 1 \\ 1 & 0 \end{bmatrix}, \begin{bmatrix} 0 & -i \\ i & 0 \end{bmatrix}, \begin{bmatrix} 1 & 0 \\ 0 & -1 \end{bmatrix}\right\}. \tag{18}$$

It is also noted that $\vec{r}$ is the position vector of the state $|P_s\rangle$ on the PS, and it can be written in spinor form as

$$|\vec{r}\,\rangle = \begin{bmatrix} \cos\left(\frac{\theta}{2}\right) \\ \sin\left(\frac{\theta}{2}\right)e^{i\phi} \end{bmatrix} \tag{19}$$

and its orthogonal polarization state has a position vector of $-\vec{r}$. It is worth noting that in normal state representation, the positions of orthogonal polarizations are represented with $r$ and $r'$ and in spinor notation, the state vectors of orthogonal polarizations are $\vec{r}$, and $-\vec{r}$. Further, the Hermitian matrix operator of polarization has the form of

$$\hat{\mathrm{H}}(r) = \begin{bmatrix} \cos\theta & \sin\theta e^{-i\phi} \\ \sin\theta e^{i\phi} & -\cos\theta \end{bmatrix} \tag{20}$$

and when it operates on a state $|A\rangle$ with a position vector $\vec{r}_A$ on PS, we have an eigen-value without transporting the state. This representation has a lot of advantages over the former one, especially when PS is used for Orbital Angular Momentum (OAM) in a similar fashion of SAM. It is very easy to compare these two angular momenta, and we can produce higher-order or hybrid PS by non-separable mixing of OAM and SAM. Also, it plays a pivotal role in the understanding of spin-orbit interaction and higher-order vector mode representation.

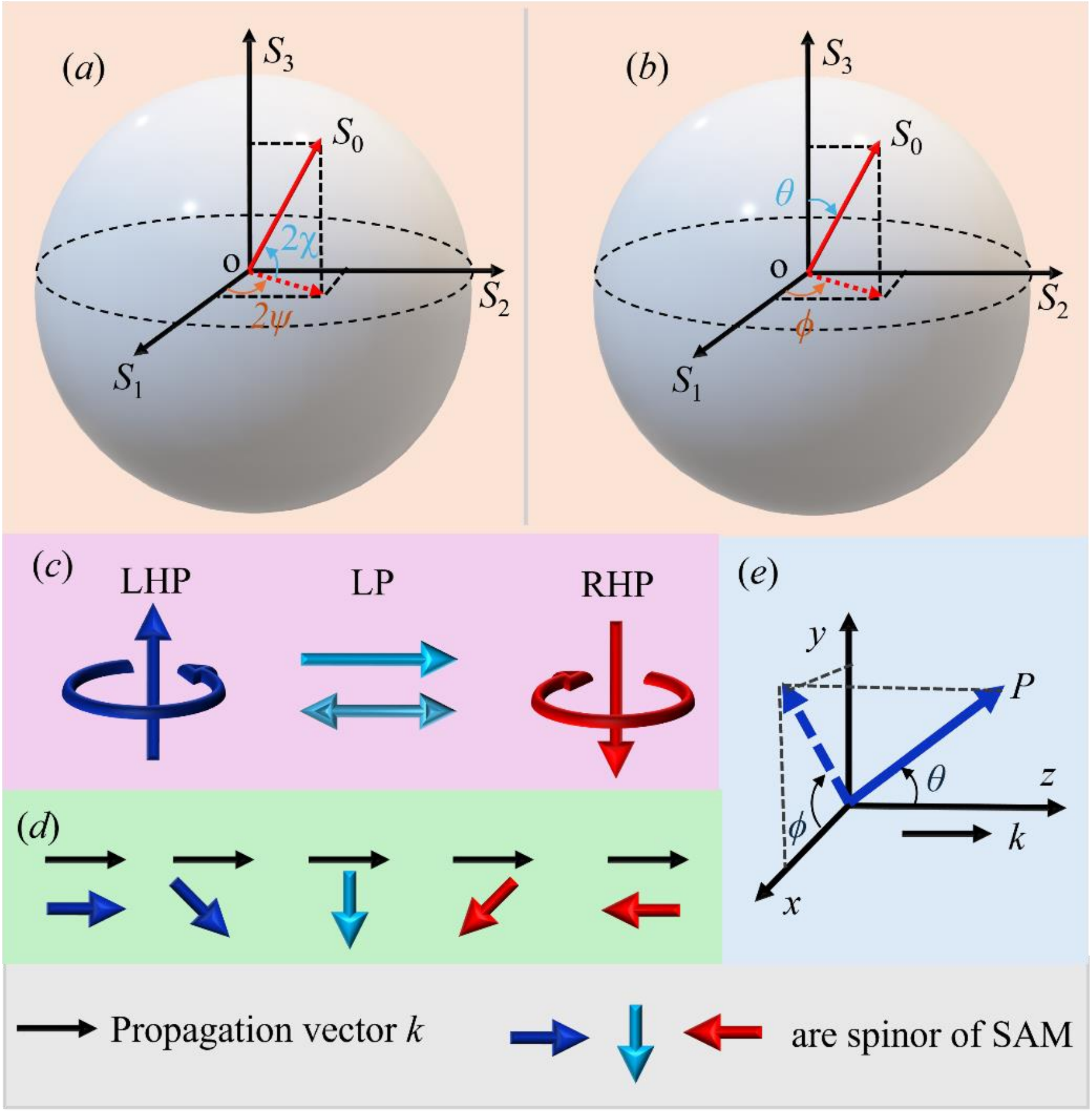

Fig. 3. Poincaré sphere with Stokes parameters as its Cartesian coordinates. (*a*) The spherical coordinates of the Poincaré sphere are the same as its conventional Jones vector representation. (*b*) The spherical coordinates of the Poincaré sphere are represented like the spherical coordinates of the Bloch sphere, and states on the sphere are represented with spinors instead of polarization. Polarization has Spin Angular Momentum, and its direction is equivalent to the direction of electron spin. (*c*) The SAM direction and corresponding polarization state are presented. (*d*) SAM can have any direction within 180° of orientation with respect to the propagation vector. (*e*) Angular position and projection of the spinor, *P* of the light wave propagating along the *z*-axis. LHP is left-hand polarization, RHP is right-hand polarization, and LP is linear polarization.

### *2.2. Pancharatnams' investigation on geometric phase*

The first seminal paper on the geometric phase was published in 1956 by Shivaramakrishnan Pancharatnam which is communicated by the renowned Indian Scientist Sir C. V. Raman [26]. Pancharatnam conducted a series of theoretical and experimental studies on the phase associated with polarization [38], aiming to develop a generalized theory for the interference of light beams in different polarization states and explore their applications. All of his calculations were carried out on the PS using spherical coordinates. Notably, he discovered an additional phase that appeared when a closed path was traced on the sphere; this phase was later recognized by M. V. Berry as the geometric phase of light in polarization, and Pancharatnam's work gained significant attention in the field.

To extract the phase between two electric field oscillations using Pancharatnam's analysis, consider two light waves drawn from a single laser source, propagating collinearly and coherently along the z-direction, with their electric fields oscillating in the *xy*-plane. Then we can write their states in terms of Dirac notation as

$$E_a = E_{0a}|A\rangle e^{i\,\delta_a}, \qquad (21a)$$

and

$$E_b = E_{0b}|B\rangle e^{i\,\delta_b}. \qquad (21b)$$

Here, $|A\rangle$ and $|B\rangle$ are polarization states of respective $E_a$ and $E_b$ waves, and can be represented on PS, and corresponding electric field amplitudes are $E_{0a}$ and $E_{ob}$ respectively. Let's consider states $|A\rangle$ and $|B\rangle$ are at a certain angles of $b/2$ and $a/2$ with respect to state $|C\rangle$ (the three states are non-collinear). Then these three states form a spherical triangle on the PS with vortices $|A\rangle$, $|B\rangle$, and $|C\rangle$. Each side of the triangle is an arc of the greatest circle of the sphere, called a geodesic arc. The states and their angular relations on a planar and spherical surface are pictorially illustrated in Fig. 4.

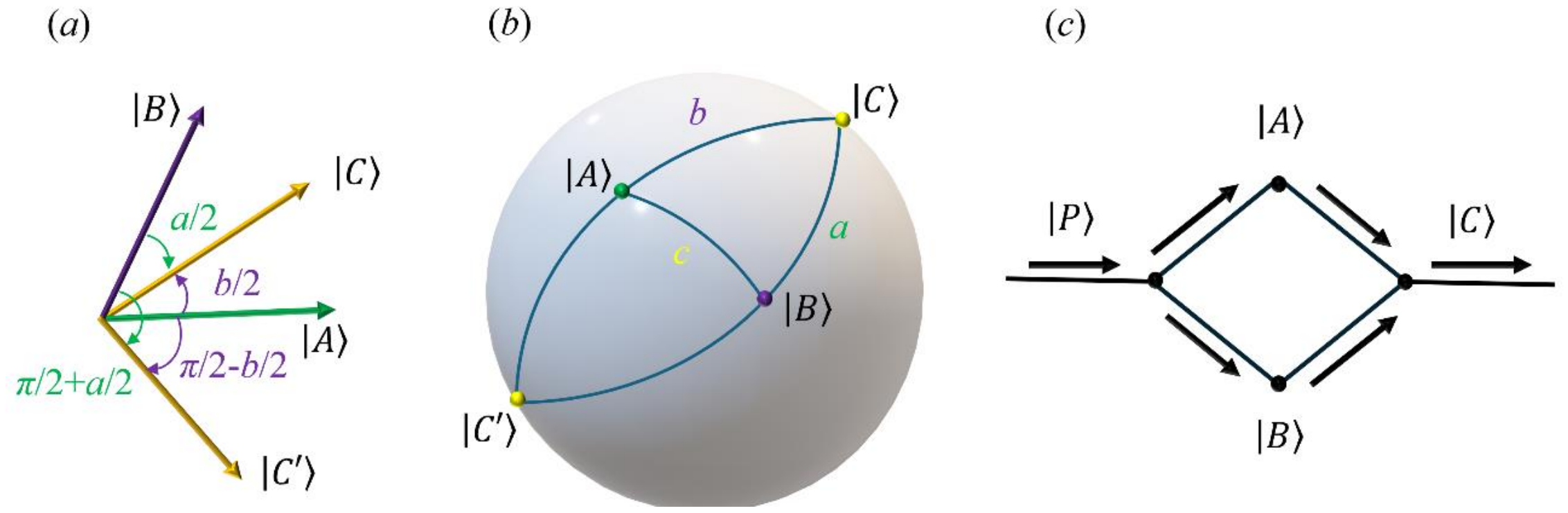

Fig. 4. Illustration of polarization states of light fields in 2D: (*a*) planar surface and (*b*) spherical surface. States $|A\rangle$ and $|B\rangle$ are positioned at respective angles of $b/2$ and $a/2$ with reference to the state $|C\rangle$. States $|A\rangle$, $|B\rangle$, and $|C\rangle$ form a spherical triangle on the Poincaré sphere with the angular separation between $|A\rangle$ and $|B\rangle$ is $c$, between $|C\rangle$ and $|B\rangle$ is $a$, and between $|A\rangle$ and $|C\rangle$ is $b$. States $|C\rangle$ and $|C'\rangle$ are orthogonal polarization states, and their angular separation on a planar surface is $\pi/2$ and on a spherical surface is $\pi$. States $|C\rangle$ and $|C'\rangle$ are antipodal points on the spherical surface. (*c*) Schematic diagram of the closed loop created in the experiment of polarization state transformation.

Here, a single laser beam is decomposed into two beams of polarization states, $|A\rangle$ and $|B\rangle$. Now the amplitude of superposition state is given by

$$E = E_a + E_b = E_{0a}|A\rangle e^{i\,\delta_a} + E_{0b}|B\rangle e^{i\,\delta_b}, \qquad (22a)$$

and its intensity is

$$I = I_1 + I_2 + 2\sqrt{I_1 I_2}\cos\left(\frac{c}{2}\right)\cos(\delta). \qquad (22b)$$

Here, $c$ is the angular separation of the states $|A\rangle$ and $|B\rangle$ on the PS and $\delta = \delta_a - \delta_b$. The $\cos^2(c/2)$ is called the similarity factor between the two interfering states, and it is essentially visibility of interference. The similarity factor between the two states on the PS can be written in terms of spherical coordinates $\psi$ and $\chi$ as

$$\cos c = \sin(2\chi_1)\sin(2\chi_2) + \cos(2\chi_1)\cos(2\chi_1)\cos(2\psi_1 - 2\psi_2). \qquad (23)$$

If there is no difference between the initial and final position along the longitude ($\psi_1 = \psi_2$) then the angular separation between the two positions is $c = \chi_1 - \chi_2$. Also, if there is no displacement along the latitude, then $c = \psi_1 - \psi_2$.

Let the two laser beams be projected onto the polarization state $|C\rangle$. Then, the phase delay acquired between the two states along the state, $|C\rangle$ is different from δ and let consider it as δ′. The intensity along the state $|C\rangle$ is given by

$$I_C = I_a \cos^2\frac{1}{2}b + I_b \cos^2\frac{1}{2}a + 2\sqrt{I_a I_b}\cos\frac{1}{2}a\,.\cos\frac{1}{2}b\,.\cos\delta'. \tag{24}$$

Now consider state $|C'\rangle$ which is antipodal of state $|C\rangle$ on the PS. Thus, the intensity of the two beams interference along the state$|C'\rangle$ is given by

$$I_{C'} = I_a \sin^2\frac{1}{2}b + I_b \sin^2\frac{1}{2}a + 2\sqrt{I_a I_b}\sin\frac{1}{2}a\,.\sin\frac{1}{2}b\,.\cos\delta''. \tag{25}$$

Here, δ″ is the phase delay between the two beams along the state, $|C'\rangle$. Pancharatnam has shown in the same paper that $\delta'' = \delta' \pm \angle C$ ($\angle C$ is the angle at point $C$ on the PS). The intensity of the initial laser beam derived from the laser source is equal to $I_C + I_{C'}$. From Eq. 24 and Eq. 25

$$I = I_a + I_b + 2\sqrt{I_a I_b}\left[\cos\frac{1}{2}a\,.\cos\frac{1}{2}b\,.\cos\delta' + \sin\frac{1}{2}a\,.\sin\frac{1}{2}b\,.\cos(\delta' \pm \angle C)\right]. \tag{26}$$

By using spherical calculations [39], Pancharatnam derived the equation for the total intensity of a light beam as

$$I = I_a + I_b + 2\sqrt{I_a I_b}\cos\frac{1}{2}c\cos\left(\delta' + \frac{1}{2}[\Delta ABC]\right). \tag{27}$$

Here, $[\Delta ABC]$ is the area of the spherical triangle formed by $|A\rangle$, $|B\rangle$, and $|C\rangle$ states, and its sign can be positive or negative for counterclockwise or clockwise on the triangle. By comparing Eq. 22$b$ and Eq. 27, we end up with the condition of $\delta' = \delta - \frac{1}{2}[\Delta ABC]$. Therefore, Eq. 24 becomes

$$I_C = I_a \cos^2\frac{1}{2}b + I_b \cos^2\frac{1}{2}a + 2\sqrt{I_a I_b}\cos\frac{1}{2}a\,.\cos\frac{1}{2}b\,.\cos\left(\delta - \frac{1}{2}[\Delta ABC]\right). \tag{28}$$

The solid angle Ω, of a unit radius sphere is equal to the corresponding area of the spherical surface and

$$I_C = I_a \cos^2\frac{1}{2}b + I_b \cos^2\frac{1}{2}a + 2\sqrt{I_a I_b}\cos\frac{1}{2}a\,.\cos\frac{1}{2}b\,.\cos\left(\delta - \frac{1}{2}\Omega\right). \tag{29}$$

Therefore, when we project any two states onto the third state, then the phase difference originated between the two states at the third state is equal to half of the spherical area/solid angle created between the three states. This is the central result in Pancharatnam's calculations. The solid angle in terms of polar angle $\theta$, and azimuthal angle $\phi$, of PS is $\Omega = (\phi_2 - \phi_1)(\cos\theta_1 - \cos\theta_2)$. Here, the angular coordinates of optical circuits on the PS are $(\theta_1, \phi_1)$ and $(\theta_2, \phi_2)$, respectively.

This phenomenon can be quantitatively investigated by passing a coherent superposition of two light beams present in respective polarization states of $|A\rangle$ and $|B\rangle$ through a polarization analyzer whose transmission axis is in $|C\rangle$ state. For example [Fig. 4($c$)], a laser beam with any arbitrary polarization state, $|P\rangle$ can be split into two laser beams with different states of polarizations of $|A\rangle$ and $|B\rangle$ by using polarization controlled optical elements [for example, a combination of Polarizing Beam Splitter (PBS) and Half-Wave Plate (HWP)]. Again, we can project both the states onto a single state of polarization $|C\rangle$ by polarization-controlled elements like a polarizer. The geometric phase acquired by the parallel transfer of the laser beam through two different polarization states will appear in the interference pattern. The shift in the fringe position as a result of the geometric phase can be experimentally observed by changing the angular position of polarization-controlled elements used in the parallel state transport.

Furthermore, the expression for geometric phase accumulated in the addition of n coherent laser beams ($|A_1\rangle, |A_2\rangle, |A_3\rangle \ldots\ldots\ldots |A_n\rangle$) with different state polarization and derived from a single laser source obtained by Pancharatnam as

$$I_C = \sum_i I_i + \sum_{i \neq j} \sqrt{I_i I_i} \cos\theta_i \,.\cos\theta_j \,.\cos\left(\delta_{ij} - \frac{1}{2}\Omega_{ij}\right). \tag{30}$$

Here, $\theta_i$ is the angle between state $|A_i\rangle$ and state $|C\rangle$, and $\delta_{ij}$ is the phase delay between the two states which are projected onto the state $|C\rangle$. The solid angle $\Omega_{ij}$ formed by states $|A_i\rangle$, $|A_j\rangle$, and state $|C\rangle$. In Fig. 5, we have given some of the closed optical circuits on the PS. In Fig. 5(*a*), we have a closed path in the elliptical shape (on a flat plane) with its major axis connecting poles (RCP and LCP points) while its minor axis is in the equatorial plane. The angular parameters are $\phi_2 - \phi_1 = \phi_0$, $\theta_1 = 0$, and $\theta_2 = \pi$. The solid angle is given by $\Omega = 2\phi_0$ and the geometric phase $\gamma_G = -\phi_0$. The closed circuit in Fig. 5 (*b*) connects one pole (RCP point) and two arbitrary points on the sphere. In this case, the solid angle of $\Omega = \phi_0$ (1 - cos$\theta_0$) and geometric phase, $\gamma_G = \phi_0$ (cos$\theta_0$ - 1) / 2. The closed circuit in Fig. 5 (*c*) is not connected to the equatorial plane and the poles of the sphere. So all points have non-zero spherical coordinates. Here, the solid angle, $\Omega = \phi_0$ (cos$\theta_1$-cos$\theta_2$), and geometric phase, $\gamma_G = \phi_0$ (cos$\theta_2$-cos$\theta_1$) / 2. The closed circuit created in the equatorial plane has the area of a hemisphere [Fig. xx (*d*)] and has the solid angle of $\Omega = 2\pi$. Therefore, the geometric phase, $\gamma_G = -\pi$.

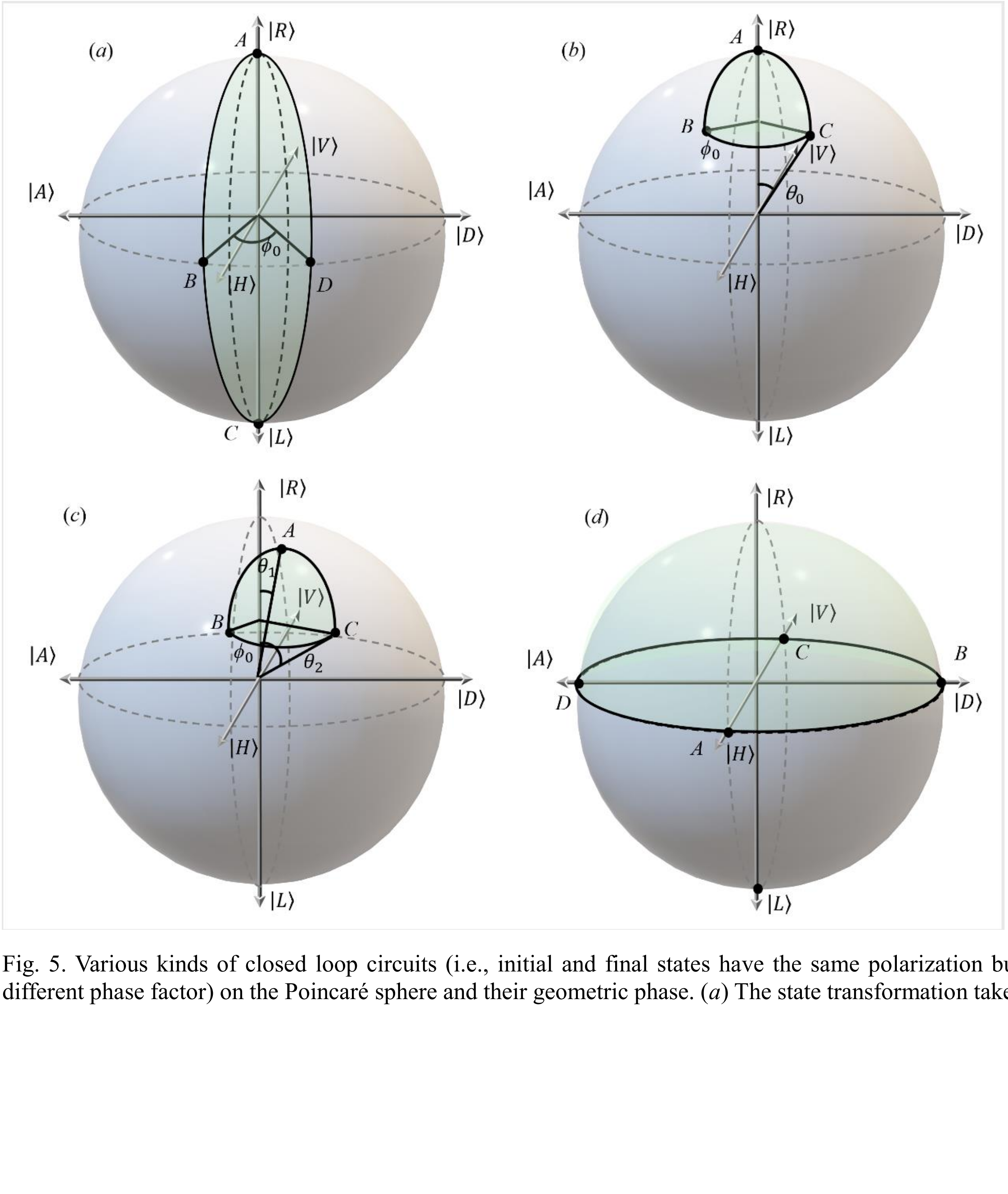

Fig. 5. Various kinds of closed loop circuits (i.e., initial and final states have the same polarization but different phase factor) on the Poincaré sphere and their geometric phase. (*a*) The state transformation takes

place in an elliptical shape with connecting poles and the geometric phase, $\gamma_G = -\phi_0$. (*b*) The optical circuit is formed with connecting north pole in spherical triangle shape and the geometric phase, $\gamma_G = \phi_0 (\cos\theta_0 - 1) / 2$. (*c*) The optical circuit, created by non-zero polar angles and azimuthal angle, has the geometric phase, $\gamma_G = \phi_0 (\cos\theta_2 - \cos\theta_1) / 2$. (*d*) The state transformation on equatorial plane in a closed path have geometric phase of $-\pi$.

### *2.3. The adiabatic phase in quantum systems by M. V. Berry*

The change in the phase of a quantum state vector under a sequence of adiabatic transformations that returns the system to its original state was first demonstrated by M. V. Berry in his 1984 seminal paper [28,40,41]; this phase is now commonly known as the Berry phase. In this scenario, when a quantum particle, such as an electron, is subjected to an adiabatically varying magnetic field and traces a closed path *C* in parameter space, returning to its initial state, it acquires an additional phase in addition to the dynamical phase. This additional phase (geometric phase) can be expressed in terms of the solid angle $\Omega(C)$ subtended by the closed path *C* on the parameter sphere, and it can be mathematically expressed as

$$\gamma_G(C) = -m_s \Omega(C). \tag{31}$$

Here, the magnetic flux in parameter space through *C* in the presence of a monopole of strength. If we consider the state vector makes a closed path in the azimuthal direction and reaches the polar position $\theta$ (i.e., $\phi_2 - \phi_1 = 2\pi$ & $\theta = 0$, $\theta_2 = \theta$) on the parameter sphere, then the geometric phase is given by

$$\gamma_G(C) = -2\pi m_s (1 - \cos\theta). \tag{32}$$

Later, Aharanov and Anandan have shown that the assumption of adiabaticity is not essential for the appearance of Berry's phase [42].

### *2.4. Geometric phase in twisted wave guide*

A few years later, the photonic counterpart of the geometric phase was theoretically proposed and experimentally demonstrated when light propagated through a helically twisted waveguide [5,43,44]. In the Berrys' formula of Eq. 31, the electrons are spin-half particles ($m_s = \pm½$) and the geometric phase is $\pm \Omega/2$. The same criteria are used for photons of unit SAM ($m_s = \pm1$) and in the case of light, the geometric phase is $\pm \Omega$. When the light passes through a helical fibre whose initial and final positions are parallel, the eigen-state of light is transported round a closed loop [Fig. 6(*a*)] and thereby acquires the geometrical phase shift, which is given by the solid angle, $\Omega$ [Fig. 6(*b*)] on the parameter sphere but not half of the solid angle ($\Omega/2$), which is predicted by Pancharatnam. In the Pancharatnam phase, we consider the change in the direction of SAM with respect to fixed propagation vector, *k* (i.e., changing the state of polarization of light wave while fixing its propagation direction) and in case of helical fibre the propagation vector itself rotate around the axis of fibre helix (the propagation vector of light wave has rotation with the constant state of polarization). The state transformation in the free space propagation with a fixed propagation vector can be considered as a unitary polarization transformation of the SU (2) group, which has three parameters on PS. In the same way, the polarization state transformation as a result of the rotation of the propagation vector about a fixed axis is understood on the sphere of directions with the SO (3) group. Therefore, the interpretation of the geometric phase in both cases is correct but is different. Two years later, another study [45] investigated the geometric phase of light arising from changes in the propagation vector in free space. The authors experimentally observed a phase shift between two beams with opposite circular polarizations, generated in a non-planar Mach-Zehnder interferometer [Fig. 6(*c*)], which agreed well with Berry's formula for the geometric phase [Fig. 6(*d*)].

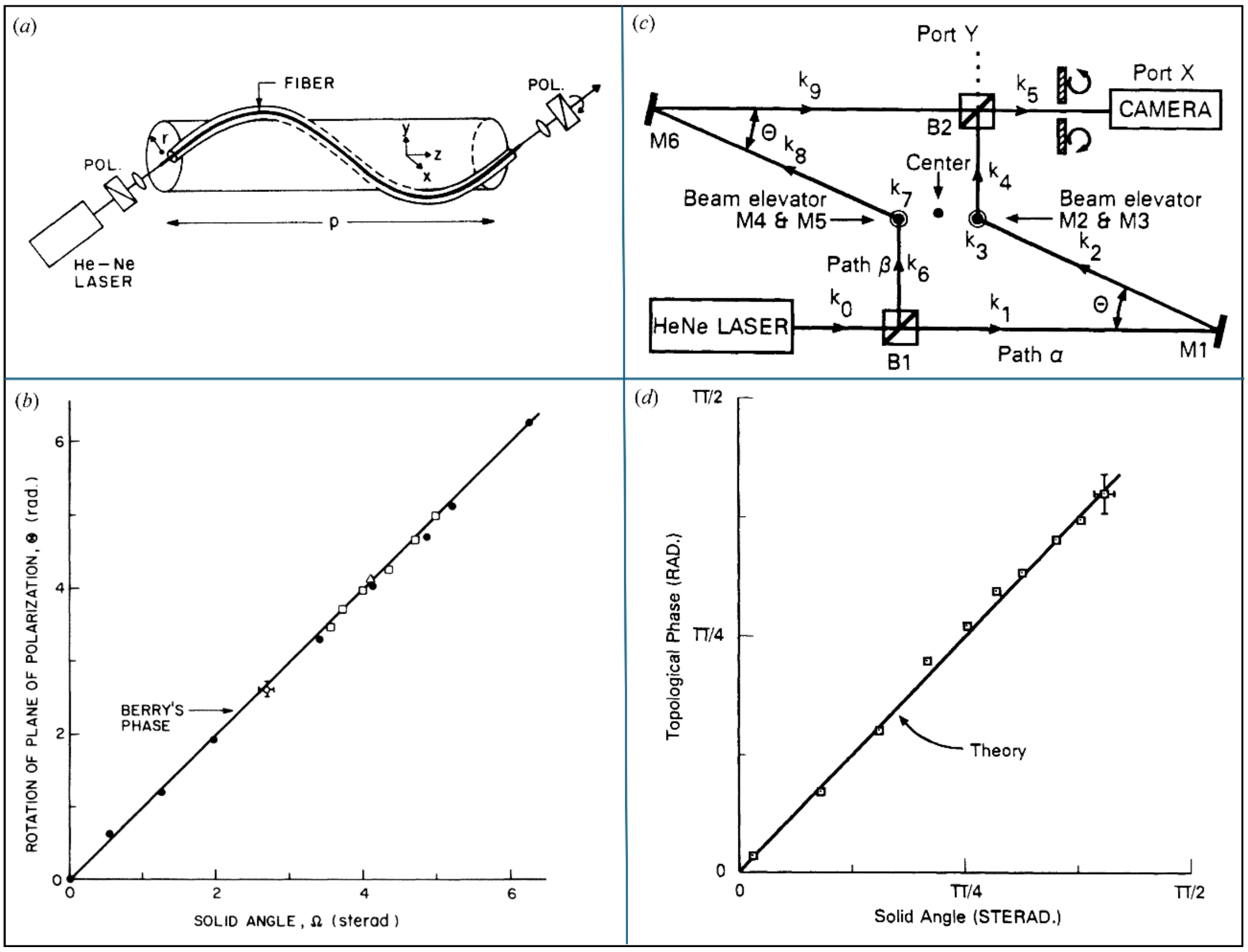

Fig. 6. (*a*) A modal experimental setup used for producing a geometric phase in the light due to the rotation of the plane of polarization of linearly polarized light while it is propagating through a single-mode fibre, which is wound into a helix. (*b*) The plot shows the experimentally observed angle of rotation of linearly polarized light verses calculated solid angle in momentum space. Here, open circles correspond to the data for uniform helices, the squares and triangles represent non-uniform helices, and solid circles represent arbitrary planar paths. The solid line corresponds to the theoretically predicted values based on Berry's formula [5]. (*c*) The schematic diagram shows the top view of a nonplanar Mach-Zehnder interferometer used for creating a geometric phase in the light field. Here, the beams in the upper half of the diagram are at a greater height than in the lower half of the diagram by 41 mm. Here, $k_i$ is the propagation vector of light. (*d*) Experimentally measured topological phase plotted with respect to the solid angle created on the parameter sphere, and it is fitted with Berrys' formula [45].

T. F. Jordan provided a simplified analysis to understand this propagation vector based geometric phase [46]. If a beam of light takes a series of propagation directions of $k_1$, $k_2$, $k_3$,…, $k_n$ and reaches the initial direction of $k_0$, and these directions ($n$ sphere of directions) take place rotation around a normal to the plane containing these propagation vectors. This scenario can be considered as $n$ geodesic arcs on a single parameter sphere whose axis is $k_0$. The solid angle created by the closed loop around the axis of rotation is the geometric phase acquired in the state transformation due to the propagation vector.

### *2.5. M. V. Berry on Pancharatnams' works (connection between Pancharatnams' and Berrys' phases)*

Two years after the publication of the Berry phase, S. Ramaseshan and R. Nityananda (1986) [47] pointed out that the special case of the Berry phase applied to the interference of polarized light had already been studied both theoretically and experimentally in a series of papers published in the 1950s by S. Pancharatnam. They showed that the phase term is proportional to the solid angle on the Poincaré sphere, which represents the states of polarization of light. Subsequently, Ramaseshan and Nityananda discussed Pancharatnam's work in the context of Berry's formulation, and following the publication of several papers highlighting Pancharatnam's contributions, the investigation of the geometric phase in polarization became widely recognized within the global optics community. In 1988, Samuel and Bhandari generalized the Berry phase based on Pancharatnam's work on the interference of polarized light [48]. One of the earliest works explicitly connecting Pancharatnam's phase to the Berry phase was published by M. V. Berry in 1987 [49]. In that study, Berry employed the algebra of spinors and 2×2 Hermitian matrices to establish a precise relationship between Pancharatnam's phase and the adiabatic phase change recently discovered in quantum systems. The key highlights of Berry's paper [49] are outlined below.

The pancharatnams' phase analysis was explained by Berry in a single sentence by the transitive rule as "if $|B\rangle$ is in phase with $|A\rangle$, and $|C\rangle$ with $|B\rangle$, then $|C\rangle$ need not be in phase with $|A\rangle$". In a mathematical expression, it becomes

$$\langle A|D\rangle = e^{-i\Omega_{ABC}/2}. \quad (33)$$

When we transport polarization state in the path of $|A\rangle \to |B\rangle \to |C\rangle \to |D\rangle$ and $|A\rangle$ and $|D\rangle$ have same position vector on the PS (i.e., $\vec{r} = \vec{r}_A = \vec{r}_D$ then $|D\rangle = |A\rangle e^{-i\Omega_{ABC}/2}$). In such a condition, $\Omega_{ABC}$ is the solid angle of the spherical triangle formed by states $|A\rangle$, $|B\rangle$ and $|C\rangle$.

When it comes to the quantum adiabatic phase, the position vector, $\vec{r}$ of a two-state quantum system is driven slowly around a circuit $C$ on the 2-sphere (Bloch sphere). The adiabatic theory based on the Schrodinger equation then shows that if the system is initially in the eigen-state $|A\rangle$ with the greater energy, it will remain in that eigen-state with a phase given by the connection of its time derivative zero [28]. Then the phase shift acquired in the circuit $C$ is given by

$$\langle A|D\rangle = e^{-i\Omega(C)/2} \quad (34)$$

with $|A\rangle$ and $|D\rangle$ have the same position vector on the Bloch sphere. Here, $\Omega(C)$ is the solid angle formed by circuit $C$ at the centre of the sphere. The close similarity achieved between Eq. 33 and Eq. 34 by Berry confirms that Pancharatnams' connection is equivalent to the Berrys' adiabatic connection. Also, the phase acquired between the states $|A\rangle$ and $|B\rangle$ which are connected by a geodesic arc, is obtained by adiabatic connection as

$$\langle A|B\rangle = \cos\phi_{AB}. \quad (35)$$

Where $\phi_{AB}$ is the angle between the states $|A\rangle$ and $|B\rangle$. It is also noted that $\langle A|B\rangle = \cos\phi_{AB}$ is real and positive if $|\phi_{AB}| < \pi$ for a shorter geodesic arc connecting the states $|A\rangle$ and $|B\rangle$.

Although the geometric phase in classical polarization and the geometric phase associated with the quantum spin of an electron are connected through the adiabatic framework, the geometric phase of polarization does not require slow evolution—a fact demonstrated by Pancharatnam through experiments with birefringent crystals.

When a laser beam present in the state $|A\rangle$ is divided into two parts without changing the state of polarization and carried out parallel transport of their states as shown in Fig. 7($a$) and ended with the same point on the 2-sphere with position vector of $\vec{r}_D$. If their final states are considered as $|D_1\rangle$ and $|D_2\rangle$ then these two states represent the same state on the 2-sphere $|D\rangle$ but different phases. Now the phase difference acquired between these two states is equal to half of the solid angle created by the closed loop of ABDC (geometric phase). In the case of light, this phase shift is the geometric phase of polarization on the PS (classical geometric phase) and in the case of an electron beam, this geometric phase is parameterized with the Bloch sphere [50] (quantum geometric phase). The quantum geometric phase, developed between two electron beams, is proportional to the magnetic flux enclosed by them, and was first predicted by Aharonov and Bohm [51,52]. Thus, the polarization phase is an optical analogue of the Aharonov-Bohm effect. As shown in Fig. 7($b$), if a laser beam present in the state $|A\rangle$ is divided into two parts without changing the state of polarization and, while the first beam is left in the state $|A\rangle$, the second beam is transported to $|B\rangle$ and then to $|C\rangle$ and back to the initial position, then its state $|A_1\rangle = |A\rangle e^{-i\Omega_{ABC}/2}$. Thus, the interference between the two beams reveals the phase difference between them as $-\Omega_{ABC}/2$.

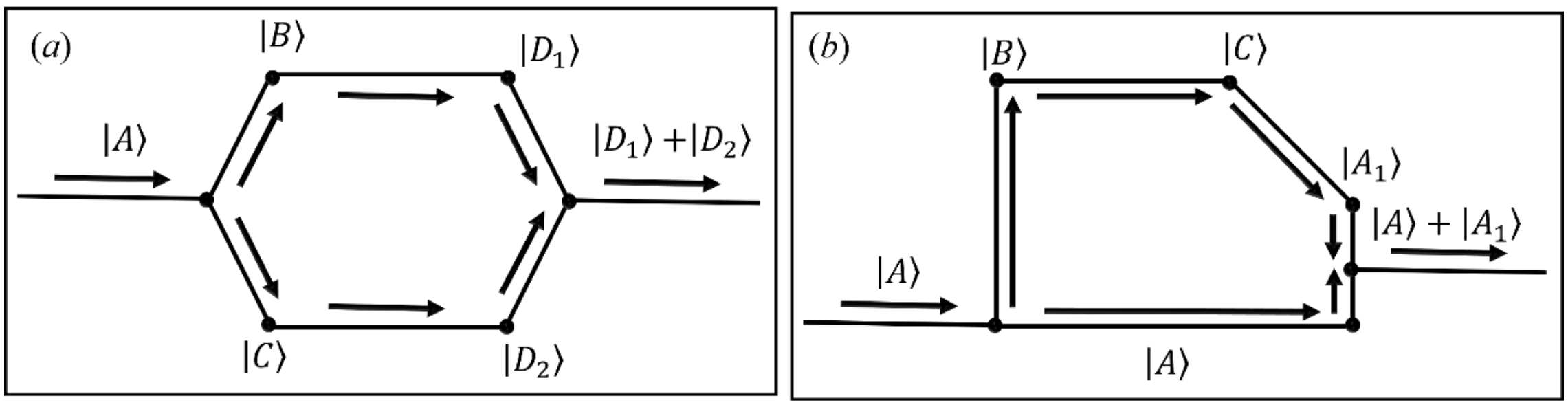

Fig. 7. Aharonov-Bohm effect with polarization of light: (a) polarization state transfer from state $|A\rangle$ to the new states $|D_1\rangle$ and $|D_2\rangle$ by two different paths. Here, the final state vectors represent same point on the 2-sphere with position vector of $\vec{r}_D$ and are separated by a phase difference. (b) An alternative scheme by transporting state $|A\rangle$ in polarization loop and reached to the same point with its state represented as $|A_1\rangle$. Also, $|A_1\rangle$ and $|A\rangle$ represent same point on the 2-sphere with position vector of $\vec{r}_A$ and are separated by a phase difference.

### *2.6. Wave plates as optical gadgets for polarization state transfer on the Poincaré sphere*

The geometric phase is acquired when a polarization state is transported from one point to another on the PS, and this phenomenon can be experimentally realized using optical components. Depending on their functionality, these components induce polarization state transformations on the PS through either unitary or non-unitary processes. The most well-known and widely used optical elements for polarization state manipulation include polarizers, wave plates, and optically active materials. Further details on crystal optics and the evolution of polarization as light propagates through optical elements fabricated from various materials can be found in Refs. [53–57]. The operator for an optical gadget in its eigen-polarization states $|0\rangle$ with phase $\phi_1$ and $|1\rangle$ with phase $\phi_2$ is in the form of

$$\hat{O} = \exp(-i\phi_0)|0\rangle\langle 0| + \exp(-i\phi_1)|1\rangle\langle 1| \quad (36)$$

and the phase difference created between the two eigen-vectors, sometimes called retardance phase $\delta = \phi_2 - \phi_1$. It is worth saying that J. Courtial has used only this operator notation in his mathematical calculations and derived a simple expression for geometric phase [58]. When the operator $\hat{O}$ acts on any polarization state of light wave, the intensity and purity of the state are preserved while it is transformed to another state, then the operator $\hat{O}$ is unitary, and it satisfies the condition $\hat{O}\hat{O}^{\dagger} = \hat{I}$ and here, $\hat{I}$ is unit matrix operator. This operator is an element of the SU (2) group. On the other side if the operator $\hat{O}$ changes the intensity of the two-state system, then it will be a non-unitary operator. This element belongs to the SL (2, C) group. The phase due to the refractive index of isotropic medium is U(1) phase and phase acquired by light while it is propagating through a varying birefringence medium is U(2) phase. Further, more details related to this topic can be found in [55].

#### *2.6.1. Polarizers*

The eigen-states of polarizers are LP states. Polarizers are non-unitary (non-eigen) operators because they attenuate one of their eigen-polarization components and therefore do not preserve the normalization of the state vector [59,60]. For example, a normalized intensity light beam of polarization state $|P(\theta_i, \phi_i)\rangle$, on the PS, passes through the polarizer, split into two orthogonal polarization states of $|P(\theta_{f1}, \phi_{f1})\rangle$ and $|P(\theta_{f2}, \phi_{f2})\rangle$. While it transmits $|P(\theta_{f1}, \phi_{f1})\rangle$ with$\cos(\alpha/2)$ amplitude fraction, it absorbs $|P(\theta_{f2}, \phi_{f2})\rangle$ with $\sin(\alpha/2)$ amplitude fraction completely. Here, α is the angle of the transmittance axis of the polarizer with respect to the incident polarization direction. From Eq. 36, we can obtain the polarizer operator, $\hat{P}(\vec{r}_0)$ which transports any spinor, $|\vec{r}\,\rangle$ to a new spinor, $|\vec{r}_0\rangle$ by the outer product of the destination state, $|\vec{r}_0\rangle\langle\vec{r}_0|$. Thus, $\hat{P}(\vec{r}_0)$ can be written in the matrix form as [37]

$$\hat{P}(\vec{r}_0) = \frac{1}{2}\begin{bmatrix} 1+\cos\theta_0 & \sin\theta_0 \cdot e^{-i\phi_0} \\ \sin\theta_0 \cdot e^{i\phi_0} & 1-\cos\theta_0 \end{bmatrix} = \frac{1}{2}(I + \vec{\sigma}\cdot\vec{r}_0) \quad (37)$$

The working principle of a polarizer on PS is as follows: the incident spinor $|\vec{r}\,\rangle$ splits into orthogonal spinors of $|\vec{r}_0\rangle$ and $|-\vec{r}_0\rangle$ which are antipodal points on the parameter sphere. The polarizer has its transmission axis along the spinor $|\vec{r}_0\rangle$ while eliminating its orthogonal spinor $-|\vec{r}_0\rangle$.

*2.6.2. Wave plates*

Wave plates are fabricated using birefringent crystals with anisotropic refractive indices [61–64]. In the absence of absorption, an incident polarization state splits into two orthogonally polarized components, which are represented on the PS as antipodal points connected by a great circle. These wave plates introduce different phase delays to the two orthogonal polarization components, resulting in a relative phase difference between them. This phase difference leads to a transformation of the polarization state as the light emerges from the wave plate. On the PS, this process is equivalent to a rotation of the sphere about the axis joining the two orthogonal polarization states, or, equivalently, a polarization state transformation about that axis. Various types of wave plates have been developed to achieve polarization state transformation on the PS. For example, a wave plate made of a linearly birefringent material decomposes an arbitrary incident polarization state into two LP components. Such wave plates can realize all SU (2) transformations. The polarization state transformation of light propagating through a birefringent medium can be quantitatively described using the Jones matrix operator, $\hat{R}(\delta)$ which belongs to the SU (2) group [65] and its mathematical form is

$$\hat{R}(\delta) = \begin{bmatrix} e^{i\delta/2} & 0 \\ 0 & e^{-i\delta/2} \end{bmatrix}. \tag{38}$$

Here, the slow-axis of the material is along the horizontal (*x*-axis), and the phase difference between the slow-axis and fast-axis is δ. The arbitrary angular position of the birefringence material can be obtained by applying the rotation operator, $\hat{\Theta}(\alpha)$ of SO (2) group and is given by

$$\hat{\Theta}(\alpha) = e^{-i\alpha\sigma_y} = \begin{bmatrix} \cos\alpha & -\sin\alpha \\ \sin\alpha & \cos\alpha \end{bmatrix}. \tag{39}$$

Therefore, the operator for a birefringence retarder whose slow-axis makes an angle α with the *x*-axis is given by

$$\hat{R}(\alpha, \delta) = \hat{\Theta}(\alpha)\hat{R}(\delta)\hat{\Theta}(\alpha)^{-1}. \tag{40}$$

This operator represents HWP and QWP for the respective phases of $\alpha = \pi$ and $\alpha = \pi/2$. The quantitative phase difference created between the two eigen-states by the linear wave plate, $\delta = 2\pi/\lambda\ (\delta_F - \delta_S)$. Here, $\delta_F$ and $\delta_S$ are the phase retardations created in the respective fast-axis and slow-axis electric field components.

*2.6.3. Operator of retarders in spinor form* [37]

Consider an ideal retarder with eigen-polarizations along its fast- and slow-axes positioned on the parameter sphere are at $\vec{r}_0$ and $-\vec{r}_0$ respectively, and corresponding phases of retarder along the positions $\vec{r}_0$ and $-\vec{r}_0$ are $-\delta/2$ and $+\delta/2$ respectively. Then the retarder operator is given by

$$\hat{R}(\vec{r}_0, \delta) = \frac{1}{2}\left[e^{-i\delta/2}\left(\hat{I} + \vec{\sigma}\cdot\vec{r}_0\right) + e^{+i\delta/2}\left(\hat{I} - \vec{\sigma}\cdot\vec{r}_0\right)\right]. \tag{41a}$$

The final result has the form of

$$\hat{R}(\vec{r}_0, \delta) = \hat{I}\cos\left(\frac{\delta}{2}\right) - i\,(\vec{\sigma}\cdot\vec{r}_0)\sin\left(\frac{\delta}{2}\right). \tag{41b}$$

From above Eq. 41, the operator of HWP is $\hat{R}_h = -i\vec{\sigma}\cdot\vec{r}_h$ and the operator of QWP in the form of $\hat{R}_q = \frac{1}{\sqrt{2}}\left[\hat{I} - i\vec{\sigma}\cdot\vec{r}_q\right]$. These linear birefringent wave plates have their spinor vector, $\vec{r}$ on the equator. Their rotation around the beam axis of $\pi$ is equal to a full rotation of $2\,\pi$ on the parameter sphere. Two polarizers or retarders whose spinor, $\vec{r}$ lie on the same latitude circle on the parameter sphere correspond to identical plates relatively rotated by half their longitude difference. In case of retardance created by wave plates, we can obtain the spherical angles of PS given in Fig. 2(*b*) in terms of the angle between the *x*-axis and the wave plate fast-axis, $\alpha$ (it must be noted that sign conversion changes when we use the slow-axis instead of the fast-axis) as

$$\psi = \frac{1}{2}\tan^{-1}[\tan 2\alpha \cos\delta], \quad (42a)$$

and

$$\chi = \frac{1}{2}\sin^{-1}[\sin 2\alpha \sin\delta]. \quad (42b)$$

Here, we consider $E_{0x} = E_x \cos\alpha$ and $E_{0y} = E_y \sin\alpha$.

*2.6.3. Optical active materials*

In optical active material, the eigen-states are circular polarizations. When a polarized light passes through an optically active material, it rotates the polarization. The rotation of polarization δα can be quantitatively understood as

$$\delta\alpha = \frac{\Delta n L \pi}{\lambda}. \quad (43)$$

Here, $L$ is the length of the interactive medium and $\Delta n = n_{RCP} - n_{LCP}$ with $n_{RCP}$ as a refractive index of RCP and $n_{RCP}$ as a refractive index of LCP. Optically active material can be used as polarization rotating optical gadget.

### *2.7. Development of generalized optical gadgets*

Each of the optical elements discussed above can access only a single path or a limited set of points on the PS for a given incident polarization state. However, this limitation can be effectively overcome by employing a selective combination of multiple optical elements [23,37,65-72].

*2.7.1. Variable circular retarder* [68]

An optical element that rotates an arbitrary polarization state on the PS about the polar axis connecting the RCP and LCP polarization states introduces a variable phase between the RHP and LHP components and is known as a variable circular retarder. The operator of the variable circular retarder is provided by the product of two HWPs

$$\hat{R}_c(\eta, \alpha) = \hat{H}\left(\eta + \frac{\alpha}{4}\right)\hat{H}(\eta). \quad (44)$$

The action of this operator on the polarization state, $|R\rangle$ can be seen as a closed loop formed by two geodesics of ABC (the first HWP transforms $|R\rangle$ to $|L\rangle$) and CDA (the second HWP transforms $|L\rangle$ to $|R\rangle$) [Fig. 8]. This closed path of the polarization is equivalent to the rotation about the axis OA by an angle α/2, and the solid angle enclosed by the loop is α. This gadget for any η is a rotation operator that rotates any point on the PS by an angle α about the circular polarization axis (i.e., η is an azimuth angle). In the present context, the angles given in the parentheses of operators refer to the angle made by fast-axis of the wave plates with respect to the *x*-axis (horizontal) in the real space.

*2.7.2. Variable linear retarder* [68]

If an optical element rotates an arbitrary polarization state on the PS about an axis lying in the equatorial plane and connecting two orthogonal polarization states, the element functions as a variable linear retarder. The operator of the variable linear retarder is synthesised with two QWPs and one HWP in a sequential order of $\hat{Q}\hat{H}\hat{Q}$ as

$$\hat{R}_l(\eta, \alpha) = \hat{Q}\left(\frac{\pi}{4} + \frac{\eta}{2}\right)\hat{H}\left(-\frac{\pi}{4} + \frac{\eta}{2} + \frac{\alpha}{4}\right)\hat{Q}\left(\frac{\pi}{4} + \frac{\eta}{2}\right). \quad (45)$$

In the present scenario, this operation can be understood through the closed loop rotation about axis OB, which is formed by three geodesics [Fig. 8]. The first geodesic curve BC path is covered with the first QWP, and the second geodesic path CDA is covered and reached to $|R\rangle$ by HWP. The final geodesic AB path is finished by the second QWP and accomplish a closed path. This time path is a rotation about OB with the same solid angle of α. The area covered by the closed loop depends on the relative angular positions of the wave plates.

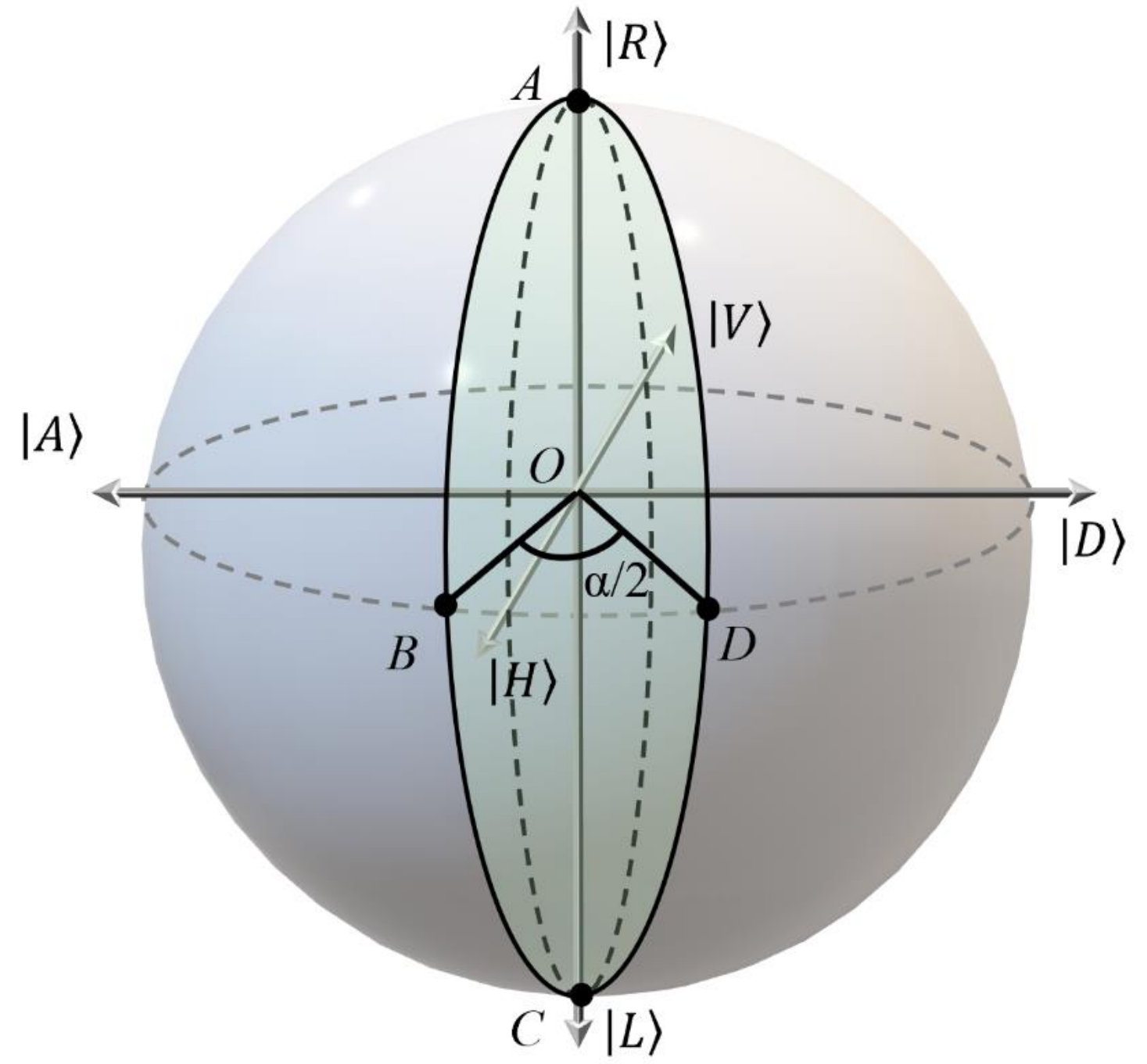

Fig. 8. The operations of variable circular and variable linear retarders can be represented on a single Poincaré sphere as a single closed loop, distinguished by different orientation axes. For a variable circular retarder, the rotation occurs about the OR axis, with both the initial and final states corresponding to right circular polarization. In contrast, for a variable linear retarder, the rotation occurs about the OQ axis, with the initial and final states corresponding to the same linear polarization state.

*2.7.3. Variable general elliptic retarder* [68]

In addition to the two special-case elements described above, Rajendra Bhandari developed a generalized optical element capable of rotating any polarization state on the PS about an arbitrarily chosen axis defined by two orthogonal elliptical polarization states, namely, a variable general elliptic retarder. The variable general elliptic retarder operator for a rotation around a point P($\theta_0$, η) by an angle α is given by the product of $\hat{Q}\hat{H}\hat{Q}\hat{H}$ as

$$\hat{R}_g(\eta,\theta,\phi) = \hat{Q}\left(\frac{\pi}{2}+\frac{\eta}{2}+\frac{\phi}{2}\right)\hat{H}\left(\frac{\theta}{2}-\frac{\pi}{4}+\frac{\eta}{2}+\frac{\phi}{2}\right)\hat{Q}\left(\frac{\eta}{2}+\frac{\phi}{2}\right)\hat{H}\left(\frac{\pi}{4}+\frac{\eta}{2}\right). \tag{46}$$

Here, $\theta$ and $\phi$ are the solutions of the Eqs. below

$$\cos\theta = -\sin\left(\frac{\alpha}{2}\right)\sin\theta_0, \tag{47a}$$

$$\sin\theta\cos\phi = \sin\left(\frac{\alpha}{2}\right)\cos\theta_0, \tag{47b}$$

and

$$\sin\theta\sin\phi = -\cos\left(\frac{\alpha}{2}\right). \tag{47c}$$

This generalized gadget becomes a variable circular retarder and a variable linear retarder for respective $\theta_0 = 0$ and $\theta_0 = \pi/2$. Further, R. Simon and N. Mukunda reduced four wave plates to three wave plates [the lowest number of wave plates necessary for completely realize SU (2) group because it is a three-parameter group] for the construction of a generalized retarder [65]. The generalized retarder can be constructed in three ways by finding a set of selective Euler angles (α, β, γ) as

$$\hat{R}_g(\alpha,\beta,\gamma) = \hat{Q}\left(\frac{\alpha}{2}+\frac{\pi}{4}\right)\hat{H}\left(\frac{\alpha+\beta-\gamma-\pi}{4}\right)\hat{Q}\left(\frac{\pi-\gamma}{4}\right) \tag{48a}$$

or

$$\hat{R}_g(\alpha,\beta,\gamma) = \hat{H}\left(\frac{\alpha+\beta-\gamma-\pi}{4}\right)\hat{Q}\left(\frac{\beta-\gamma}{2}+\frac{\pi}{4}\right)\hat{Q}\left(\frac{\pi}{4}-\frac{\gamma}{2}\right) \tag{48b}$$

or

$$\hat{R}_g(\alpha,\beta,\gamma) = \hat{Q}\left(\frac{\pi}{4}+\frac{\alpha}{2}\right)\hat{Q}\left(\frac{\alpha+\beta}{2}+\frac{\pi}{4}\right)\hat{H}\left(\frac{\alpha+\beta-\gamma-\pi}{4}\right). \tag{48c}$$

*2.7.4. Generalized polarizer in spinor notation* [37]

By a sequential combination of QWP, polarizer, and QWP, we can construct a generalized polarizer whose eigen-polarizations are elliptical. In this combination, the two QWPs must have orthogonal eigen-polarizations of $\vec{r}_q$ and $-\vec{r}_q$. If the angle between the first QWP ($\vec{r}_q$) and the polarizer ($\vec{r}_p$) is α, then the operator of the generalized polarizer is

$$\hat{P}_g = \hat{R}_q\hat{P}\hat{R}_q = \frac{1}{2}\left[\hat{I} + \cos(\alpha\vec{\sigma}\cdot\vec{r}_q) + \sin(\alpha\sigma_z)\right]. \tag{49}$$

*2.7.5. Generalized retarder in spinor notation* [37]

In a similar fashion of above discussion, the operator of a generalized retarder can be constructed by the sequential product of QWP, HWP, and QWP. If the angle between the first QWP and HWP is α (|α| < π) and the angle between HWP and the second QWP is β (|β| < π), then the operator is given by

$$\hat{R}_g = \hat{R}_{q1}\hat{R}_h\hat{R}_{q2} = -\frac{i}{2}\left[\hat{I} - i\vec{\sigma}\cdot\vec{r}_{q2}\right]\vec{\sigma}\cdot\vec{r}_h\left[\hat{I} - i\vec{\sigma}\cdot\vec{r}_{q1}\right] \tag{50}$$

and the parameters of the generalized retarder (retardance phase, $\delta_g$ and polar angle on the parametric sphere, $\theta_g$) are provided by

$$\delta_g = -2\cos^{-1}\left[\frac{1}{2}(\cos\alpha + \cos\beta)\right], \tag{51a}$$

and

$$\theta_g = \cos^{-1}\left\{\frac{\sin\left(\frac{\alpha}{2}\right)\cos\left(\frac{\alpha}{2}\right) + \sin\left(\frac{\beta}{2}\right)\cos\left(\frac{\beta}{2}\right)}{\left[\left(\sin^2\left(\frac{\alpha}{2}\right) + \sin^2\left(\frac{\beta}{2}\right)\right)\left(\cos^2\left(\frac{\alpha}{2}\right) + \cos^2\left(\frac{\beta}{2}\right)\right)\right]^{\frac{1}{2}}}\right\}. \tag{51b}$$

The elliptical eigen-polarizations of the generalized element can be smoothly varied on the parameter sphere by controlling the relative angles between the wave plates. In the special case of $\alpha = \beta$, $\cos\theta_g = \pm 1$, the general retarder has circular polarizations as eigen-states with arbitrary rotation of $\delta_g = -2(\pi - \alpha)$ and for $\alpha = -\beta$, $\cos\theta_g = 0$, it became linear retarder with arbitrary rotation of $\delta_g = 2(\pi - \alpha)$.

*2.7.6. Generalized optical retarder by linear retarder and optical active material* [23,66,67]

When a wave plate exhibits both linear birefringence and optical activity, the incident polarization state is decomposed into elliptically polarized components, which serve as the eigen-polarizations of the compound optical element. G. N. Ramachandran and S. Ramaseshan developed a straightforward theoretical framework based on the PS to describe and understand the evolution of light polarization as it propagates through optically active birefringent materials.

If light with an elliptical polarization represented by point P on the PS passes through an optically active birefringent medium of length *l*, its polarization state changes to point Q on the sphere [Fig. 9]. This rotation occurs about the axis $RR'$, which lies in the plane defined by the horizontal, vertical, right-circular, and left-circular polarization states. Let $\delta_0$ denote the phase retardance per unit length due to the birefringent retarder, and $\rho_0$ represent the rotation of polarization per unit length induced by the optically active medium. The angular position of the axis of rotation is given by

$$2\Theta = \tan^{-1}\left(\frac{2\rho_0}{\delta_0}\right). \tag{52}$$

The rotation taken by the polarization of light about $RR'$ in the unit propagation distance is given in terms of the parameters of preference and the optically active rotation is

$$\Delta_0 = \sqrt{\delta_0^2 + \rho_0^2}. \tag{53}$$

Therefore, the rotation taken by the polarization in its walk from point $P$ to point $Q$ is $\Delta = \Delta_0\, l$. It is also noted that $\Delta_0$ it has an extra term in Eq. 53, however, it can successfully be neglected in any practical applications due to its negligible contribution. In case of pure birefringence, $\rho_0 = 0$ and $2\Theta = 0$. The PS rotation takes place around the axis connecting horizontal and LP states. If the material is optically active but has no birefringence, then $\delta_0 = 0$ and $2\Theta = \pi/2$. It means the sphere rotation takes place about the axis connecting circular polarization states. In this case, the rotation took place along the *EPF* arc.

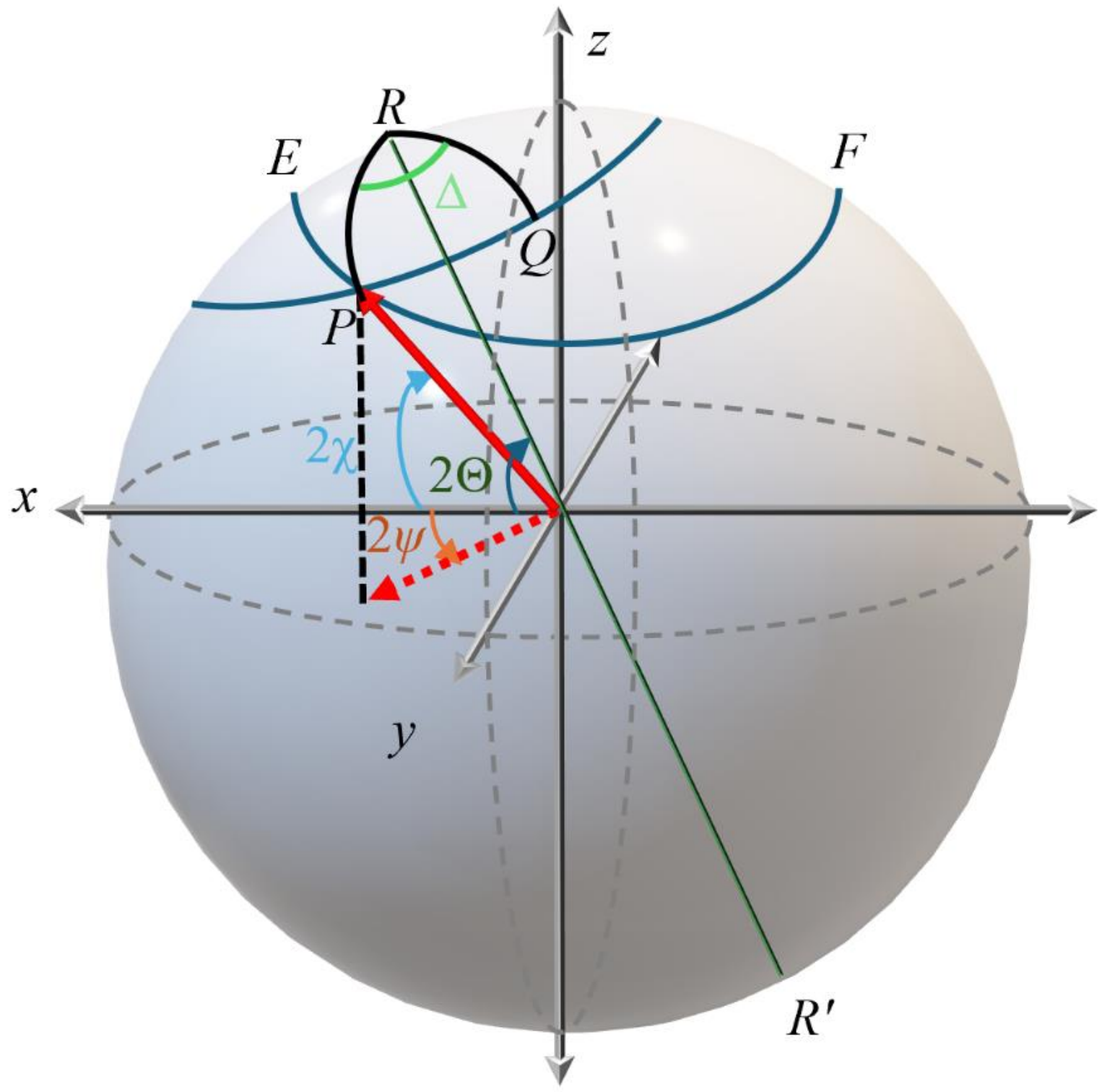

Fig. 9. Polarization state transformation on the Poincaré sphere by passing polarized light through an optical gadget that has both birefringence and optical activity.

*2.7.7. Phase maps with optical gadgets*

R. Bhandari and T. Dasgupta [73,74] successfully employed the optical elements described above to traverse paths on the PS and construct optical circuits [Fig. 10(*a*)]. Two examples illustrate how these gadgets were used to develop circuits and analyze phase evolution. The authors applied a simple classical interferometric technique to measure the phase change, which was plotted as a function of the angle β, corresponding to the rotation of the half-wave plate (HWP). In circuit *ACDA*, the path is completed through successive operations with the optical elements. The observed phase change as a function of β exhibits a triangular-shaped phase map. Here, *AC* and *DA* are geodesic curves, while *CD* is a non-geodesic curve representing rotation around the polar axis. The same phase change was observed for any arbitrary value of θ; the current phase map corresponds to θ = 120° [Fig. 10(*b*)]. Another circuit, *CBEDC*, exhibits a null phase [Fig. 10(*c*)]. This circuit consists of two geodesic arcs, *BC* and *DE*, and two non-geodesic arcs, *CD* and *BE*. Notably, these non-geodesic arcs become geodesic when θ = 90°.

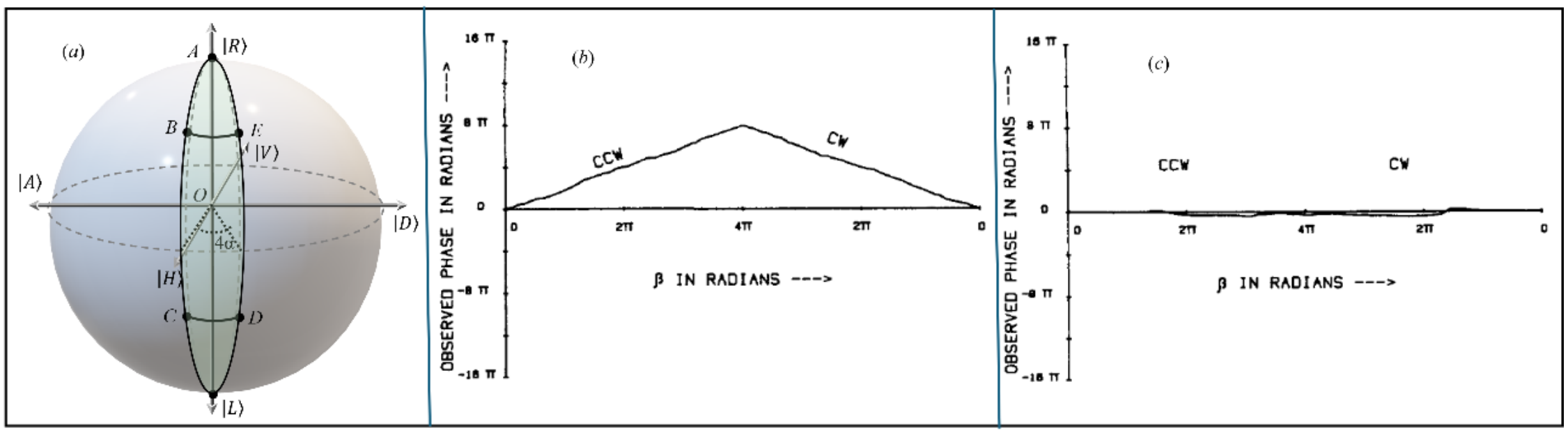

Fig. 10. (*a*) Optical paths traced on the Poincaré sphere by optical gadgets. The observed phase change in the path of the optical circuit created on the sphere is plotted as a function of angle $\beta$ in (*a*) and (*b*) for two different cases. (*b*) Phase map of the *ACDA* circuit created on the Poincaré sphere. The initial polarization state of light is prepared in the RCP state *A* (0°, 0°). First, this state is taken to *C* located at (120°, 0°) by a linear gadget $\hat{R}_l(90^0, 120^0)$ and then to the state *D* (120°, 4$\beta$) by means of a circular gadget $\hat{R}_C(4\beta)$. Finally, the optical path on the sphere is completed in a closed loop of *ACDA* by moving to the initial state *A* (0°, 4β) with the aid of a linear gadget$\hat{R}_l(-90^0 + 4\beta, 120^0)$. The polar coordinate of points *C* and *D* is 120°; the results are independent of polar coordinate and experimentally verified. (*c*) Phase map of the *CBEDC* circuit created on the Poincaré sphere. The initial state *C* is prepared by taking the state of (90°, 0°) to (135°, 0°) with the QWP *Q* (22.5°). The circuit *CBEDC* is completed by four consecutive operators: (i) *Q* (45°), (ii) $\hat{R}_C(4\beta)$, (iii) Q (45° + 4$\beta$), and (iv) $\hat{R}_C(-4\beta)$. (CCW is counter-clockwise direction, and CW is clockwise direction) [73]

***2.8. Geometric phase creation with optical gadgets and its measurement through interferometric methods***

*2.8.1. Geometric and dynamical phases*

When an initial polarization state $|A_i\rangle$ is transferred to a final state of $|A_f\rangle$ by wave plates, then the phase acquired between the two states is given by

$$\gamma_T = arg\langle A_i | A_f \rangle. \tag{54}$$

The total phase $\gamma_T$ acquired in this process is the sum of the geometric phase $\gamma_G$ and the dynamical phase $\gamma_D$ . The optical paths traced on the PS may be either geodesic or non-geodesic arcs. While geodesic arcs contribute only to the geometric phase, non-geodesic arcs contribute to both geometric and dynamical phases [73]. As illustrated in Fig. 11, consider a point *P* ($\theta_0$, $\phi_0$) on the PS. A small rotation producing an arc *AB* generates a non-geodesic closed path, with the arc positioned at an angle $\beta$ relative to the rotation axis *OP*. The angle subtended by the arc *AB* at point *P* is denoted by $\delta$. Rajendra Bhandari has used the Aharonov-Anandan model to derive an expression for the dynamical phase accumulated on the PS while moving from one point to another point with rotation about the axis *OP* as [68]

$$\gamma_D = -(C_1^2 - C_2^2)\delta/2 \qquad (55a)$$

$$\Rightarrow \gamma_D = -\frac{\delta}{2}\cos\beta. \qquad (55b)$$

Here, $C_1$ and $C_2$ are amplitudes/weight-factors of base states (for example, cos $\theta/2$ and sin $\theta/2$ in Eq. 15). When β = π/2, the arc becomes A′B′ and the dynamic phase $\gamma_D = 0$. In addition to the above path, we can also have a path with $\beta = 0$ (*PQ* in Fig. 11), i.e., arbitrary rotations about each of the nodes of the circuit when the state of the beam is represented with that node. The dynamical phase contributed by this path is minus half the rotation angle. The fundamental difference between the geometric phase and the dynamical phase in polarized light can be summarized as follows. The geometric phase is path-dependent on the PS and possesses a topological character; it is additive and unbounded. In contrast, the dynamical phase arises from the birefringence of the optical medium and is defined modulo 2π.

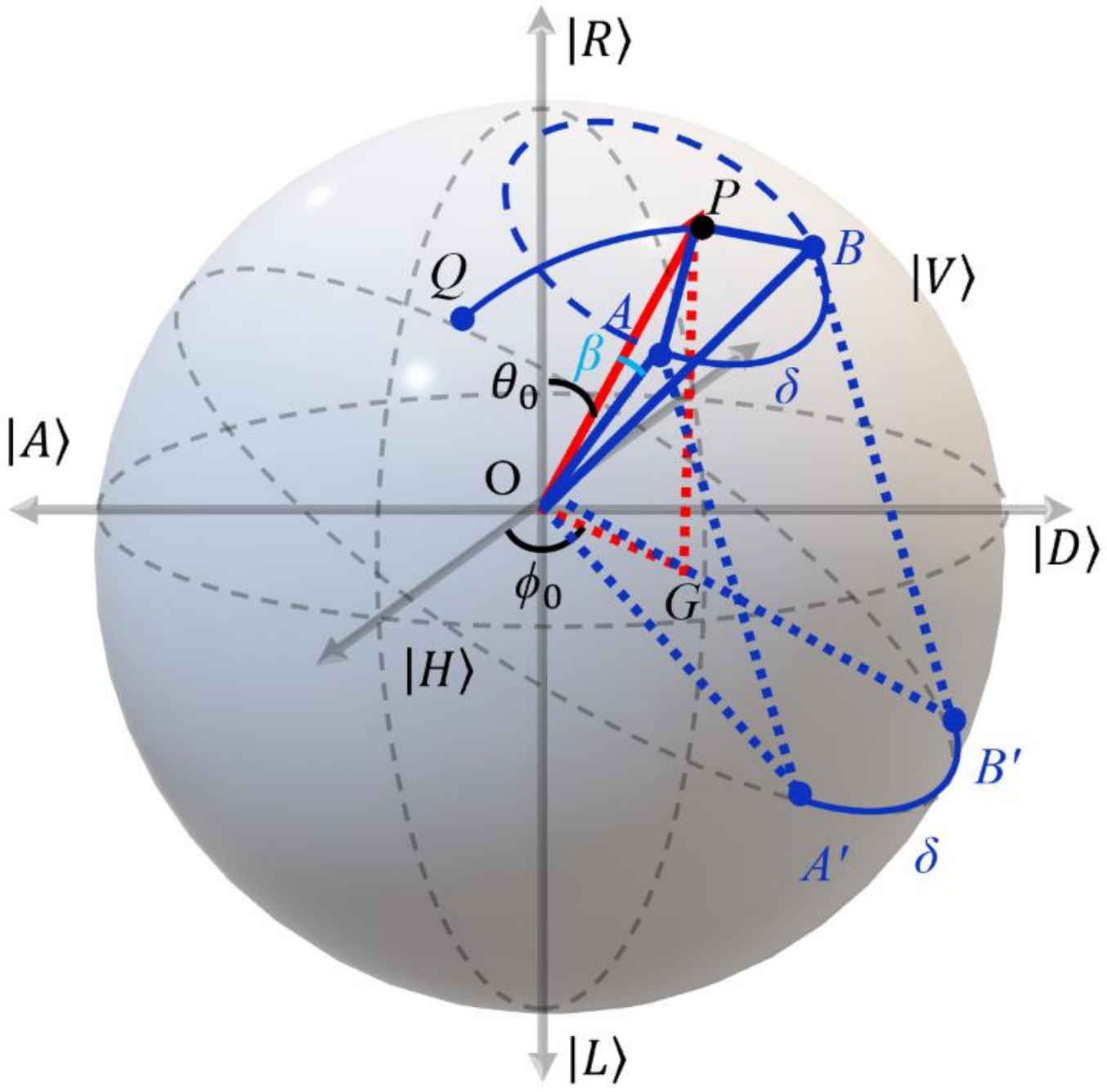

Fig. 11. Rotation around and from the point *P* on the Poincaré sphere and evolution of phase.

An example of the presence of geometric phase and dynamical phase in a single closed loop on the parametric sphere is given in Fig. 12. In this example, *AB* and *CA* are geodesic arcs, and *BC* is a non-geodesic arc. The solid angle created by this closed loop at the centre, $\Omega = \phi_0$ (1-cosθ), and the geometric phase is $\gamma_G = \phi_0 / 2$ (cosθ-1). The dynamical phase is $\gamma_D = \phi_0 / 2$ cosθ. Therefore, the total phase, $\gamma = \phi_0 / 2$ (2cosθ-1). Here, the rotation around *R* by an angle equal to the sum of the solid angle and minus twice the dynamical phase, which is equal to $\phi_0$, i.e., the angle between the two geodesic arcs.

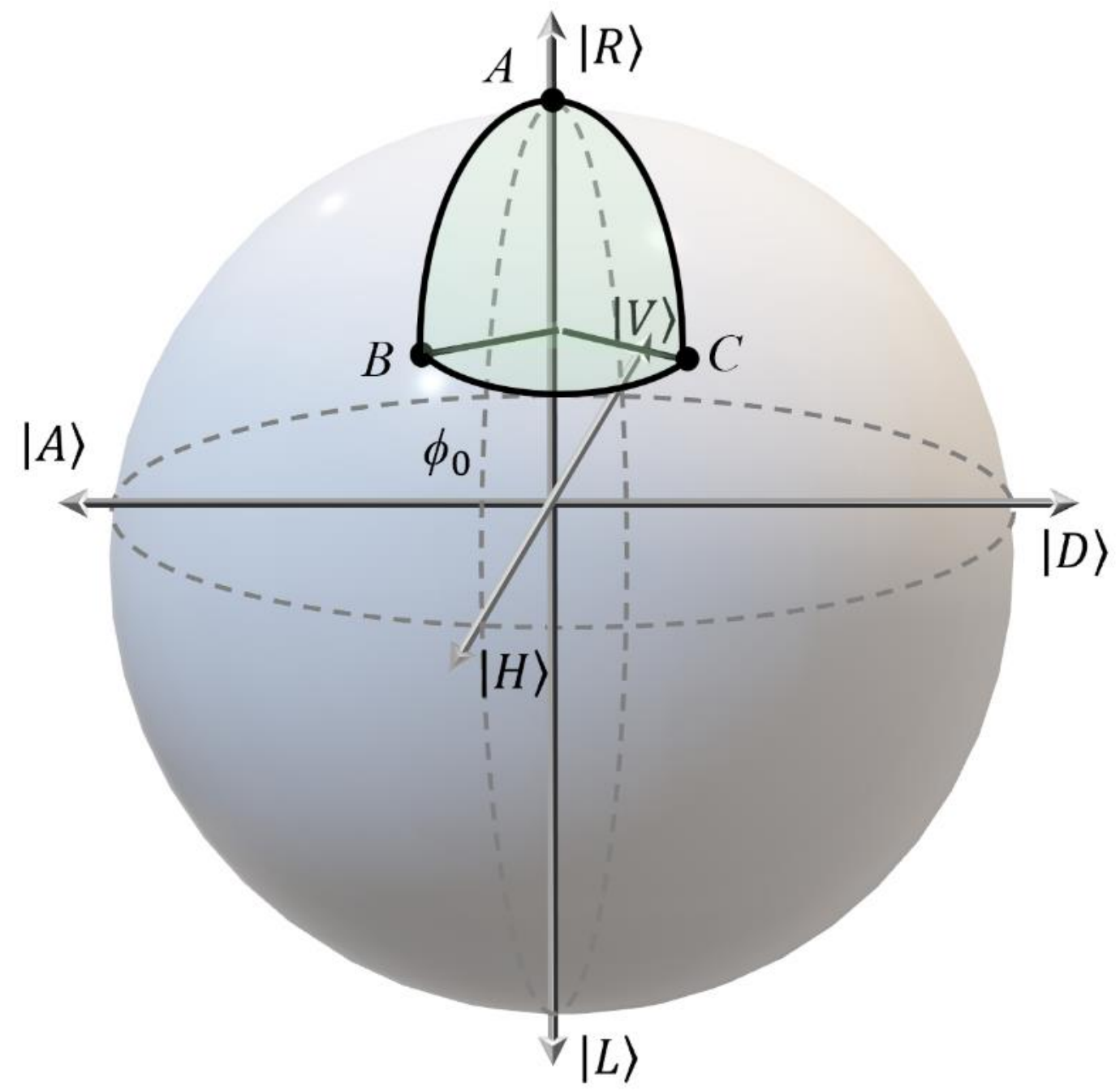

Fig. 12. A closed path on a Poincaré sphere has both geometric phase and dynamical phase.

*2.8.2. Early experiments on geometric phase measurements*

In 1988, R. Bhandari and J. Samuel [75] experimentally demonstrated Pancharatnam's phase using an amplitude-division interferometer. In their setup, the polarization state of one laser beam was taken around a closed circuit on the PS [Fig. 13(a)] using two quarter-wave plates (QWPs) and a polarizer, while the second beam remained in the same polarization state. The interference between the two beams was then used to extract the geometric phase accumulated on the PS.

The polarization of the beam in the first arm, which is horizontally polarized ($x$-polarized) (position $A$), is passed through the first QWP whose fast-axis is oriented at an angle of $\pi/4$ with respect to the $x$-axis. After the first QWP, the polarization becomes RCP, which corresponds to position B. The RCP beam is converted to LP oriented at angle $\alpha/2$ with the horizontal by the second QWP, whose fast-axis makes an angle $\alpha/2-\pi/4$ with the $x$-axis. Then the position of the light on PS is at $C$, which is angularly separated from position $A$ by an angle $\alpha$. The polarization state of the beam from $C$ can be brought back to point $A$ by a linear polarizer whose transmission axis is horizontal [Fig. 13($b$)]. The geometric phase acquired by the beam one is equal to half of the solid angle created by the $ABC$ spherical triangle circuit, and which is equal to $\alpha/2$. The topological phase extracted from the interference is equal to the theoretically predicted half-solid angle created by the circuit on the PS and verified for several angular positions of the second QWP while the QWP position is fixed [Fig. 13($c$)]. Here, the circuit is completed by two unitary transformations (QWP1 & QWP2) and one non-unitary transformation (polarizer). Therefore, R. Bhandari and J. Samuel showed that the assumption of unitarity in the time evolution of the system is not essential for the appearance of the geometric phase.

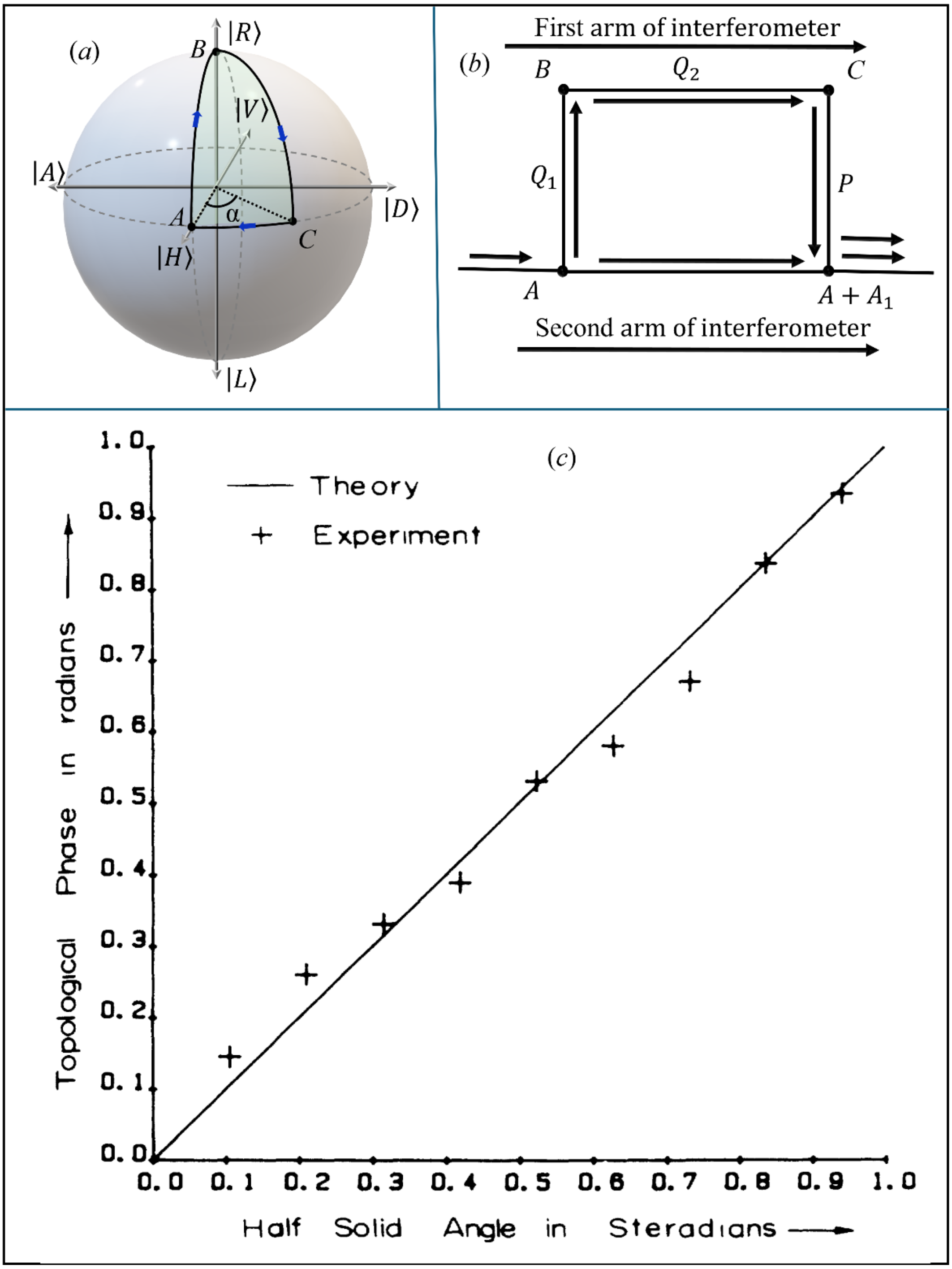

Fig. 13. (*a*) The optical circuit is traced on the Poincaré sphere by two quarter-wave plates and one polarizer. (*b*) Phase map of the closed circuit created in the interferometer. Here, $A$ and $A_1$ represent the same point on the Poincaré sphere but differ by phase. $Q_i$ is $i^{\text{th}}$ quarter-wave plate, and $P$ is the polarizer (*c*) Plot between half-solid angle and topological phase [75].

In the same year, Simon and his co-workers [76] also experimentally measured the Pancharatnam phase of light by a Michelson interferometer experiment with two unitary transformations (two QWPs) and covered the full solid angle on the PS [Fig. 14(*a*)]. In the experiment, the laser beam in one arm is used as a reference source, and in the second arm, they have used two QWPs. The initial polarization of the laser beam is linear and horizontal direction (point $A$). The fast-axis of the first QWP is inclined at an angle $\pi/4$ with the polarization of the laser beam, and the polarization after the first QWP is RCP (point *B*). The second QWP (positioned at an angle of $\beta$ with respect to the original polarization) changes the RCP into LP at an angle of $\beta$ with respect to the horizontal (point *C*). This state, after reflecting from the mirror and being transmitted back through QWP2, changes to LCP (point *D*). Further, this LCP transformed to horizontal polarization and reaches to initial position *A* [Fig 14*b*)]. The interference of this beam with the reference beam produces interference where the extra phase accumulated in the second beam is measured due to unitary transformations. The angle $\beta$ of the second QWP angularly opens the circuit by $2\beta$. The intensity transmitted through a pinhole of a photodiode as a function of $\beta$ is plotted in Fig 14(*c*). The measured geometric phase at interference is equal to $2\beta$ and is the same as half the solid angle created by spherical elliptic circuit *ABCD* on the parameter sphere. In the present experimental configuration, the initial state of *x* - polarization and *y* - polarization produce respective geometric phases of $+2\beta$ and $-2\beta$.

Further, investigated the frequency shift in the laser beam due to the evolution of geometric phase [77]. To do this, authors rotated the second QWP at uniform angular velocity $\Omega = d\beta/dt$ while fixing the angular position of the first QWP. The geodesic arc *DAB* is governed by the first QWP and has remained constant. The second geodesic arc *BCD* is controlled by rotating second QWP around the polar axis *BD* with angular velocity 2 Ω. As a result, the change in the geometric phase is linear in time and provided by $\gamma_G(t) = \gamma_G(0) \pm 2\,\Omega$. Here + and – signs correspond to the respective *x* - polarization and *y* – polarization states of the initial laser beam. The light with angular frequency ω changes to $\omega' = \omega \mp 2\,\Omega$ after one complete round trip. Here, the minus (plus) sign refers to *x* (*y*) polarization. The change in the frequency of light Δω is investigated as a function of the rotational frequency of the second QWP Ω [Fig. 14(*d*)].

*2.8.3. Dynamic phase effect on experimental findings of geometric phase*

H. Schmitzer *et al.* theoretically modelled and experimentally demonstrated the contribution of the dynamical phase arising from optical elements to the geometric phase using a simple Michelson interferometer [78]. From their analysis, when incident light with polarization state $|P_i\rangle$ passes through an optical wave plate with eigen-polarization states $|A\rangle$ and $|A'\rangle$, the polarization state evolves as the light propagates through the crystal, and the output polarization state after passing through the crystal is given by

$$|P_o\rangle = \cos\left(\frac{\widehat{PA}}{2}\right)|A\rangle + \cos\left(\frac{\widehat{PA'}}{2}\right)|A'\rangle. \quad (56)$$

Now, if we incorporate the explicit effect of birefringence on the polarization, then the polarization state after propagating through the crystal by *z* distance is provided by

$$|P_o(z)\rangle = exp[-i\pi z(n_S + n_F)/\lambda]\left[\cos\left(\frac{\widehat{PA}}{2}\right)exp\left(\frac{i\delta}{2}\right)|A\rangle + \cos\left(\frac{\widehat{PA'}}{2}\right)exp\left(\frac{-i\delta}{2}\right)|A'\rangle\right] \quad (57)$$

Here, $\delta = 2\pi z\,(n_S - n_F)/\lambda$ is the retardation phase of the wave plate applied to the light field. The polarization can further be expressed in spinor form as

$$|P_o(z)\rangle = exp[-i\pi z(n_S + n_F)/\lambda]exp\left[\frac{i\delta}{2}\vec{n}\vec{\sigma}\right]|P_i\rangle. \quad (58)$$

Where, $\vec{n}$ is the unit vector in the direction of axis. From the above expressions, it is clearly visualized that the output polarization carries information about the birefringence of the interacting medium, and it can be extracted when we experimentally analyse the output polarization with reference to the input polarization. Indeed, the wave plate produces not only geometric phase but also dynamical phase. To experimentally verify their analysis, the authors used a Michelson interferometer with inserting one HWP with its fast-/slow-axis at an angle $\alpha$ with respect to the incident polarization in one of the arms. The polarization passes through the HWP twice, and it rotates in a closed circle. The solid angle created by this closed loop is $2\pi(1-\cos\alpha)$. Therefore, geometric phase is given by

$$\gamma_G = -\pi\,(1 - \cos\alpha). \quad (59)$$

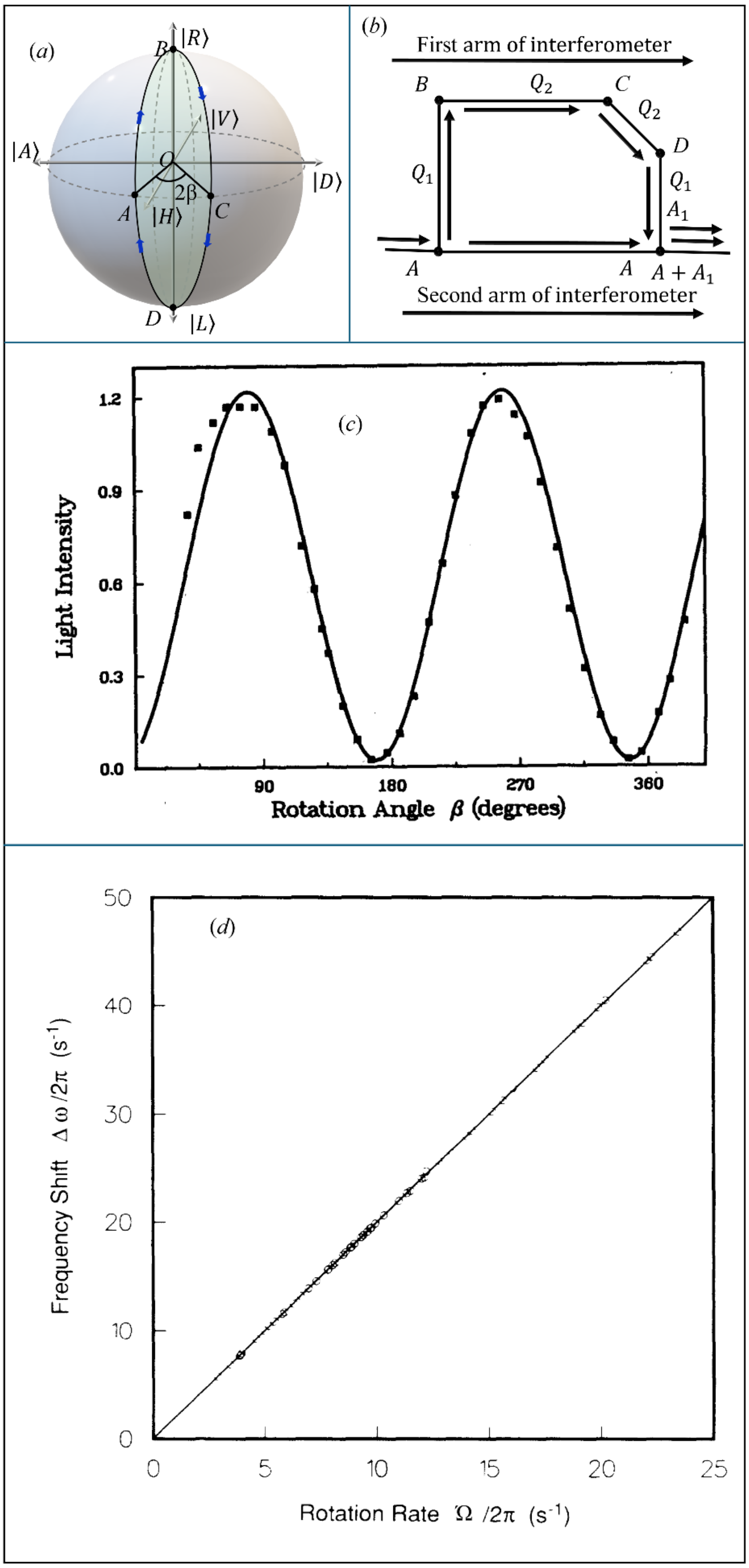

Fig. 14. (*a*) The optical circuit is traced on the Poincaré sphere by unitary transformations of two quarter-wave plates. (*b*) Phase map of the closed circuit created in the Michelson interferometer. $Q_i$ is $i^{th}$ quarter-wave plate. Here, $A$ and $A_1$ represent the same point on the Poincaré sphere but differ by phase. (*c*) Intensity of interference pass through pinhole as a function of rotation angle of second quarter-wave plate [76]. (*d*) Frequency shift in the light field as a function of the rotation angle of the second quarter-wave plate [77].

From the above Eq., the geometric phase with respect to the HWP angle $\alpha$ should vary in cosine form as shown in Fig. 15 (dashed line). However, the experimentally measured fringe position was constant with angle $\alpha$ (scatters). This is exactly matched with the theory of total phase. The total phase acquired by the light beam in this arm with respect to the light beam in the second arm is given by

$$\gamma_T = \frac{-\pi d(n_S + n_F)}{\lambda} + \pi \cos\alpha + \pi\,(1 - \cos\alpha) \tag{60}$$

In the above expression, the total phase is constant and independent of angle $\alpha$. Here, the geometric phase and dynamical phase vary in exactly opposite directions and cancel. The solid line in the plot is from Eq. 60.

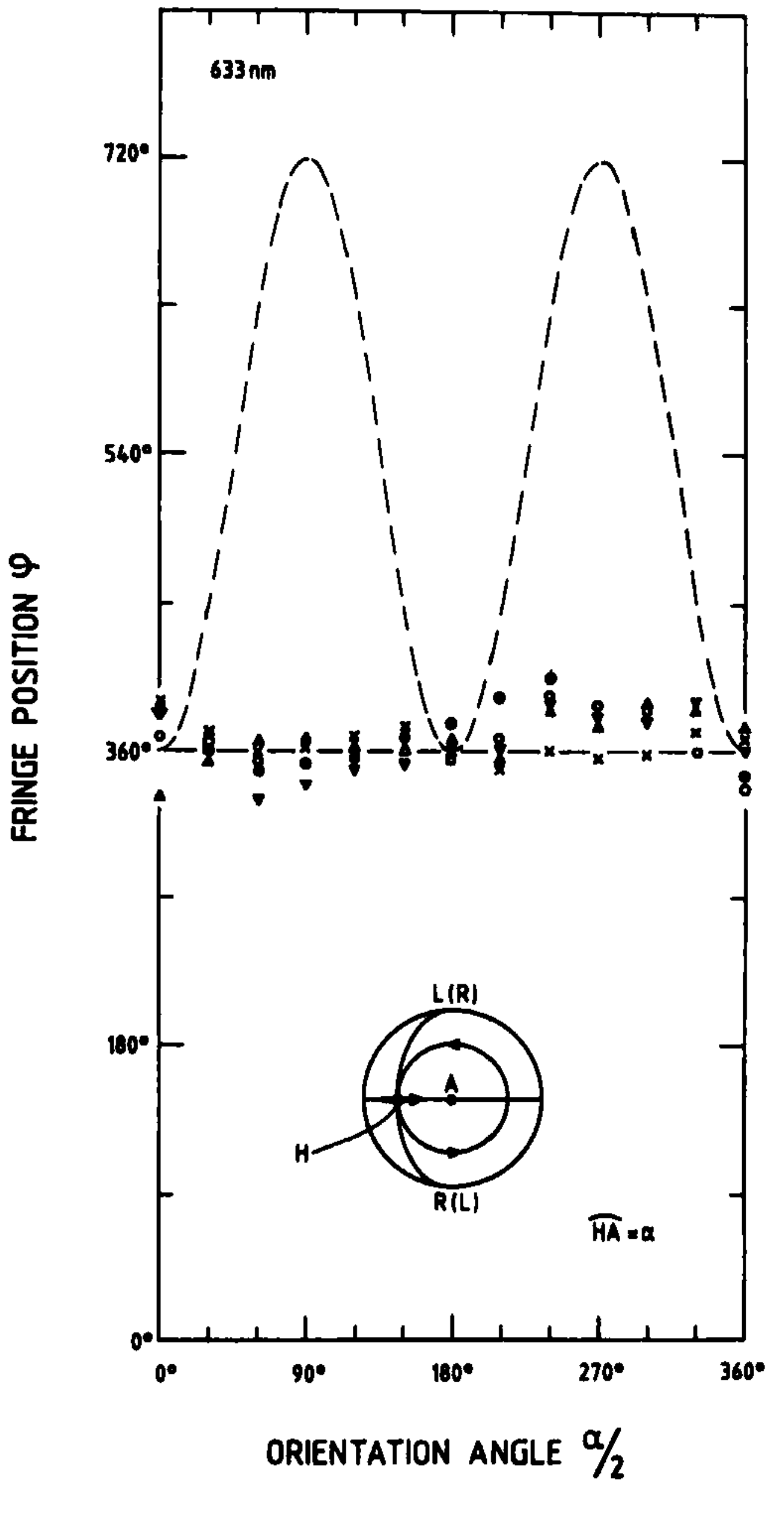

Fig. 15. Fringe shifts of interference in the Michelson interferometer for a closed non-geodesic path on the Poincaré sphere (The paths traced on the spherical surface are provided in the inset) [78].

*2.8.4. Experimental analysis of geometric phase by spinor notation*

M. V. Berry and S. Klein employed spinor notation to theoretically derive expressions for the geometric phase and compared the results with experimentally measured values for two cases: (i) polarizers and (ii) retarders [37]. The geometric phase $\gamma_G$ and the dynamical phase $\gamma_D$ are given by

$$\gamma_D = -\frac{1}{2}\int_0^{2\pi} \vec{r}(\alpha).\vec{r}_0(\alpha)d\alpha, \tag{61a}$$

and

$$\gamma_G = -\frac{\Omega}{2}. \tag{61b}$$

Here, $\alpha$ parametrizes the state, $\vec{r}$ and $\vec{r}_0(t)$ (a piecewise continuous function) is the axis of the crystal plate being traversed, and $\Omega$ is the solid angle of the loop made by $\vec{r}_0(t)$ on the sphere.

*Twisted stacks of polarizers:* When the polarization state, $|A_i\rangle$ is passed through $N$-stack of polarizers and reached to final state of $|A_f\rangle$ and points $\vec{r}_n$ uniformly placed around a latitude circle ($\theta_n = \theta$ and $\phi_n = 2\pi n/N$ with $0 \le n \le N$), then

$$\langle A_i|A_f\rangle = a_N e^{i\gamma_N} \tag{62}$$

with amplitude

$$a_N = \sqrt{\left[1 - \sin^2\theta \sin^2\left(\frac{\pi}{N}\right)\right]^N}, \tag{63a}$$

and phase

$$\gamma_N = -\pi + N\tan^{-1}\left[\cos\theta \tan\left(\frac{\pi}{N}\right)\right]. \tag{63b}$$

The closed circuit on the sphere can be regarded as the loop made of $N$ arcs of great circles [Fig. 16(*a*)]. The state transformation carries through only one of the eigen-polarizations of the polarizer. The area of the loop on the sphere is estimated by elementary spherical geometry and shown that the geometric phase $\gamma_G = \gamma_N$ and dynamic phase $\gamma_D = 0$. For a certain large number of polarizers stack, the geometric phase becomes

$$\gamma_N = -\pi(1 - \cos\theta). \tag{63c}$$

To experimentally realize the geometric phase, a 4-stack of polarizers has been used in one of the arms of the Mach-Zehnder interferometer, and to compensate dynamical phase and attenuation in the light, suitable non-polarizing elements used in the second arm of the interferometer. The interference obtained at the end of the beam combiner was observed through an optical microscope, and it shows the fringe shift by π due to polarizers with respect to the interference created without any polarizers [Fig. 16(*b*)].

*Twisted stacks of retarders:* Let us consider $N$-stack of retarders with their fast-axis eigen-polarizations $\vec{r}_n$ on the latitude circle of radius $\theta$, longitudes $\phi_n = 2\pi\,(n - 1/2)\,/\,N$; ($1 \le n \le N$), and eigen-phases $\pm\delta/2$. This stack can be considered as a single retarder with $\vec{r'}$ and $\delta'$. When the polarization state $|A_i\rangle$ with spinor $\vec{r'}$ passes through this $N$-stack of retarders and reached to final state of $|A_f\rangle$ then

$$|A_f\rangle = |A_i\rangle e^{i\gamma}. \tag{64}$$

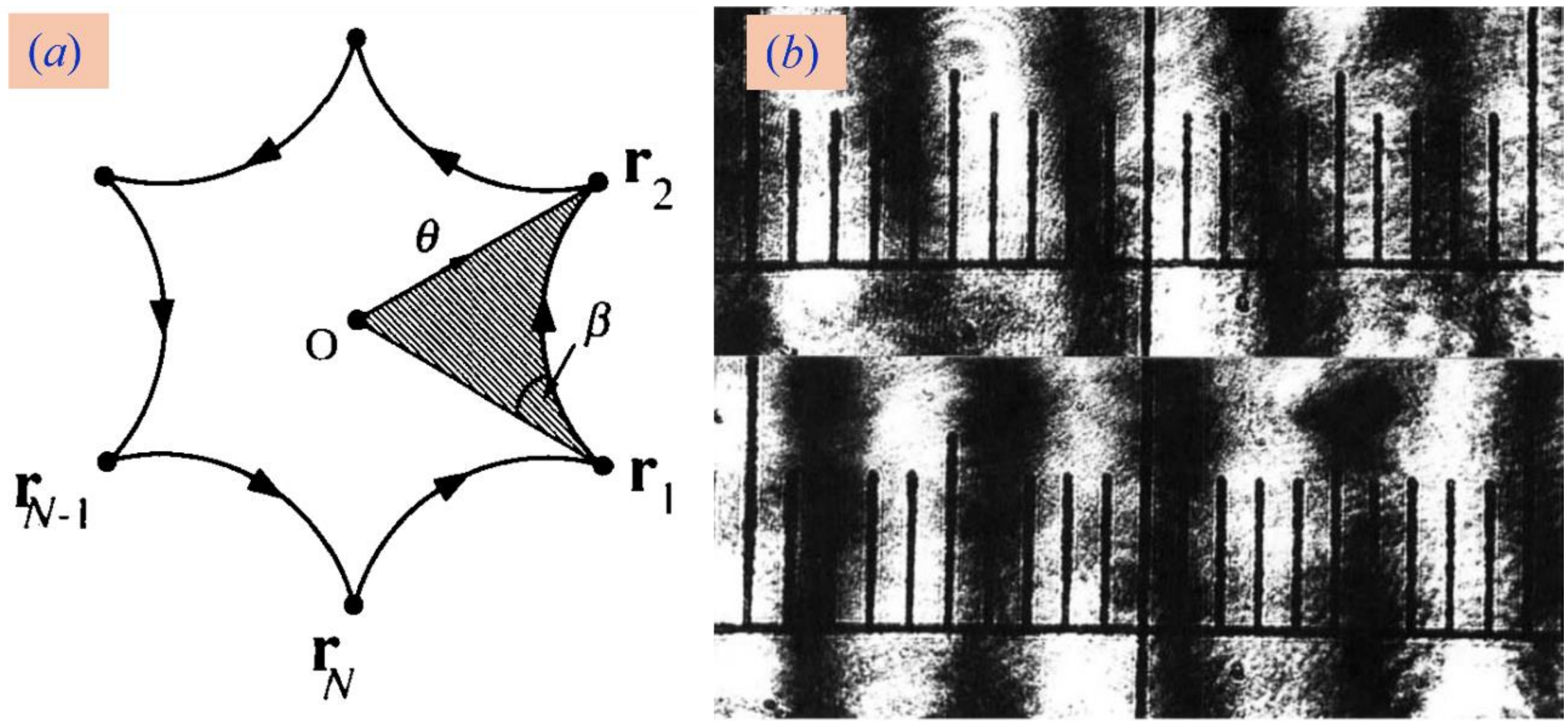

Fig. 16. ($a$) Walk on the Poincaré sphere by passing light of polarization, $|A_i\rangle$ through N-stack of polarizers whose transmitting spinor vectors are $\vec{r}_1$, $\vec{r}_2$, $\vec{r}_3$,….. $\vec{r}_N$, and reaching a final state $|A_f\rangle$ formed a closed loop (i.e., $\vec{r}_1 = \vec{r}_N$). Here, the sphere is projected on a plane, with its north pole represented by O. ($b$) Geometric phase shift of π between fringes from a twisted four-stack of polarizers (top) and an untwisted reference four-stack (bottom); the divisions on the scale are 100 μm apart [37].

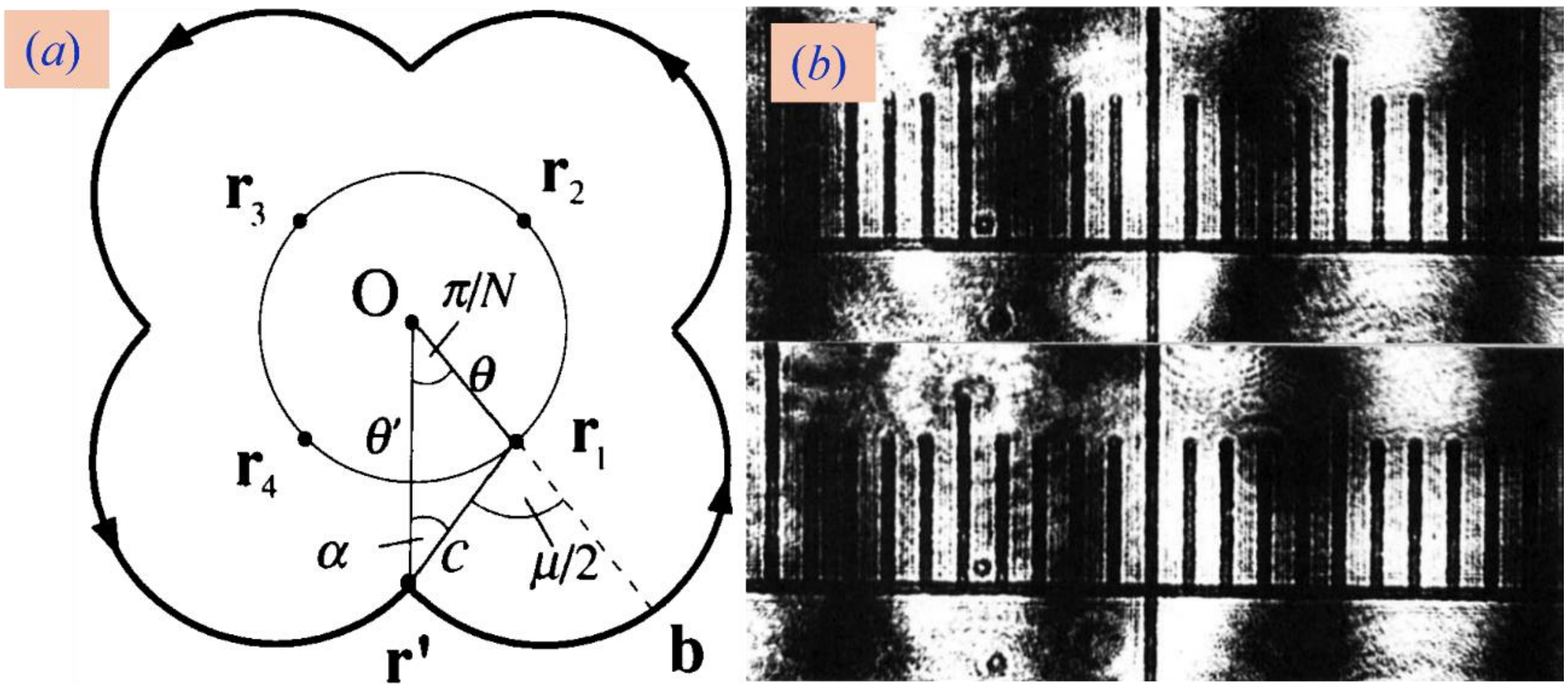

Fig. 17. ($a$) Walk on the Poincaré sphere by passing light of polarization, $|A_i\rangle$ through N-stack of retarders whose transmitting spinor vectors are $\vec{r}_1$, $\vec{r}_2$, $\vec{r}_3$,….. $\vec{r}_N$, and reaching a final state $|A_f\rangle$ formed a closed loop (i.e., $\vec{r}_1 = \vec{r}_N$). The initial polarization is $\overrightarrow{r'}$. Here the sphere is projected on a plane, with its north pole represented by O. ($b$) Phase shift of π/3 between fringes from a twisted four-stack of quarter-wave plates (top) and an untwisted reference four-stack (bottom); the divisions on the scale are 100 μm apart [37].

The path on the parametric sphere made of small circles of radius $c$ with state transformation through two eigen-polarizations and the phase shift, $\gamma = -\delta'/2$ [Fig. 17($a$)]. Indeed, each retarder covers two spherical triangles on the parametric sphere. After straightforward calculations on the total area covered by the polarization state in its transformation, the geometric phase is given by

$$\gamma_G = -\pi + \frac{N\delta}{2}\cos c - N\alpha, \qquad (65a)$$

and the dynamic phase is given by

$$\gamma_D = -\frac{N\delta}{2}\cos c. \tag{65b}$$

Here, $\alpha$ is the angle O$r'r_1$. The total phase acquired after reaching the final state, $\gamma = -\pi - N\alpha$. Authors employed four QWPs in a Mach-Zehnder interferometer to experimentally realize the geometric phase [Fig. 17(*c*)].

*2.8.5. Sagnac interferometer for geometric phase measurement*

P. Hariharan and M. Roy [79] employed a Sagnac interferometer to simultaneously generate two optical circuits on the PS and experimentally measure the associated geometric phase. In a Sagnac interferometer, the optical path lengths traversed by the two interfering beams are identical, except for a phase difference introduced by the polarizing beam splitter (PBS). As a result, dynamical phase differences can be effectively eliminated, allowing the measured phase to be purely geometric. The trajectories of the light beams on the PS and within the interferometer are illustrated in Fig. 18(*a*) and Fig. 18(*b*), respectively.

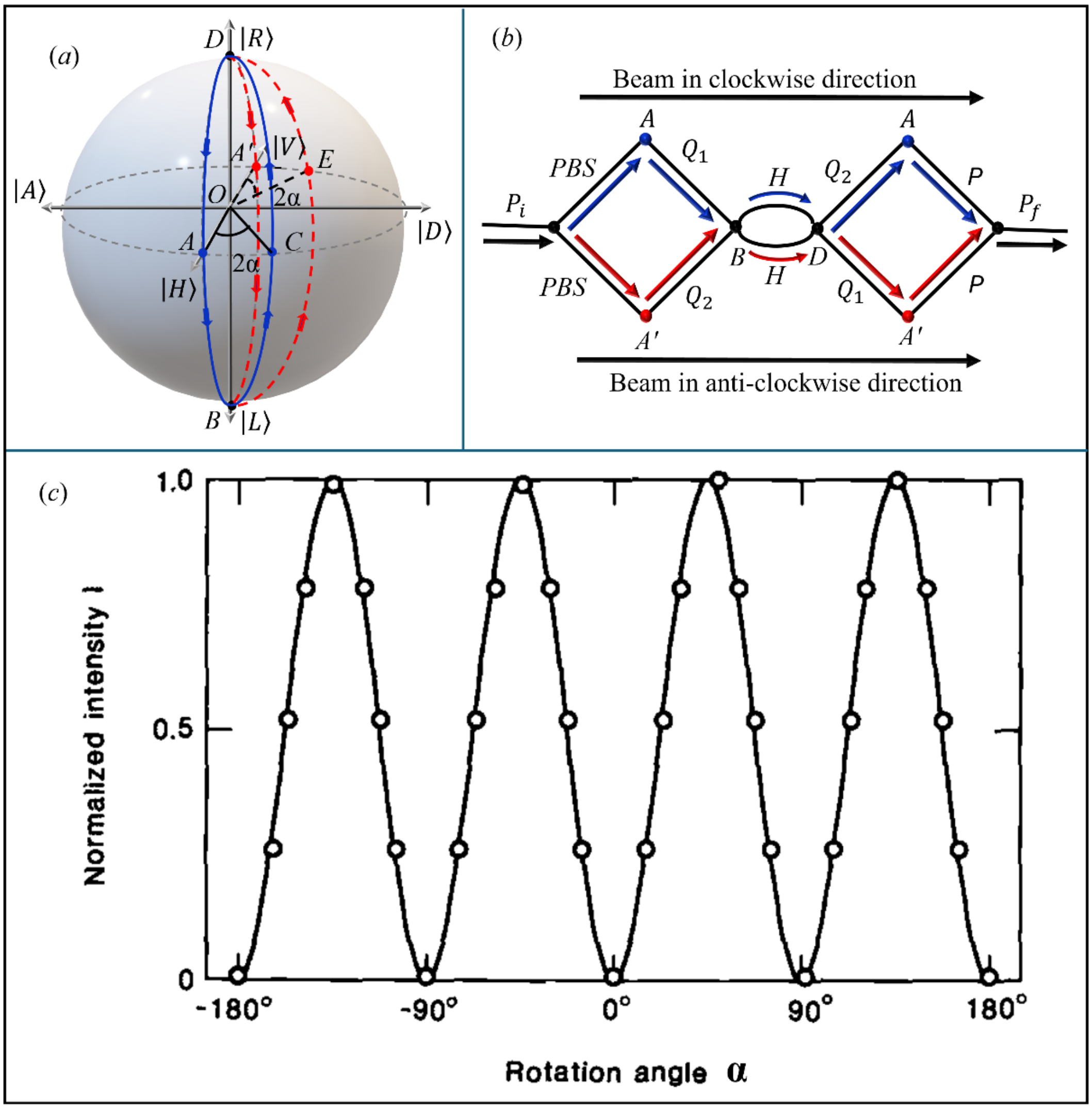

Fig. 18. (*a*) Optical circuits created on the Poincaré sphere by the Sagnac interferometer, which has three wave plates in the order of quarter-wave plate, half-wave plate, and quarter-wave plate, and the corresponding path diagram is shown in (*b*). Here, PBS is a polarizing beam splitter, $Q_i$ is a quarter-wave plate, *H* is a half-wave plate, and *P* is a polarizer. (*c*) Normalized intensity of the interferometer as a function of the rotation angle of the half-wave plate. The sinusoidal curve corresponds to the relation $\Delta\phi = 4\alpha$ [79].

In the Sagnac interferometer, the input laser beam, after passing through the PBS, is split into p- and s-polarized components that propagate in opposite directions and recombine at the same PBS after traversing identical optical path lengths. The p-polarized beam originating at point *A* propagates in the anticlockwise direction and passes through a second quarter-wave plate (QWP), whose fast-axis is oriented at −45° with respect to the p-polarization, converting it into LCP, corresponding to the south pole (point *B*). This LCP state is then transformed into RCP, labelled as point *D*, by a half-wave plate (HWP) whose axis is oriented at an angle $\alpha$ relative to the axis of the first QWP. This path touches the equator at an angle 2 $\alpha$ (point *C*) with respect to point *A*. Finally, the second QWP bring backs to the LP of point *A* and forms a closed loop of *ABCDA* on the PS. The resultant geometric phase in this loop is 2 $\alpha$. In a similar fashion to the above, the s-polarized light reflected from PBS (point *A'*) travels in the clockwise direction, and its polarization completes the *A'BEDA'* circuit on PS after passing through three wave plates. This time, the circuit opened with -2 $\alpha$ angle. As a result, the beam experiences the phase shift of -2 $\alpha$. The geometric phase difference created between the two beams after PBS is 4 $\alpha$. The geometric phase effect experimentally demonstrated by measuring the intensity of interference as a function of HWP angle $\alpha$ [Fig. 18(*c*)]. Here, both the circuits start at different positions on PS, and polarization states travel in different directions, touching each other at polar points, and finally back to their initial position.

### *2.8.6. Youngs' dual-pinhole interference experiment for geometric phase measurement*

Recently, A. Hannonen *et al.* theoretically proposed and experimentally demonstrated a method for measuring the geometric phase using two pinhole laser beams with non-uniform polarization distributions [80,81]. The traditional Young's double-pinhole experimental configuration was modified to produce a periodic variation of polarization across the transverse plane. Within a single period, the polarization undergoes an incremental phase variation, forming a closed loop on the PS. A set of $n$ movable pinholes was periodically positioned at locations $x_n$ (implemented using digital diffractive elements), and measurements were subsequently performed at a separate observation plane. These three planes are located at Fourier planes with the aid of normal convex lenses. The magnitude of the geometric phase at the measurement plane is given by

$$|\gamma_G| = \pi - \pi \frac{|S_{01} - S_{02}|}{|\vec{P}_1 - \vec{P}_2|}. \tag{66}$$

Here, $\vec{P}_i = [S_{0i}\, S_{1i}\, S_{2i}\, S_{3i}]^T, i \in [1,2]$ is stokes vector of $i^{\text{th}}$ pinhole beam on the PS. In the experimental data, the total phase is contributed by both the geometric phase and the dynamical phase. After subtraction of the dynamical phase, the geometric phase is given by

$$\gamma_G = \pm\pi - \sum_{n=1}^{N-1} \gamma(x_n, x_{n-1}). \tag{67}$$

Here, the upper (lower) sign corresponds to $S_{01} / S_{02} < 1$ ($S_{01} / S_{02} > 1$). Through this experiment, it is shown that the geometric phase not only depends on the polarization state but also on the relative intensities of the pinhole beams [Fig. 20(*a*)].

### *2.8.7. Geometric phase due to eigen-states of a single birefringent material*

P. Kurzynowaski et al. theoretically proposed a method based on Jones matrices to understand the geometric phase due to a single birefringent material and experimentally confirmed their theoretical predictions [82]. When an initial polarization state $|A_i\rangle$ is split into two beams. The first one is the reference beam, and the second one passes through the birefringent wave plate of the operator $\hat{R}_\delta(\alpha)$(wave plate rotated at an angle $\alpha$ with reference to the initial polarization of light). The retardance phase along the slow-axis of refractive index, $n_S$ is $\delta_S$ and along the fast-axis of refractive index, $n_F$ is $\delta_F$.The average phase retardance is given by $\bar{\delta} = (\delta_S + \delta_F)/2$, and the difference in the retardance is $\Delta\delta = \delta_S - \delta_F$. The final state, $|A_f\rangle$ after the wave plate is given by

$$|A_f\rangle = \hat{R}_\delta(\alpha)|A_i\rangle. \tag{68}$$

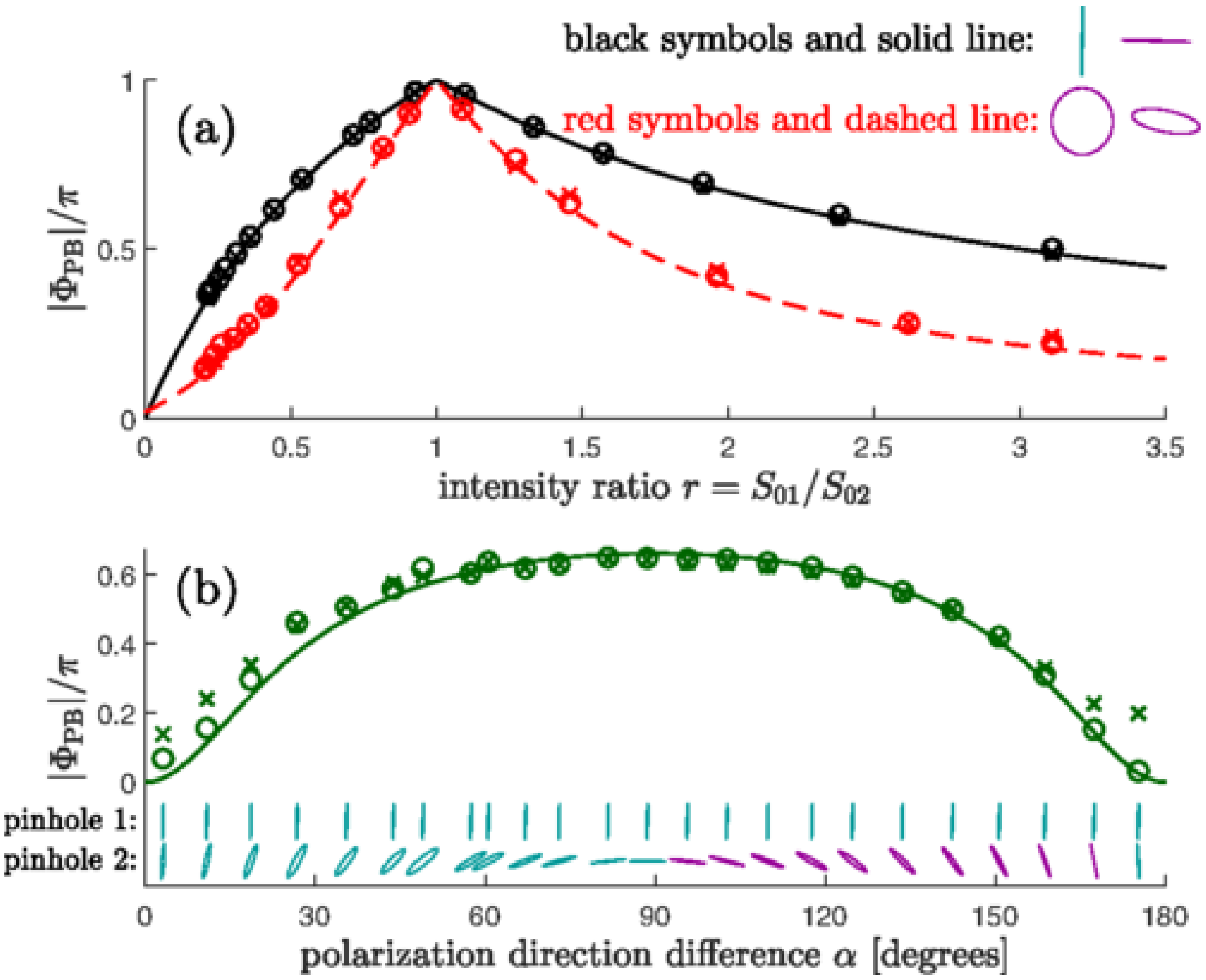

Fig. 19. Experimentally measured and theoretically obtained normalized values of the Pancharatnam-Berry phase through Youngs' two-pin hole beams interference. (a) Polarization states of the pinhole fields are fixed, but the intensity ratio varies. In the black plot, the $y$ and $x$ linearly polarized fields in pinholes 1 and 2, respectively. The red data is for circular and elliptical right-hand polarizations in pinholes 1 and 2, respectively (with different ellipticities). (b) Intensity ratio ($r \approx 0.49$) and the polarization state in pinhole 1 ($y$ linear) are fixed, and the polarization state in pinhole 2 rotates. In all cases, circles and crosses correspond to the measured results, while solid lines present the theoretical values [81].

The projection of the initial state on the final state gives the interference, which contains phase information

$$\langle A_i | A_f \rangle = e^{i\bar{\delta}} \left( \cos\frac{\Delta\delta}{2} + i \sin\frac{\Delta\delta}{2} \cos 2\alpha \right). \tag{69}$$

The phase retardance created between the initial and final polarization states is given by

$$\gamma_T = \bar{\delta} + arctan \left( \tan\frac{\Delta\delta}{2} \cos 2\alpha \right). \tag{70}$$

Here,

$$\gamma_D = \bar{\delta}, \tag{71a}$$

and

$$\gamma_G = arctan \left( \tan\frac{\Delta\delta}{2} \cos 2\alpha \right). \tag{71b}$$

The dynamical phase is determined by the average of the retardances accumulated along the two eigen-states of the birefringent medium and is independent of the orientation of the medium. However, as discussed in the previous subsections, when an optical circuit on the PS is formed by non-geodesic arcs, the dynamical phase becomes dependent on the orientation of the wave plate. In contrast, the geometric phase depends on both the retardance values and the orientations of the wave plates. As illustrated in Fig. 20($a$), light propagating through a birefringent medium split into two orthogonal polarization components. One state is along the fast-axis, $|F\rangle$ and the other state is along the slow-axis, $|S\rangle$. After passing through the medium, they combine and form a single state of light beam whose polarization is determined by the characteristics of the birefringent material, i.e., incident light polarization on the material is split into two polarization states at the input facet, and these two states are parallel transported in the medium and then combined at the output facet to form a single polarization state. Therefore, we have a closed circuit in the medium which is open at $|A_i\rangle$ state and closed at $|A_f\rangle$. This single circuit is split into two spherical triangles. First triangle formed along the fast-axis ($|A_i\rangle \rightarrow |F\rangle \rightarrow |A_i\rangle$ ) and the solid angle given by

$$\Omega_F = \Delta\delta - 2arc\tan\left(\tan\frac{\Delta\delta}{2}\cos 2\alpha\right) \tag{72}$$

with dynamical phase

$$\gamma_D = \delta_F, \tag{73a}$$

and geometric phase

$$\gamma_G = -\frac{\Omega_F}{2}. \tag{73b}$$

The second triangle formed along the slow-axis ($|A_i\rangle \rightarrow |S\rangle \rightarrow |A_i\rangle$ ) with the solid angle given by

$$\Omega_S = \Delta\delta + 2arc\tan\left(\tan\frac{\Delta\delta}{2}\cos 2\alpha\right) \tag{74}$$

with dynamical phase, $\gamma_D = \delta_S$, and geometric phase, $\gamma_G = \frac{\Omega_S}{2}$. The sum of these phases equal to the total phases given in Eq. 70. The total geometric phase can be written in terms of individual solid angles of spherical triangles as

$$\gamma_G = \frac{\Omega_S - \Omega_F}{4}. \tag{75}$$

From this Eq. we can infer that the geometric phase is half of the sum of the geometric phases accumulated along the fast- and slow- axes. The solid angle of the spherical quadrangle formed between the states $|A_i\rangle$, $|A'_i\rangle$, $|A'_f\rangle$, and $|A_f\rangle$ equal to the difference of the above two triangles. Thus, the solid angle of this rectangle is given by $\Delta\Omega = \Omega_S - \Omega_F$. When the incident light polarization is at exactly 45° with either fast-axis or slow-axis (i.e., $\alpha$ = 45°) lead to equal tringles and there will not be any quadrangle, which means no geometric phase is present. Thus, the geometric phase observed in any light beam passing through a birefringent medium with respect to the reference beam (sibling beam) acquires the geometric phase $\Delta\Omega/4$. The geometric phase between the two parallel transported light beams can be obtained by observing the interference at the detector [Fig. 20($b$)]. Linearly polarized light at $\alpha$ = 45° by a polarizer passes through a horizontally rotated elliptical Wollaston Compensator (EWC) split the incident LP state into its eigen-polarization states (elliptical polarization states). These two states again combined to form a single state by analyzer whose transmission axis is orientated at $\alpha_A$. As a result, we have a closed circuit on the polarization PS with its opening angle 2 $\alpha_A$ [Fig. 20($c$)]. The intensity pattern after the analyzer is given by

$$I(x) \propto 1 + v.\cos[\Delta\delta(x) + \gamma_G]. \tag{76}$$

Here, $\Delta\delta(x)$ is phase rotation created by EWC. The geometric phase is given by

$$\gamma_G = arc\tan(\tan 2\alpha_A \sin 2\,\theta_F) = \frac{\Omega}{2} = \frac{\Omega_S}{2} + \frac{\Omega_F}{2}. \tag{77}$$

Here, $\theta_F$ is the ellipticity angle of eigen-states of EWC. In the present case, the geometric phase is given by the sum of the geometric phases of the individual laser beams.

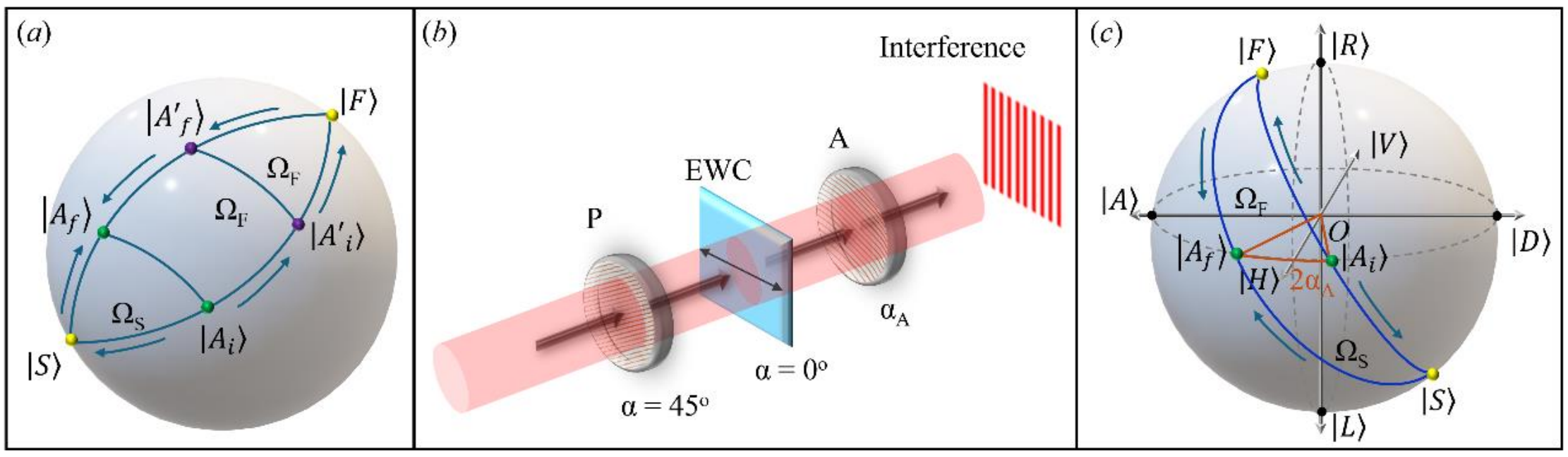

Fig. 20. (*a*) Splitting and parallel transport of the incident polarization state of light via eigen-states of birefringent material and ending up with a single state and forming a closed circuit on the parameter sphere. (*b*) Schematic diagram of experimental setup used for creation and detection of geometric phase due to parallel transport of polarization state in the wave plates. (*c*) Closed loop created on the polarization Poincaré sphere through parallel transport of initial polarization through eigen-polarization states of the elliptical wave plate. P is a polarizer, EWC is an elliptical Wollaston Compensator, A is analyzer, and $\Omega_i$ is the solid angle. Here, S stands for slow-axis, and F stands for fast-axis.

### *2.8.8. Geometric phase due to non-cyclic polarization changes*

In the previous examples, we discussed the geometric phase generated by cyclic polarization transformations composed of geodesic and non-geodesic arcs. The geometric phase can also be quantitatively defined for non-cyclic transformations, in which the initial and final states on the parameter sphere are different. T. van Dijk *et al.* [83] experimentally measured the PB phase arising from such non-cyclic polarization transformations on the Poincaré sphere. The generalized state of the PS is taken as

$$|P_g\rangle = \cos\theta_g|0\rangle + \sin\theta_g e^{i\phi_g}|1\rangle \tag{78}$$

with $0 \leq \theta_g \leq \pi/2$, and $-\pi \leq \phi_g \leq \pi$. As shown in Fig. 21(*a*), let state $|P_A\rangle$ correspond to point *A* pass through a linear polarizer whose transmission axis is at an angle of $\alpha_1$ with the positive *x*-axis, and the output state reaches point *B*, which is present at the equator. Now this state passes through QWP whose axis at $\alpha_1 - \pi/4$ to transform into LCP (point *C*). Again, this state passes through one more polarizer whose transmission axis is at $\alpha_2$ with respect to the positive *x*-axis. As a result, the state transforms to point *D*. Finally, this transforms to RCP and reach to final point *E*. This state transformation can be mathematically expressed as

$$|P_E\rangle = \hat{Q}\left(\frac{\pi}{2}, \alpha_2 - \frac{\pi}{4}\right)\hat{P}(\alpha_2)\hat{Q}\left(-\frac{\pi}{2}, \alpha_1 - \frac{\pi}{4}\right)\hat{P}(\alpha_1)|P_A\rangle. \tag{79}$$

By using Jones matrix formulism, the geometric phase achieved in this configuration is given by

$$\gamma_G = \arg\left[T(A, \alpha_1)e^{i(2\alpha_2 - \alpha_1)}\right] \tag{80}$$

with transmission matrix of initial state

$$T(A, \alpha_1) = \frac{\cos\theta_A\cos\alpha_1 + \sin\theta_A e^{i\phi_A}\sin\alpha_1}{|\cos\theta_A\cos\alpha_1 + \sin\theta_A e^{i\phi_A}\sin\alpha_1|}. \tag{81}$$

In order to investigate and compare the geometric phase due to cyclic and non-cyclic changes in the polarization state transformations, four different conditions are discussed. The first case is the initial state, and the final states are the same and are at the north pole. In this case, the geometric phase $\gamma_G = 2(\alpha_2 - \alpha_1)$. Now, the second polarizer transmitted axis is fixed at $\alpha_2 = 0$, and the geometric phase is investigated as a function of first polarizer angle, $\alpha_1$ [blue plot in Fig. 21(*b*)]. We can clearly see that the geometric phase is double the polarizer angle, and the variation is linear. When we consider the initial state on the upper hemisphere with Stokes vector (0.99, −0.14, 0.07), the change in the geometric phase is nonlinearly dependent on the polarizer angle [red plot in Fig. 21(*b*)]. The sudden jump in the geometric phase can be seen at the first polarizer angle, $\alpha_1 = 90^\circ$. At this position, the points *A* and *B* are antipodal, and a singularity in the geometric phase is created. When *A* is at the south pole, the geometric phase, $\gamma_G = 2\alpha_2$, and there is no effect of the first polarizer on it. Hence, the geometric phase is zero line with respect to the first polarizer angle [blue plot in Fig. 21(*c*)]. If we consider the point *A* on the lower hemisphere but not at the pole with Stokes vector (0.93, 0.23, −0.28), then the geometric phase varies nonlinearly with respect to the first polarizer angle [red plot in Fig. 21(*c*)].

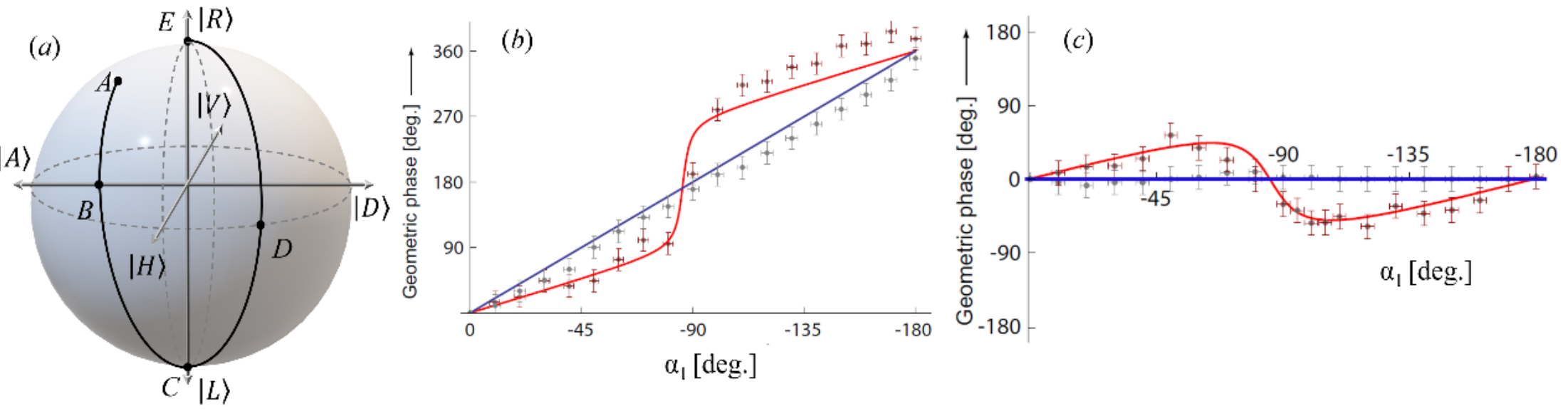

Fig. 21. Tracing a non-cyclic path on the on the Poincaré sphere by passing polarized light through a sequence of wave plates. (*a*) Non-closed path *ABCDE* of polarized light on the Poincaré sphere. Geometric phase accumulated on the Poincaré sphere while tracing the path of *ABCDE* as a function of angular position of first polarizer ($\alpha_1$) for four cases: (*b*) when the initial state *A* coincides with the north pole (blue curve), and when *A* lies between the equator and the north pole (red curve) and (*c*) when the initial state *A* coincides with the south pole (blue curve), and when *A* lies between the equator and the south pole (red curve). In all four cases, $\alpha_2 = 0$. Solid lines corresponding to theoretical curves obtained by Eq. 81 for experimental conditions [83].

## 3. Geometric phase in spatially structured modes

In the preceding section, we presented a detailed discussion of the theoretical framework and key experimental results related to the geometric phase arising from polarization transformations. All of those discussions were based on the spin/polarization PS. In an analogous manner, a geometric phase can also be observed in transformations of the transverse phase of higher-order laser modes. Such spatial transformations can be quantitatively represented on the orbital/modal PS, which was first theoretically proposed by M. J. Padgett and J. Courtial [84]. In this section, we introduce the modal PS and compare it with the polarization PS, as the strong analogy between the two facilitates a clear understanding of the geometric phase in modal space. Mode transformations between Laguerre–Gaussian (LG), Hermite–Gaussian (HG), and Hermite–Laguerre–Gaussian (HLG) modes allow access to all points on the modal PS [85]. In polarization optics, all polarization states can be constructed either by superposition of LP states (linear basis) or by superposition of circular polarization states (circular basis). More generally, any pair of orthogonal elliptical polarization states may serve as a basis; however, the linear and circular cases are highlighted here for ease of comparison with spatial modes. In an analogous fashion, a wide variety of spatial modes can be constructed on the modal PS through superposition of either HG modes or LG modes. The mathematical expression for the HG mode is given in rectangular coordinates (*x*, *y*) as

$$HG_{m,n}(x,y,z) = \frac{1}{w(z)\sqrt{2^{m+n-1}\pi m!\,n!}} H_m\left(\frac{\sqrt{2}x}{w(z)}\right) H_n\left(\frac{\sqrt{2}y}{w(z)}\right) exp\left(-\frac{r^2}{w^2(z)}\right) exp\left(-\frac{ikr^2}{2R(z)}\right) exp(-ikz)\, exp[i\Phi(z)] \tag{82}$$

with $N = m + n$ is the mode number of HG modes, and $\Phi(z)$ is the Gouy phase. At position *z*, $w(z)$ is the Gaussian spot size and $R(z)$ is the radius of curvature. The fractal phase variation along the *x* and *y* axes produces rectangular phase variation [Fig. 22]. The fractals along the *x* and *y* directions can have infinite values, i.e., $m = 0, 1, 2, 3, \ldots, \infty$ and $n = 0, 1, 2, 3, \ldots \infty$. The equation for the LG mode in cylindrical coordinates ($r$, $\varphi$) is given by

$$LG_{p,\ell}(r,\varphi,z)=\frac{1}{w(z)}\sqrt{\frac{2p!}{\pi(p+|\ell|)!}}\left(\frac{\sqrt{2}r}{w(z)}\right)^{|\ell|}L_p^{\ell}\left(\frac{2r^2}{w^2(z)}\right)exp\left(-\frac{r^2}{w^2(z)}\right)$$
$$exp\left(-\frac{ikr^2}{2R(z)}\right)exp(-ikz)\,exp(-i\ell\varphi)\,exp[i\Phi(z)]. \tag{83}$$

Here, the order of the LG mode is defined in terms of its mode number $N = 2p + \ell$. The parameters $p$ and $\ell$ represent the respective radial and azimuthal indices. The transverse phase variation of LG modes along radial and azimuthal direction are quantitatively expressed in respective radial index, $p$ = 0, 1, 2, 3, …, ∞ and azimuthal index, $\ell$ = - ∞, …, -3, -2, -1, 0, 1, 2, 3, …., ∞. The azimuthal phase variation around the beam axis produces a phase singularity at the centre of the LG mode or on the beam axis [Fig. 22]. The radial phase variation creates circular intensity rings around the beam axis. The HG and LG modes have an infinite number of modes; however, they form a Hilbert space with completeness of their bases, i.e., $\sum_{m=0}^{\infty}\sum_{n=0}^{\infty}|HG_{m,n}\rangle\langle HG_{m,n}|=I$ and $\sum_{p=0}^{\infty}\sum_{\ell=-\infty}^{\infty}|LG_{p,\ell}\rangle\langle LG_{p,\ell}|=I$. By utilizing their completeness, we can write LG modes in terms of HG modes [86,87]

$$|LG_{p,\ell}\rangle=\sum_{m=0}^{\infty}\sum_{n=0}^{\infty}|HG_{m,n}\rangle\langle HG_{m,n}|LG_{p,\ell}\rangle \tag{84}$$

and HG modes in terms of LG modes

$$|HG_{m,n}\rangle=\sum_{p=0}^{\infty}\sum_{\ell=-\infty}^{\infty}|LG_{p,\ell}\rangle\langle LG_{p,\ell}|HG_{m,n}\rangle. \tag{85}$$

The generalized expression for the projection of one basis mode on the other basis mode is given by

$$\langle HG_{m,n}|LG_{p,\ell}\rangle=\langle LG_{p,\ell}|HG_{m,n}\rangle^{*}=\begin{cases} i^m b\left(\frac{N+\ell}{2},\frac{N-\ell}{2},n\right) & : \ 2p+|\ell|=m+n \\ 0 & : \ 2p+|\ell|\neq m+n \end{cases} \tag{86}$$

$$b(m',n',n)=\sqrt{\frac{(m'+n'-n)!\,n!}{2^{m'+n'}m'!\,n'!}}\frac{1}{n!}\frac{d^{n'}}{dt^{n'}}[(1-t)^{m'}(1+t)^{n'}]_{t=0}. \tag{87}$$

This polynomial coefficient provides the connection between HG and LG modes. The relation between LG and HG modes is, in terms of their mode indices, given by $p$ = min ($m$, $n$) and $\ell$ = | $m$-$n$ |. Hence, we can transform HG to LG modes and vice versa through the mode converters. For a given mode number $N$, we have multiple HG and LG modes, and this degeneracy in the modes increases with the mode number. The major advantage of HG and LG beams is that they are eigen-modes of the laser cavity and have a self-similarity property, i.e., their structure is propagation independent, and variation in their intensity follows the scaling factor. Furthermore, we can obtain the matrix elements for the basis transformations between HG and LG modes for any particular mode number $N$ [86]. The elements of the matrix that change the HG basis to the LG basis at mode number $N$ are given by

$$[B_{HG\to LG,N}]_{i,j}=\langle LG_{p_i,\ell_i}|\,HG_{m_j,n_j}\rangle \tag{88a}$$

and the LG basis transformation to the HG basis is provided by

$$[B_{LG\to HG,N}]_{i,j}=\langle HG_{m_i,n_i}|\,LG_{j,\ell_j}\rangle \tag{88b}$$

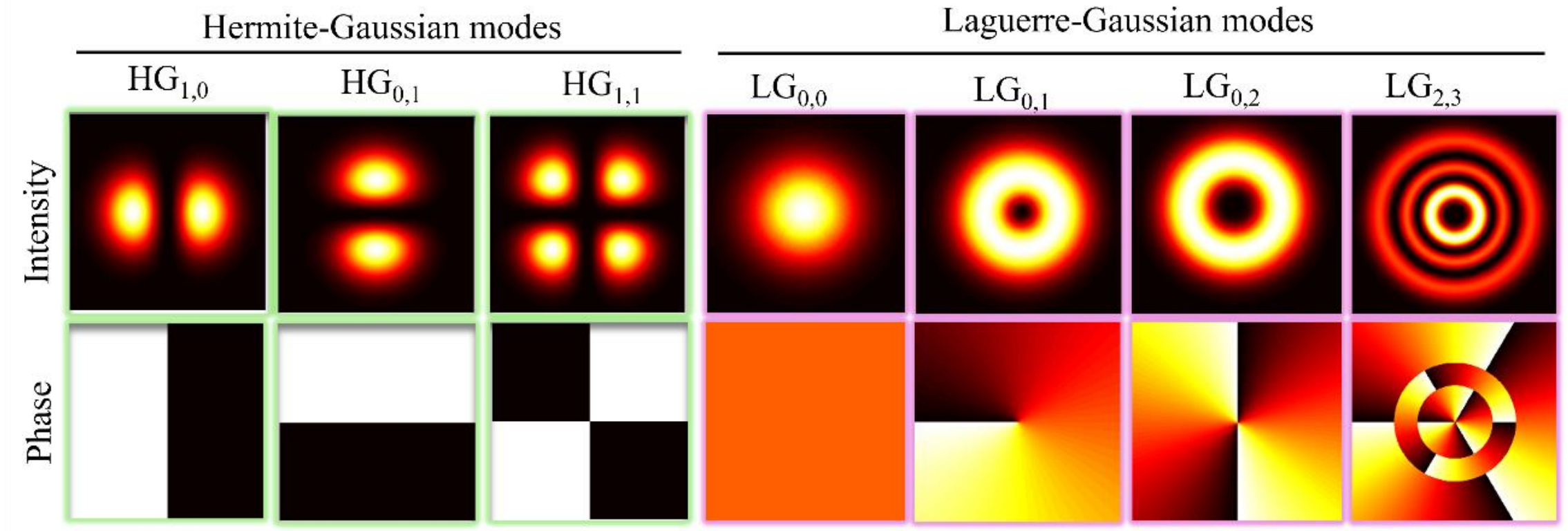

Fig. 22. Phase and intensity profiles of Hermite-Gaussian and Laguerre-Gaussian modes.

From the above discussion, it is evident that while polarization states form a 2D Hilbert space, structured spatial modes generated via transverse phase modulations occupy an infinite-dimensional Hilbert space. In practical applications, the dimensionality can be limited by selecting a finite set of modes with a given mode number *N*. Spatial modes exhibit behavior analogous to polarization modes, with the key difference being that polarization is inherently 2D, whereas spatial modes can span *N* dimensions. Understanding higher-order spatial modes can be challenging; however, a direct comparison is possible for lower-order modes with $N$=1 (Fig. 23). Specifically, the first-order Hermite–Gaussian (HG) modes consist of $HG_{0,1}$ and $HG_{1,0}$, which are equivalent to horizontal (H) and vertical (V) polarization states. Similarly, the diagonal and anti-diagonal polarization states have their counterparts in corresponding superpositions of these lower-order HG modes. On the other side, the LG modes with mode number $N = 1$ have two states of $LG_{0,-1}$, and $LG_{0,1}$ and correspond to R and L in the polarization states. Therefore, we can write all the states of LG and HG modes in terms of HG modes when we consider HG modes as basis vectors, and the treatment is the same even when we consider LG modes as basis vectors.

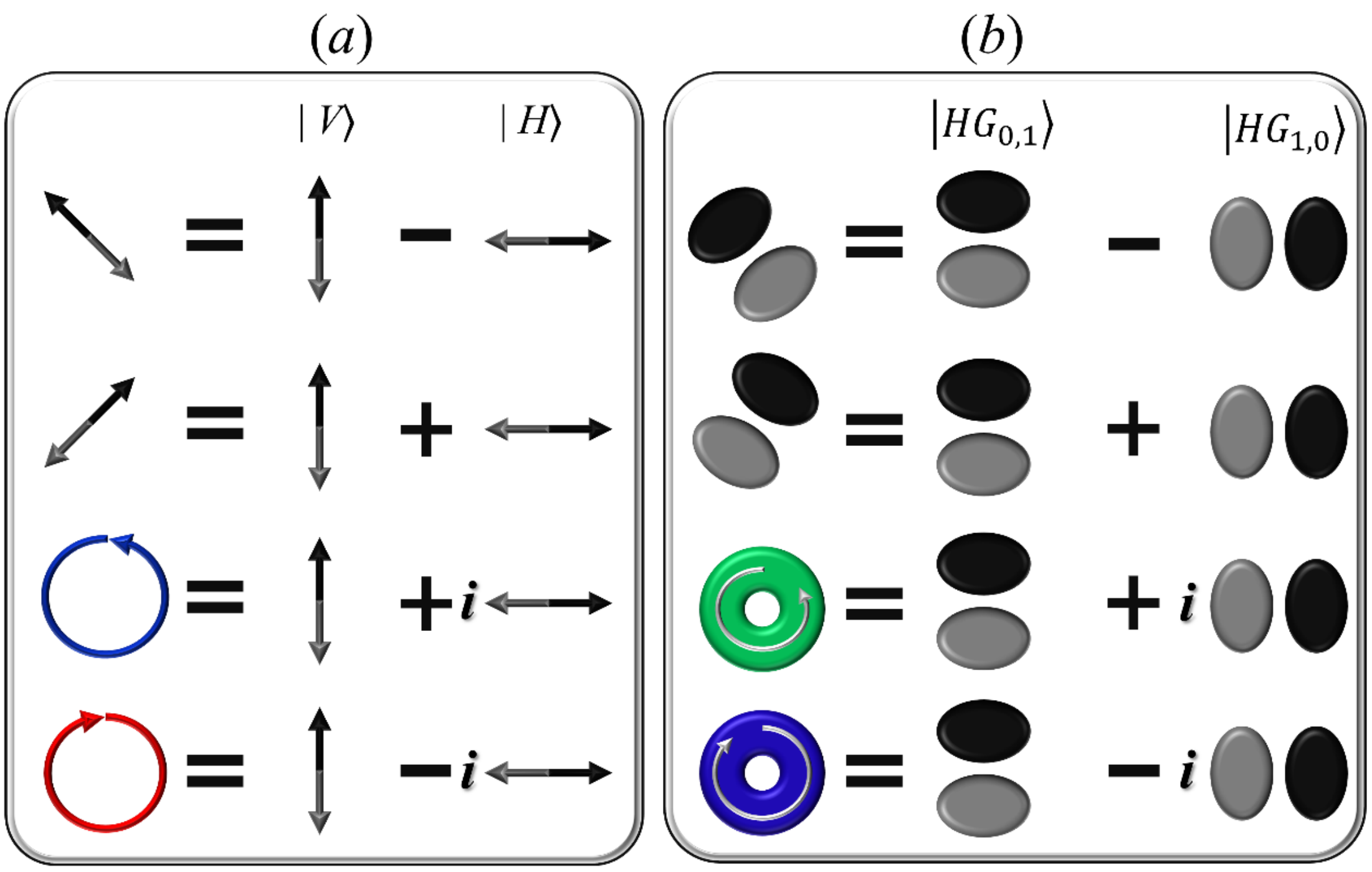

Fig. 23. Comparison between (*a*) polarization modes and (*b*) lower-order spatial modes. Linear polarizations are equivalent to HG modes, and circular polarizations are equivalent to LG modes.

### 3.1. Modal Poincaré sphere

We can construct the modal PS of order $N$ by the superposition of all the LG modes whose mode number is equal to $N$ as [88]

$$|P_{N,\ell}(\theta,\phi)\rangle = \sum_{\frac{\ell'}{2}=-\frac{N}{2}}^{\frac{N}{2}} e^{-\frac{i}{2}\ell'\phi} d^{\frac{N}{2}}_{\frac{\ell}{2},\frac{\ell'}{2}}(\theta) LG_{N,\ell'}. \tag{89}$$

Here, $d^{j}_{m',m}(\theta)$ is a Wigner function and the LG mode at $z = 0$ in terms of its mode number $N$ instead of radial index $p$ can be written as

$$\mathrm{LG}_{N,\ell}(r,\varphi) = \frac{i^{|\ell|-N}}{w}\sqrt{\frac{2^{|\ell|+1}\left(\frac{N-|l|}{2}\right)!}{\pi\left(\frac{N+|\ell|}{2}\right)!}}\, e^{\frac{r^2}{w^2}}\left(\frac{r}{w}\right)^{|\ell|} e^{i\ell\varphi} L^{\ell}_{\frac{N-|\ell|}{2}}\left(\frac{2r^2}{w^2}\right). \tag{90}$$

Here, spherical angles of the parameter sphere are constrained by $0 \le \theta \le \pi$, $0 \le \phi \le 2\pi$. The poles of the modal PS represent LG modes, while the modes at the equator are HG modes. The continuous transformation between LG and HG modes gives rise to new modes called HLG modes, and their phase is elliptical. This phase can be tuned to rectangular (HG) or circular (LG) [89]. Therefore, HLG modes are generalized Gaussian modes, which are equivalent to elliptical polarization states in polarization PS. For a given mode number, we can have multiple modal PSs. For $N = 1$, we have one modal PS, which has a superposition of two LG modes, i.e., $|P_{1,1}(\theta,\phi)\rangle = f(\mathrm{LG}_{1,-1}, \mathrm{LG}_{1,1})$ [Fig. 24(*a*)]. Which is the same for $N = 2$, but modal PS has a superposition of three LG modes i.e., $|P_{2,2}(\theta,\phi)\rangle = f(\mathrm{LG}_{2,-2}, \mathrm{LG}_{2,0}, \mathrm{LG}_{2,2})$. In these two cases, we have a single parameter sphere with zero radial index of LG modes, and HG modes are 1D. In case of $N = 3$, we have two modal PSs. One is from the mode indices of $(p, \ell) \equiv \{(0, -3), (1, -1), (1, 1), (0, 3)\}$ and is given by $|P_{3,3}(\theta,\phi)\rangle = f(\mathrm{LG}_{3,-3}, \mathrm{LG}_{3,-1}, \mathrm{LG}_{3,1}, \mathrm{LG}_{3,3})$. The second modal PS constructed by the combination of mode indices $(p, \ell) \equiv \{(1, -1), (1, 1)\}$ is $|P_{3,1}(\theta,\phi)\rangle = f(\mathrm{LG}_{3,-1}, \mathrm{LG}_{3,1})$. While in the first modal PS, we have a superposition of four LG modes, in the second modal PS, we have two LG modes in the superposition. $N = 3$, we have a nonzero radial index and produce 2D HG modes at the equator. For $N = 4$, we have Modal PS with $p = 0$ and $\ell = \pm 4$, which corresponds to a transformation between helical wave-front and 1D HG mode [Fig. 24(*b*)]. For the same mode number, we can construct another modal PS with $p = 1$ and $\ell = \pm 2$, where the radial index of the LG mode transforms to a 2D HG mode [Fig. 24(*c*)]. Furthermore, when we investigate higher-order mode numbers, we encounter multiple numbers of modal PS with complex textures. Even though we have multiple modes in the superposition, irrespective of order, each of the poles of the sphere has a single LG mode, and these LG modes at the poles have the opposite vorticity. From this, we can conclude that for $N > 1$, the HLG modes do not correspond to the linear combination of the two LG modes represented by the poles, but it is a superposition of all the LG modes used in the construction of modal PS. The modal PS, which is constructed with the LG modes with the same mode number, have same Gouy phase. Therefore, the intensity and phase distribution in the modal PS are preserved with the propagation.

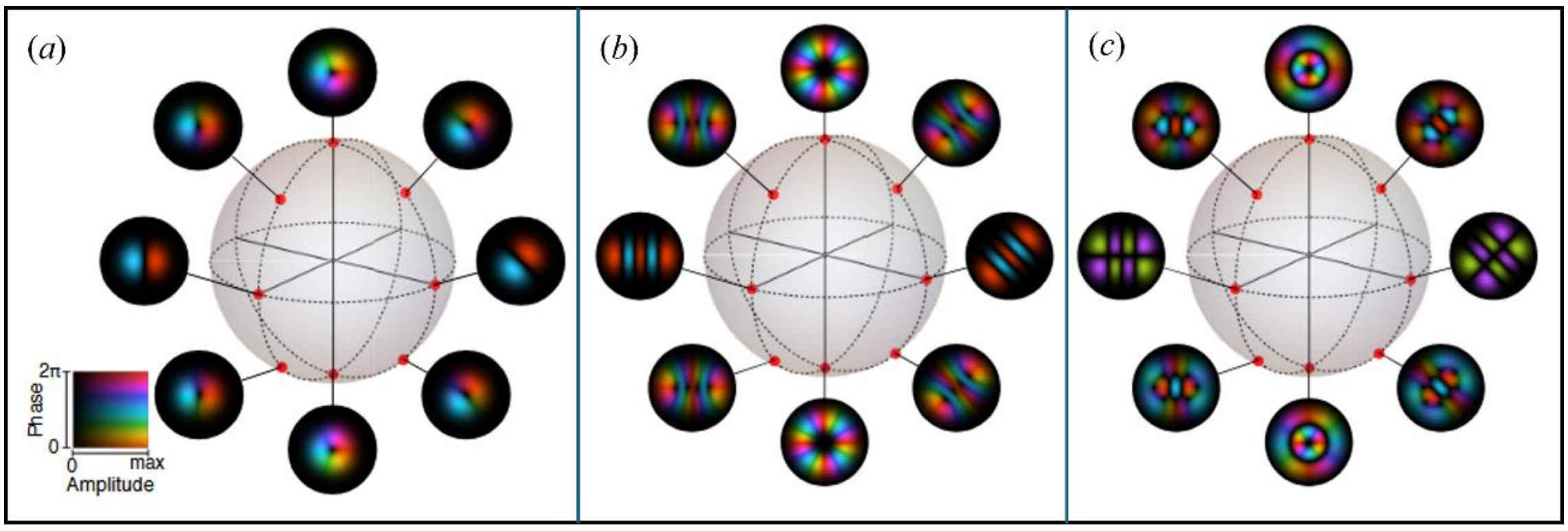

Fig. 24. Orbital/Modal Poincaré sphere of (*a*) order $N = 1$, (*b*) order $N = 4$ with $\ell = 4$, and (*c*) order $N = 4$ with $\ell = 2$ [88].

From the above discussion, the most intuitive way to understand the modal PS is by comparing the first-order modal PS with the polarization PS. It is worth noting that M. J. Padgett and J. Courtial were the first to introduce the modal PS, constructed using spatial modes with mode number $N$=1, and compared it directly with the polarization PS [Fig. 25]. The handedness of polarization in the polarization PS corresponds to the helicity of the transverse phase in the modal PS. In the polarization PS, laser modes exhibit uniform polarization, phase, and amplitude, with polarization serving as the control parameter. In contrast, in the modal PS, the modes maintain uniform polarization, but their phase and amplitude, governed by the transverse phase, are non-uniform, making the transverse phase the control parameter in this context.

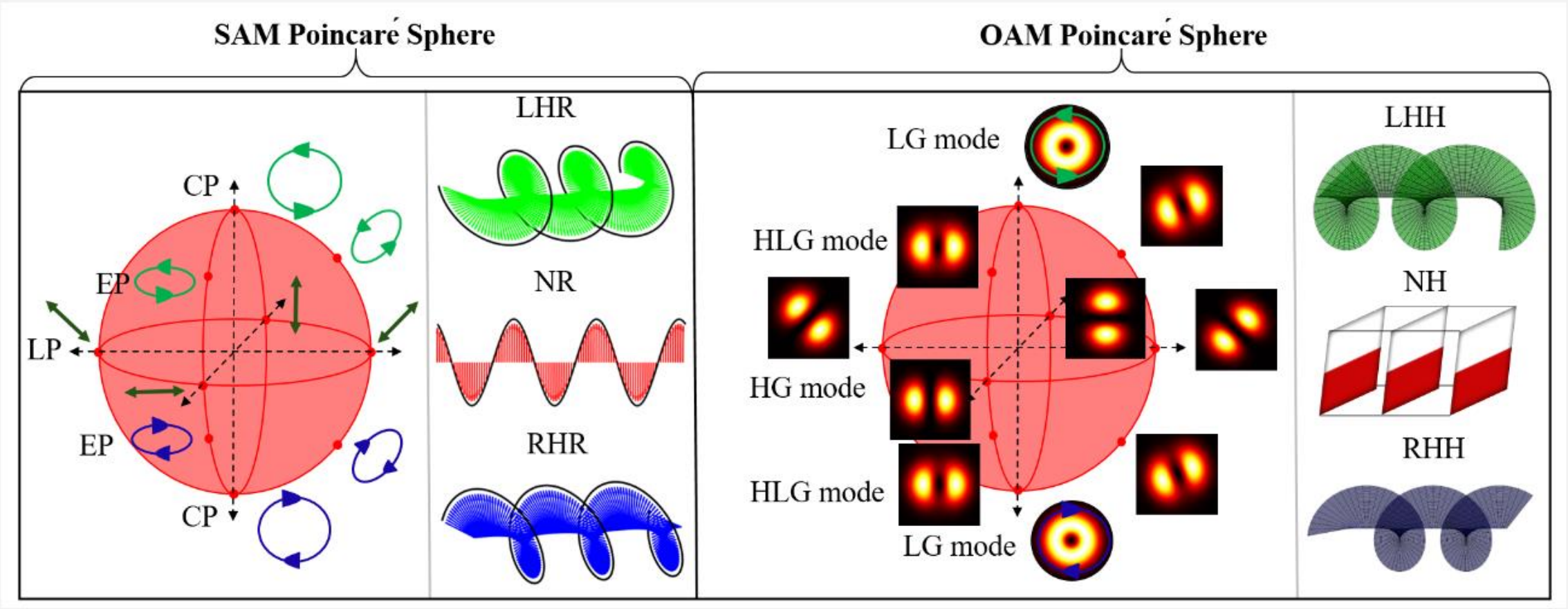

Fig. 25. Comparison between spin/polarization Poincaré sphere and orbital/Modal Poincaré sphere. LHR: left-handed rotation, RHR: right-handed rotation, NR: no rotation, CP: circular polarization, EP: elliptical polarization, LP: linear polarization, LHH: left-handed helicity, RHH: right-hand helicity, NH: no helicity, LG: Laguerre-Gaussian, HG: Hermite-Gaussian, HLG: Hermite-Laguerre-Gaussian.

### 3.2. Optical gadgets for spatial mode conversion

One approach to generating all modes on the modal PS is through mode conversion of a Gaussian beam using suitable digital phase holograms and spiral phase plates [Fig. 26(*a*)] [90–92]. More recently, we reported the direct generation of first-order modal PS modes from a laser cavity via off-axis pumping, followed by an increase in their order to two through intra-cavity frequency doubling [93,94]. In addition, Hermite–Gaussian (HG) and Laguerre–Gaussian (LG) modes generated directly from the laser cavity can be transformed into other states on the modal PS using astigmatic mode converters (AMCs). Such AMCs can be readily constructed using two identical cylindrical lenses aligned parallel to each other with their axes oriented in the same direction. Two types of AMCs are commonly employed. First one is π/2-AMC (the separation between the two cylindrical lenses is $\sqrt{2}f$), which creates mode transformation between HG and LG modes [Fig. 26(*b*)]. The second AMC is called π-AMC (the separation between the two cylindrical lenses is $2f$), which transforms the handedness of helicity of the LG mode to opposite handedness [Fig. 26(*c*)]. Another optical gadget is an optical rotator, which can produce a geometric phase by rotating the transverse profile of light fields. For instance, we can rotate the laser mode around its propagation axis by a Dove prism [Fig. 26(*c*)] and we can produce geometric phase [95]. In the mode conversion, the mode number is preserved, and all the modes have the same fundamental mode with beam waist and Raleigh range characteristics. The eigen-modes of AMCs are HG modes. Further, more details related to the construction and functionality of the AMCs can be found in [86,96,97]. In references [86,98], the matrix operators developed for spatial mode converters in a similar treatment of polarization mode conversion operators developed with Jones matrices. The spatial mode with mode number $N$ has $N$+1 column vector. For example, $HG_{m,n}$ mode with complex amplitude coefficient $a_{m,n}$ can be expressed as $|HG_{m,n}\rangle = \left(a_{N,0}\ a_{N-1,0}\ a_{N-2,0} \dots \dots a_{0,N-1}\ a_{0,N}\right)^T$. The operator of spatial mode with mode number $N$ is in the dimension of ($N$+1) × ($N$+1), and it is diagonal. The $\theta$ phase mode converter in the HG basis for mode number $N$ is given by

$$\hat{C}_{HG,N}(\theta) = diag\left[e^{-\frac{iN\theta}{2}}, e^{-i\left(\frac{N}{2}+1\right)\theta}, \ldots, e^{\frac{iN\theta}{2}}\right] \quad (91a)$$

and this mode converter can also be written in terms of the LG basis as

$$\hat{C}_{LG,N}(\theta) = B_{HG\rightarrow LG,N}\hat{C}_{HG,N}(\theta)B_{LG\rightarrow HG,N}. \quad (91b)$$

The function of the mode converter in both bases is the same, and we can omit the name of the mode in the notation, i.e., $\hat{C}_N(\theta)$. Also, the best way to understand these mode converters is in the HG mode basis, which we follow in this review. When we use a mode converter, it is necessary to know its angular position $\phi$, as like in the case of wave plates used in polarization mode conversion. Thus, sometimes we write the mode converter operator as $\hat{C}_N(\theta,\phi)$. For $N$=0 (0$^{th}$ order), the mode converter is idempotent [$\hat{C}_0(\theta)$=1], and its operation does not effect on the mode structure (wave plates are idempotent operators for spatial structured modes). The 1$^{st}$ and 2$^{nd}$ order mode converters are in the matrix form as

$$\hat{C}_1(\theta) = \begin{bmatrix} e^{-\frac{i\theta}{2}} & 0 \\ 0 & e^{\frac{i\theta}{2}} \end{bmatrix}, \text{ and } \hat{C}_2(\theta) = \begin{bmatrix} e^{-i\theta} & 0 & 0 \\ 0 & 1 & 0 \\ 0 & 0 & e^{i\theta} \end{bmatrix}, \quad (92)$$

respectively. The well-known and mostly used mode converters are π/2-AMC [$\hat{C}_N(\pi/2)$] and π-AMC [ $\hat{C}_N(\pi)$ ] mode converters. The operator for π/2-AMC [Fig. 26($b$)] is $\hat{C}_N(\pi/2) = e^{-\frac{iN\pi}{4}}diag[1, -i, -1, i, 1, \ldots]$ and in the full form it is given by

$$\hat{C}_N(\pi/2) = e^{-\frac{iN\pi}{4}} \begin{bmatrix} 1 & 0 & 0 & 0 & 0 & 0 & \cdots & \cdots & \cdots \\ 0 & -i & 0 & 0 & 0 & 0 & \cdots & \cdots & \cdots \\ 0 & 0 & -1 & 0 & 0 & 0 & \cdots & \cdots & \cdots \\ 0 & 0 & 0 & i & 0 & 0 & \cdots & \cdots & \cdots \\ 0 & 0 & 0 & 0 & 1 & 0 & \cdots & \cdots & \cdots \\ 0 & 0 & 0 & 0 & 0 & -i & \cdots & \cdots & \cdots \\ \cdots & \cdots & \cdots & \cdots & \cdots & \cdots & \cdots & \cdots & \cdots \\ \cdots & \cdots & \cdots & \cdots & \cdots & \cdots & \cdots & \cdots & \cdots \\ \cdots & \cdots & \cdots & \cdots & \cdots & \cdots & \cdots & \cdots & \cdots \end{bmatrix}. \quad (93)$$

When a single HG mode with mode number $N$, which is oriented at 45° with respect to the mode converter axis, passes through π/2-AMC, the HG mode is split into a superposition of all HG modes with mode number $N$, and cylindrical lenses produce a phase shift between them and deliver the corresponding LG mode with mode number $N$. The operator for π-AMC [Fig. 26($c$)] is $\hat{C}_N(\pi) = (-i)^N diag[1, -1, 1, -1, \ldots]$ and in the full form is

$$\hat{C}_N(\pi) = (-i)^N \begin{bmatrix} 1 & 0 & 0 & 0 & 0 & 0 & \cdots & \cdots & \cdots \\ 0 & -1 & 0 & 0 & 0 & 0 & \cdots & \cdots & \cdots \\ 0 & 0 & 1 & 0 & 0 & 0 & \cdots & \cdots & \cdots \\ 0 & 0 & 0 & -1 & 0 & 0 & \cdots & \cdots & \cdots \\ 0 & 0 & 0 & 0 & 1 & 0 & \cdots & \cdots & \cdots \\ 0 & 0 & 0 & 0 & 0 & -1 & \cdots & \cdots & \cdots \\ \cdots & \cdots & \cdots & \cdots & \cdots & \cdots & \cdots & \cdots & \cdots \\ \cdots & \cdots & \cdots & \cdots & \cdots & \cdots & \cdots & \cdots & \cdots \\ \cdots & \cdots & \cdots & \cdots & \cdots & \cdots & \cdots & \cdots & \cdots \end{bmatrix}. \quad (94)$$

Further, we can produce desired rotation in the spatial mode structure around their axis by passing them through dove prism [Fig. 26($d$)]. The dove prism not only rotates the spatial structure but also changes the handedness of the helical wave-front. The rotation operator of the spatial mode of mode number $N$ is represented with $\hat{R}_N(\phi)$ and is given in LG basis as

$$\hat{R}_{LG,N}(\phi) = diag\left[e^{-i\phi\ell_0}, e^{-i\phi\ell_1}, \ldots, e^{-i\phi\ell_N}\right], \quad (95a)$$

and this mode rotator can also be written in terms of HG basis as

$$\hat{R}_{HG,N}(\phi) = B_{LG\to HG,N}\hat{R}_{LG,N}(\phi)B_{HG\to LG,N}. \quad (95b)$$

The simplest and best way to understand the rotation mode converters is in the LG mode basis due to circular symmetry, which we follow here. For $N = 0$ (0th order), the mode rotator is idempotent [$\hat{R}_0(\phi)$=1]. The 1st and 2nd order mode converters are in the matrix form as

$$\hat{R}_1(\phi) = \begin{bmatrix} e^{i\phi} & 0 \\ 0 & e^{i\phi} \end{bmatrix}, \text{ and } \hat{R}_2(\phi) = \begin{bmatrix} e^{i2\phi} & 0 & 0 \\ 0 & 1 & 0 \\ 0 & 0 & e^{-i2\phi} \end{bmatrix}. \quad (96)$$

Like a polarizer, we can have a mode filter that can pass an elective mode while rejecting all other modes [99]. Order tunable mode filters can be developed using digital holograms. For instance, the mode filter for $HG_{N\text{-}2,2}$ is given by

$$\hat{F}_{N-2,2} = \begin{bmatrix} 0 & 0 & 0 & 0 & 0 & 0 & \cdots & \cdots & \cdots \\ 0 & 0 & 0 & 0 & 0 & 0 & \cdots & \cdots & \cdots \\ 0 & 0 & 1 & 0 & 0 & 0 & \cdots & \cdots & \cdots \\ 0 & 0 & 0 & 0 & 0 & 0 & \cdots & \cdots & \cdots \\ 0 & 0 & 0 & 0 & 0 & 0 & \cdots & \cdots & \cdots \\ 0 & 0 & 0 & 0 & 0 & 0 & \cdots & \cdots & \cdots \\ \cdots & \cdots & \cdots & \cdots & \cdots & \cdots & \cdots & \cdots & \cdots \\ \cdots & \cdots & \cdots & \cdots & \cdots & \cdots & \cdots & \cdots & \cdots \\ \cdots & \cdots & \cdots & \cdots & \cdots & \cdots & \cdots & \cdots & \cdots \end{bmatrix}. \quad (97)$$

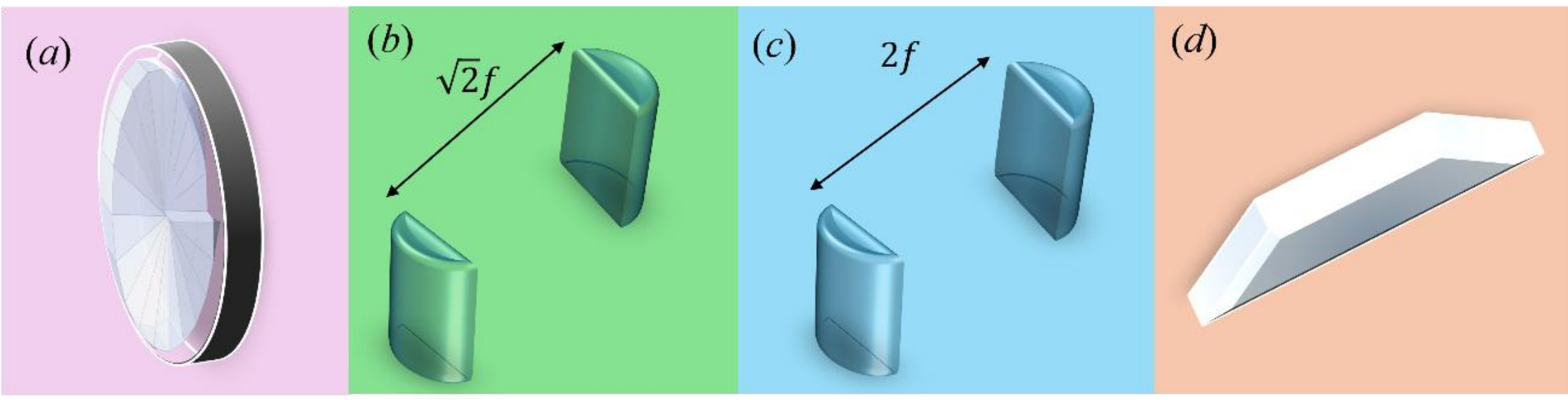

Fig. 26. Various kinds of spatial mode converters: (*a*) Spiral phase plate (SPP): converts a Gaussian mode into an LG mode with an arbitrary azimuthal index with zero radial index. Further, it transforms LG mode into hollow Gaussian (HoG) mode, which have plane wave front with annular intensity distribution. (*b*) π/2 astigmatic mode converter (π/2-AMC): It makes a transformation between HG and LG modes. (*c*) π astigmatic mode converter (π-AMC): it transforms the handedness of the LG modes from one sign to another sign. (*d*) Dove prism: transforms the sign of the handedness of LG modes, and also it can rotate the structured mode.

The AMCs can be easily understood in terms of their functionality by comparing the mode conversion of lower-order spatial modes (mode number $N$ = 1) with polarization mode converters where the dimensionality is equal [Fig. 27]. The π/2-AMC is equivalent to the QWP [Fig. 27(*a*)]. The incident LP at 45° with respect to the fast-/slow -axis of the QWP converted into corresponding circular polarization, and it is true vice versa. When the cylindrical axis of the mode converter is horizontal/vertical, the π/2-AMC converts $HG_{01}$ mode oriented at diagonal/anti-diagonal (at 45°/135°) into $LG_{0,1}$ and vice versa, i.e.,

$$HG_D = \frac{1}{\sqrt{2}}\left[HG_{1,0} + HG_{0,1}\right] \overset{\hat{C}_1\left(\frac{\pi}{2}\right)}{\longleftrightarrow} \frac{1}{\sqrt{2}}\left[HG_{1,0} + iHG_{0,1}\right] = LG_{0,1}, \quad (98a)$$

and

$$HG_A = \frac{1}{\sqrt{2}}\left[HG_{1,0} - HG_{0,1}\right] \overset{\hat{C}_1\left(\frac{\pi}{2}\right)}{\longleftrightarrow} \frac{1}{\sqrt{2}}\left[HG_{1,0} - iHG_{0,1}\right] = LG_{0,-1}. \quad (98b)$$

The second converter $\pi$-AMC is equivalent to HWP [Fig. 27(*b*)]. While HWP transfers the polarization state between LCP and RCP, the $\pi$-AMC acts as a mode transformer between the LG mode with left-hand helicity and the LG mode with right-hand helicity, i.e.,

$$LG_{0,1} = \frac{1}{\sqrt{2}}\left[HG_{1,0} + iHG_{0,1}\right] \overset{\hat{C}_1(\pi)}{\longleftrightarrow} \frac{1}{\sqrt{2}}\left[HG_{1,0} - iHG_{0,1}\right] = LG_{0,-1}. \quad (99)$$

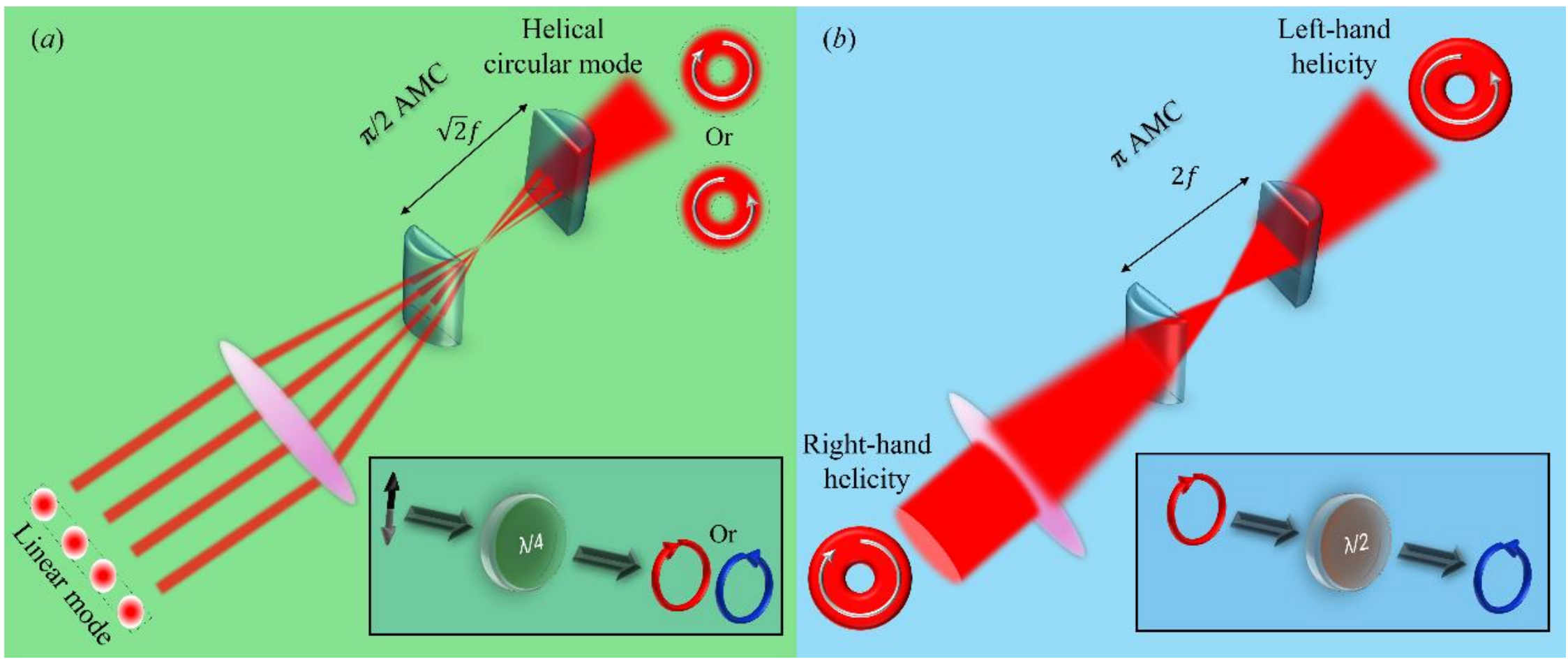

Fig. 27. (*a*) π/2 and (*b*) π astigmatic mode converters used for the spatial mode transformation on the modal Poincaré sphere. Equivalent polarization mode converters are provided in the insets.

### *3.3. Geometric phase creation in scalar structured modes by spatial mode converters*

S. J. Van Enk was the first person to theoretically show that when we create a closed circuit on the modal PS, the transported scalar structured mode acquires a geometric phase that depends on the OAM, and in mathematical form is given by [101]

$$\gamma_G(C) = -\ell\, \Omega(C) \quad (100)$$

We can trace this closed optical circuit on the modal PS using mode converters. The analysis on the parameter sphere is quite similar to the polarization PS. For example, as shown in Fig. 28, we can trace the same closed circuit on both modal and polarization PS. Two π-AMCs of $\hat{C}_1(\pi, 0)$ and $\hat{C}_1(\pi, \alpha)$ (the second π-AMC is rotated at an angle α with respect to the first π-AMC) acted on the north pole and produced a closed path around the axis $\overleftrightarrow{NS}$ of modal PS. Here, first π-AMC, $\hat{C}_1(\pi, 0)$ transports the structured mode from the north pole to the south pole, and the second π-AMC, $\hat{C}_1(\pi, \alpha)$ again transports the structured mode from the south pole to the north pole with the path angularly separated at α with respect to the first path and produced closed circuit [Fig. 28(*b*)].

This closed circuit is also shown in Fig. 29(*c*) as ABDCA. The opening angle of the circuit is equal to $2\alpha$. In the case of polarization PS, the same circuit can be created by two HWPs of $\hat{R}_h(0)$ and $\hat{R}_h(\alpha)$. The geometric phase created by both circuits is $\gamma_G = 2\alpha$. From this, infer that in both cases, the frequency shift created in the light is given by twice the rotation frequency of the optical component [84]. Further, we can create half of the above circuit by two π/2-AMCs and two dove prims [Fig. 29(*a*)]. The first π/2-AMC angularly positioned as $\hat{C}_1\left(\frac{\pi}{2}, -\frac{\pi}{4}\right)$ transports the north pole vortex beam with right-hand helicity to the horizontal HG mode ($HG_{01}$) present equator (point B). Two parallel aligned dove prisms with angularly separated by $\alpha/2$ are form a rotating operator $\hat{R}_2\left(\frac{\alpha}{2}\right)$. This operator rotates the HG mode and moves to point *C* and, provides 2 $\alpha$ angular separation from point B. The second π/2-AMC angularly positioned as $\hat{C}_1\left(\frac{\pi}{2}, \frac{\pi}{4} - \alpha\right)$, transported the mode back to the north pole from point C. This closed loop has a geometric phase of $\gamma_G = \alpha$ [Fig. 29(*b*) and Fig. 29(*c*)]. The geometric phase in the structured modes was experimentally first demonstrated by G. F. Brand in mm waves [102], and later it was demonstrated in optical modes by E.

J. Galvez et al. [6]. The controlled change in the geometric phase for the above mentioned two circuits was experimentally measured by E. J. Galvez et al. as a function of $\alpha$ are given in Fig. 29($d$).

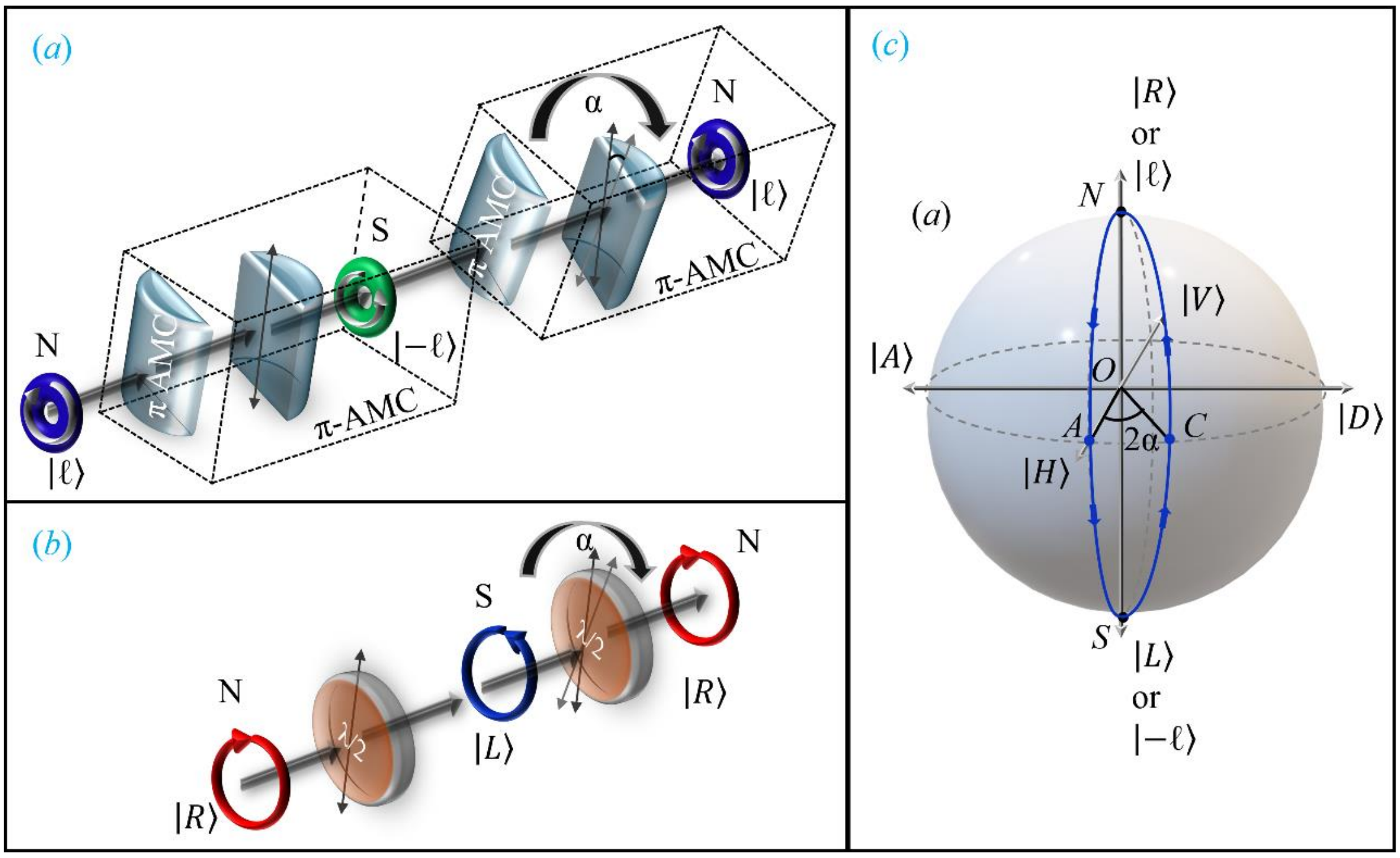

Fig. 28. An optical circuit created on the parameter sphere by ($a$) a spatial mode converter and ($b$) a polarization mode converter. The corresponding evolution of geometric phase on the parameter sphere is illustrated in ($c$). The polarization Poincaré sphere and modal Poincaré sphere are represented on a single parameter sphere with the aid of their equivalence. The double arrow at AMC is its principal axis and represents the focusing line of the cylindrical lens [84].

T. Malhotra et al. experimentally measured the geometric spatial modes without any interference technique [103]. In this report, they blocked some part of the initial beam and the trajectories described by the centroid of the remaining intensity under mode transformations created by anisotropic quadratic phase masks projected on Spatial Light Modulators (SLMs) [Fig. 30]. The numerical analysis of the evolution of the mode is characterized by a ray family. The generalized Gaussian beams can be represented in terms of a ray family in which each ray is specified by the values of two periodic parameters, τ and $\eta$. At a given transverse plane, the rays corresponding to all values of $\tau$, for fixed $\eta$ [104,105]. In analogy with polarization, this ellipse of rays corresponds to a point on the PS. Also, the second parameter, η, represents a closed loop over the sphere, referred to here as the Poincaré path (PP). Therefore, each structured mode is represented with an extended path on the modal PS instead of a point. The ray structure follows the intensity distribution. For example, the ray structure has rectangular symmetry in HG modes and circular symmetry in LG modes. As shown in Fig. 30($a$), the optical path initiated with HG mode on the left side and traced a closed circuit with opening angle, α, by racing its final state, which is the same as the initial state. Now the ray distribution in both the initial and final states is the same, but their colour is different due to the accumulation of geometric phase, which is given in terms of mode number as

$$\gamma_G(C) = (N - 2n)\alpha. \tag{101}$$

The shift in the centroid of the quarter of the 2D HG mode took place while tracing a closed path on the modal PS, and it is sufficient to estimate the geometric phase [Fig. 30($c$)]. The geometric phase achieved with this technique was well matched with the phase measured through the interference method [Fig. 30($b$)]. J. Courtial et al. have shown that the rotation of an OAM beam at an angular frequency of Ω by dove prism operator can produce a frequency shift of ℓ Ω [106]

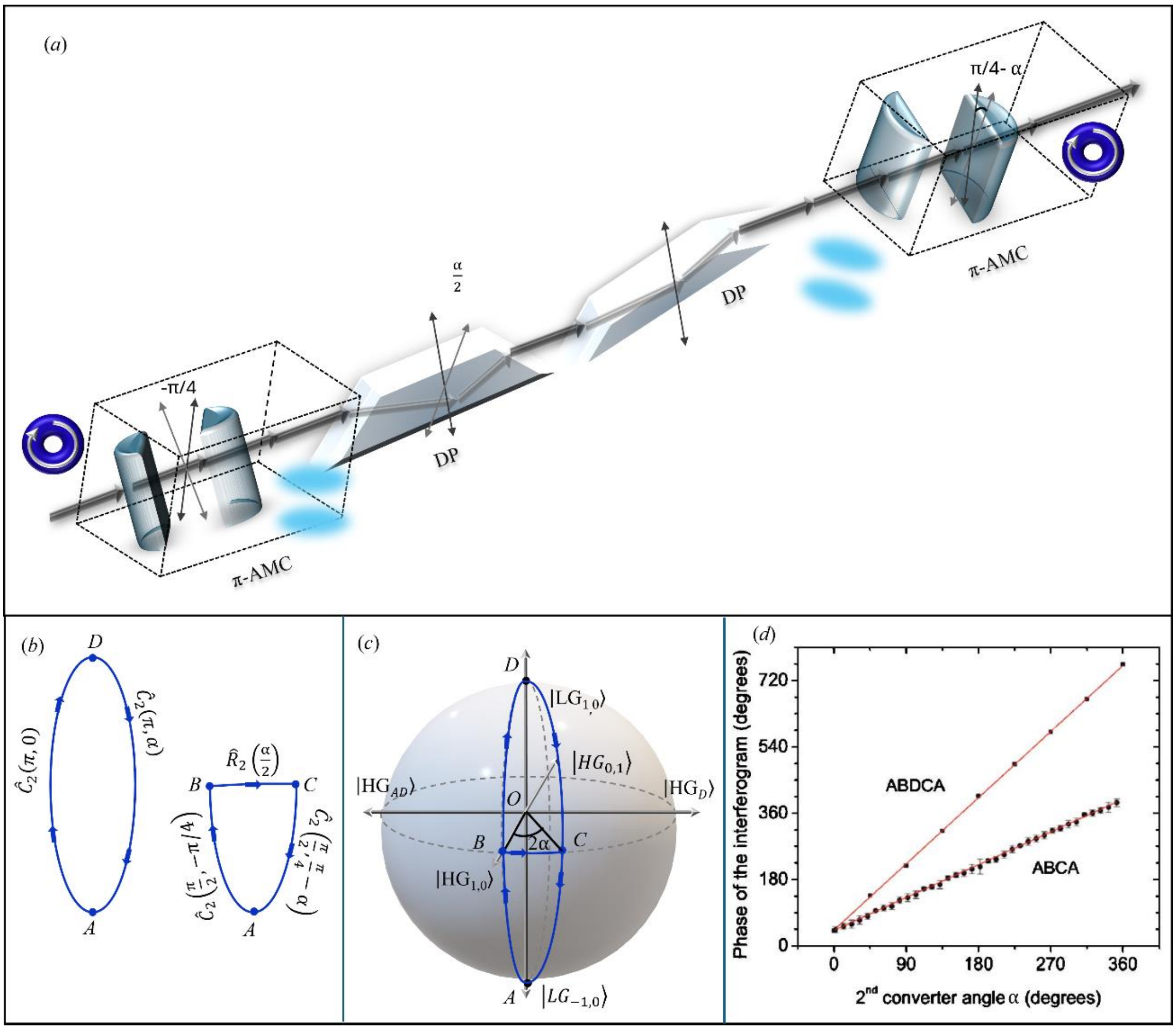

Fig. 29. An optical gadget created by the combination of two π-astigmatic mode converters (π-AMCs) and two dove prisms (DPs), used for creating a closed circuit of ABCA on the modal Poincaré sphere. (*a*) The closed loops are created by two different optical gadgets. (*c*) The closed optical paths of *ABCA* and *ABDCA* on the modal Poincaré sphere. (*d*) The experimentally measured geometric phase in both loops for different angles of rotation of the second mode converter [6].

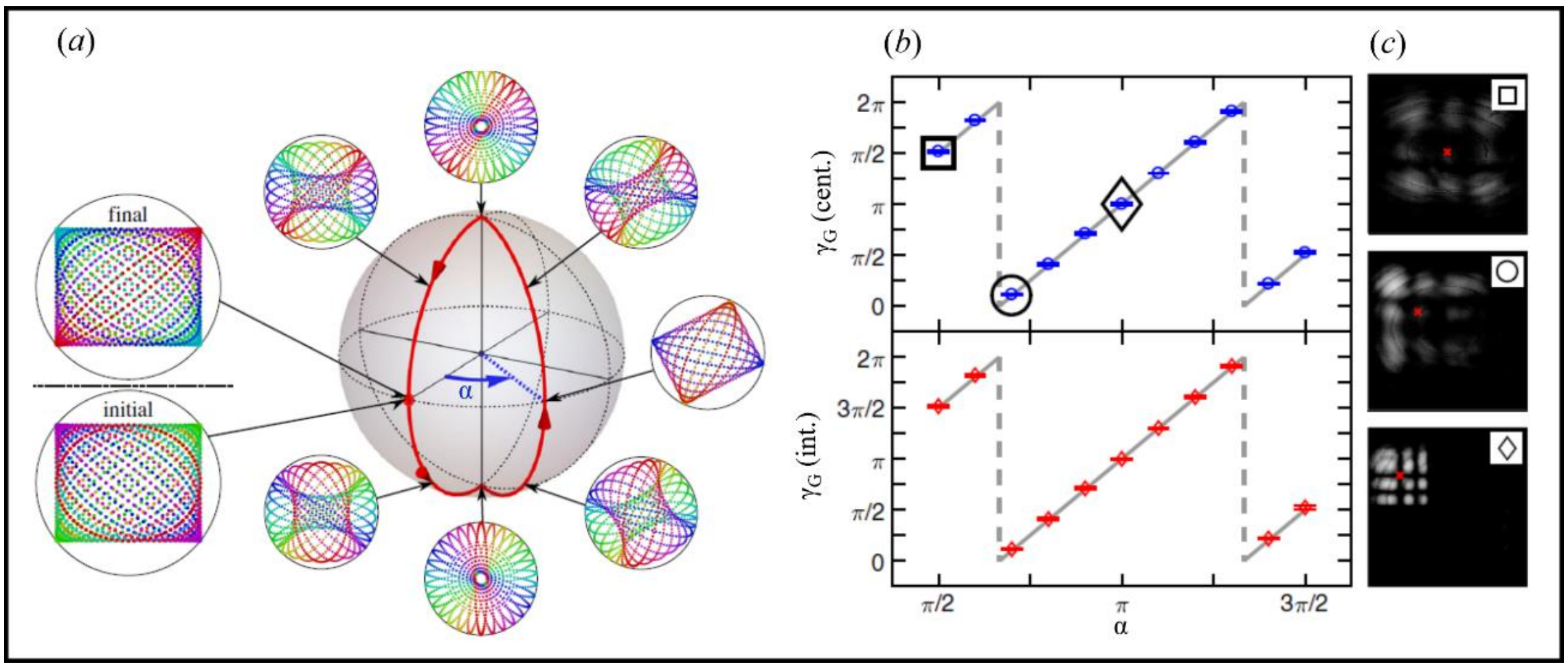

Fig. 30. (*a*) Ray distribution representation of spatial mode transformation in a closed geodesic circuit of a modal spot on the modal Poincaré sphere. The ray distributions of spatial modes at different stages along the path are shown inside the circles, where points of equal color correspond to equal values of η. The initial and final ray distributions (directly to the left of the modal PS) have the same rectangular shape and correspond to the same HG mode, but the different color distribution reveals the cycling of the rays that gives rise to the geometric phase. (*b*) Experimentally measured geometric phase, $\gamma_G$ as a function of α using the centroid of the blocked beam (circles, top panel) and interferometric measurements (diamonds, bottom panel). The non-interferometric results are wrapped onto [0, 2$\pi$) for comparison with the interferometric ones. The experimental values are fitted with the theoretical plot shown in gray color. (*b*) Measured intensity of the transformed blocked beam (plotted on a log scale) with centroids as red crosses, for different *α*. The symbols in the insets correspond to the markers in (*b*) [103].

## 4. Geometric phase in vector modes (modes with spatially inhomogeneous polarization and phase)

### *4.1. Construction of Poincaré sphere for vector beams*

In the previous sections, we discussed the origin and experimental measurement of the geometric phase in polarization and phase. In a similar way, we can create a geometric phase by mode transformation of vector beams on a 2D parameter sphere called the higher-order/hybrid PS [107]. Vector beams have non-uniform polarization with certain polarization singularities and can also have transverse phase variation. We can easily construct an Eq. for a higher-order/hybrid PS by superposition of two orthogonal LG modes with different OAM present in circular polarization basis (it is true for any two orthogonal polarization states). In this case, the PS Eq. becomes

$$|P(\theta,\phi)\rangle = \cos\theta LG_{\ell_{R},p_R}(r,\phi,z)e^{-\frac{i\phi}{2}}|R\rangle + \sin\theta LG_{\ell_{L},p_L}(r,\phi,z)e^{\frac{i\phi}{2}}|L\rangle. \qquad (102)$$

This equation is a composite of polarization and modal PSs and is formed by the combination of Eq. 14 and Eq. 89 (for 2D). Here, we used (2$\theta$, 2$\phi$) instead of ($\theta$, $\phi$) to make it convenient with the original article [108]; in either case, the results are the same. Eq. 102 represents the higher-order PS given in Fig. 31 for $p_R = p_L = 0$ with $\ell_R = - \ell_L$. In this scenario, the poles have oppositely handed circular polarized LG modes with equal and opposite helicity. At the equator, we have equal contributions from the polar modes and produce a non-uniform LP distribution. Based on the distribution of LP at the equator, we can classify the higher-order PS into cylindrical vector higher-order PS depicted in Fig. 31(*a*) [109] and π vector higher-order PS depicted in Fig. 31(*b*) [110]. The transformation between these two PSs can be obtained either by interchanging the circular polarizations or by interchanging the LG modes at the poles. All the points on the PS other than poles and equator will have elliptical polarizations with their major axis following the LP distribution at the equator. In this context, the geometric phase depends on the total angular momentum (TAM) number (sum of SAM number and OAM number), and it is given by

$$\gamma_G(C) = -(\ell + \mathrm{s})\Omega(C)/2. \qquad (103)$$

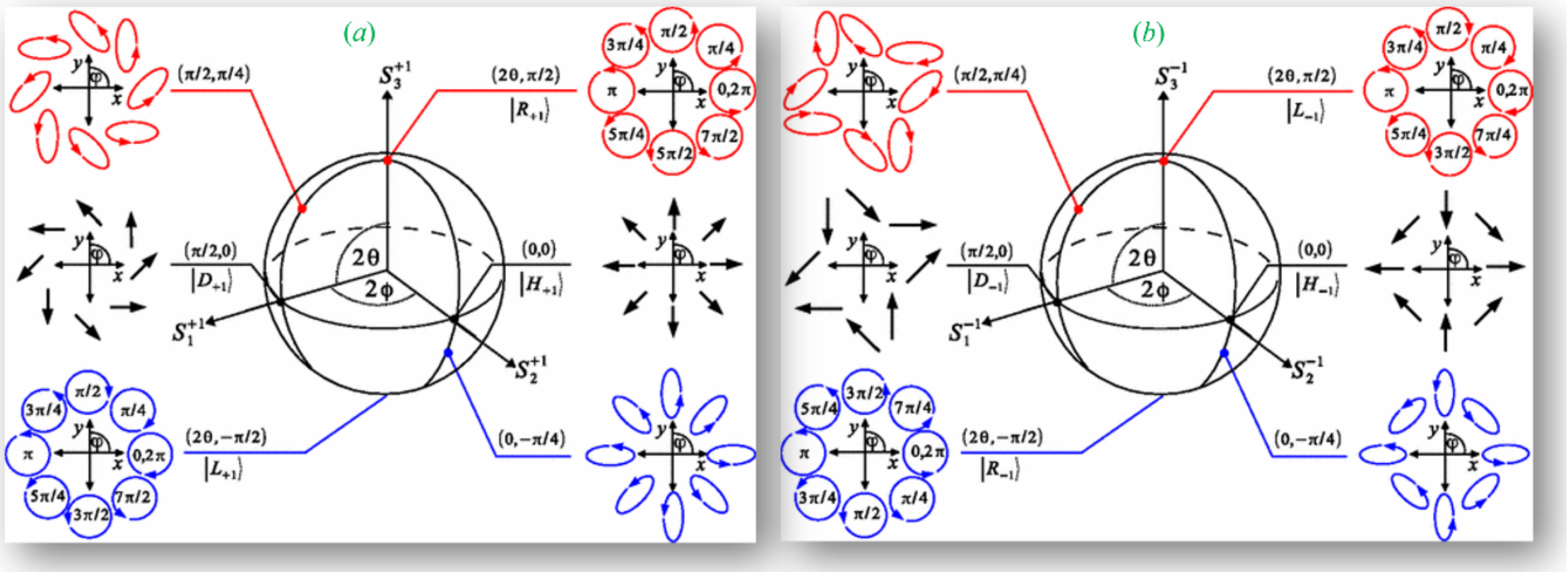

Fig. 31. Construction of a higher-order Poincaré sphere by vector modes. Poles have uniform orthogonal circular polarizations with orthogonal helical wave fronts. The modes at the equator have non-uniform linear polarization distributions with petal phase structure. All other points have non-uniformly distributed elliptical polarizations with a certain helical wave-front. higher-order Poincaré sphere with (*a*) cylindrical vector modes and (*b*) π vector modes at equator [108].

### *4.2. Optical gadgets for mode transformation*

In the polarization PS, wave plates serve as the primary optical gadgets, whereas in the modal PS, mode converters are employed to manipulate spatial modes. The higher-order PS is a composite sphere formed by combining the polarization and modal PSs; consequently, the corresponding optical gadgets are constructed using a combination of an AMC and a wave plate. L. Allen et al. extended the operator formalism developed for scalar modes to vector modes by simultaneously considering both the spatial and polarization degrees of freedom of structured laser beams [98]. As discussed in the previous subsection, any vector beam can be generated by superimposing two scalar structured modes with orthogonal polarizations. By passing such superposed modes through polarization and spatial mode converters, the spatial structure and polarization of each constituent mode can be independently transformed. This allows the transfer of both polarization and spatial mode from one point to another on the higher-order PS. A spatial mode with mode number $N$ has two orthogonal polarizations, which can be represented by a $2(N+1)$-dimensional column vector. For instance, $HG_{m,n}$ mode with complex amplitude coefficient $a_{m,n}$ having two orthogonal polarizations in the horizontal and vertical, can be expressed as

$$|HG_{m,n}\rangle = \left(a_{N,0}(\leftrightarrow)\; a_{N,0}(\updownarrow)\; a_{N-1,0}(\leftrightarrow)\; a_{N-1,0}(\updownarrow)\; \dots\dots\dots\; a_{0,N-1}(\leftrightarrow)\; a_{0,N-1}(\updownarrow)\; a_{0,N}(\leftrightarrow)\; a_{0,N}(\updownarrow)\right)^{T}. \tag{104}$$

The matrices of vector mode convert have the dimensions of $2(N+1) \times 2(N+1)$, with each term of the mode repeated along the diagonal so that both polarizations are re-phased. In this scenario, the operator for π/2-AMC, $\hat{C}_N(\pi/2)$ is given by

$$\hat{C}_N(\pi/2) = e^{-\frac{iN\pi}{4}} \begin{bmatrix} 1 & 0 & 0 & 0 & 0 & 0 & 0 & 0 & \dots \\ 0 & 1 & 0 & 0 & 0 & 0 & 0 & 0 & \dots \\ 0 & 0 & -i & 0 & 0 & 0 & 0 & 0 & \dots \\ 0 & 0 & 0 & -i & 0 & 0 & 0 & 0 & \dots \\ 0 & 0 & 0 & 0 & -1 & 0 & 0 & 0 & \dots \\ 0 & 0 & 0 & 0 & 0 & -1 & 0 & 0 & \dots \\ 0 & 0 & 0 & 0 & 0 & 0 & i & 0 & \dots \\ 0 & 0 & 0 & 0 & 0 & 0 & 0 & i & \dots \\ \dots & \dots & \dots & \dots & \dots & \dots & \dots & \dots & \dots \end{bmatrix} \tag{105a}$$

and the operator for π-AMC, $\hat{C}_N(\pi)$ is given by

$$\hat{C}_N(\pi) = (-i)^N \begin{bmatrix} 1 & 0 & 0 & 0 & 0 & 0 & \cdots & \cdots & \cdots \\ 0 & 1 & 0 & 0 & 0 & 0 & \cdots & \cdots & \cdots \\ 0 & 0 & -1 & 0 & 0 & 0 & \cdots & \cdots & \cdots \\ 0 & 0 & 0 & -1 & 0 & 0 & \cdots & \cdots & \cdots \\ 0 & 0 & 0 & 0 & 1 & 0 & \cdots & \cdots & \cdots \\ 0 & 0 & 0 & 0 & 0 & 1 & \cdots & \cdots & \cdots \\ \cdots & \cdots & \cdots & \cdots & \cdots & \cdots & \cdots & \cdots & \cdots \\ \cdots & \cdots & \cdots & \cdots & \cdots & \cdots & \cdots & \cdots & \cdots \\ \cdots & \cdots & \cdots & \cdots & \cdots & \cdots & \cdots & \cdots & \cdots \end{bmatrix}. \qquad (105b)$$

In addition to the above two operators, we have similar operators for the spatial mode filter, mode rotator, QWP, and HWP. One of the famous optical gadgets is a spin-orbit converter (SOC), which is formed by a combination of $\pi$-AMC and HWP [Fig. 32]. This optical gadget can transfer the state on the higher-order PS between the north and south poles with a closed geodesic path. Two SOC optical gadgets used in the experimental measurement of the geometric phase are given in Fig. 33 ($a$).

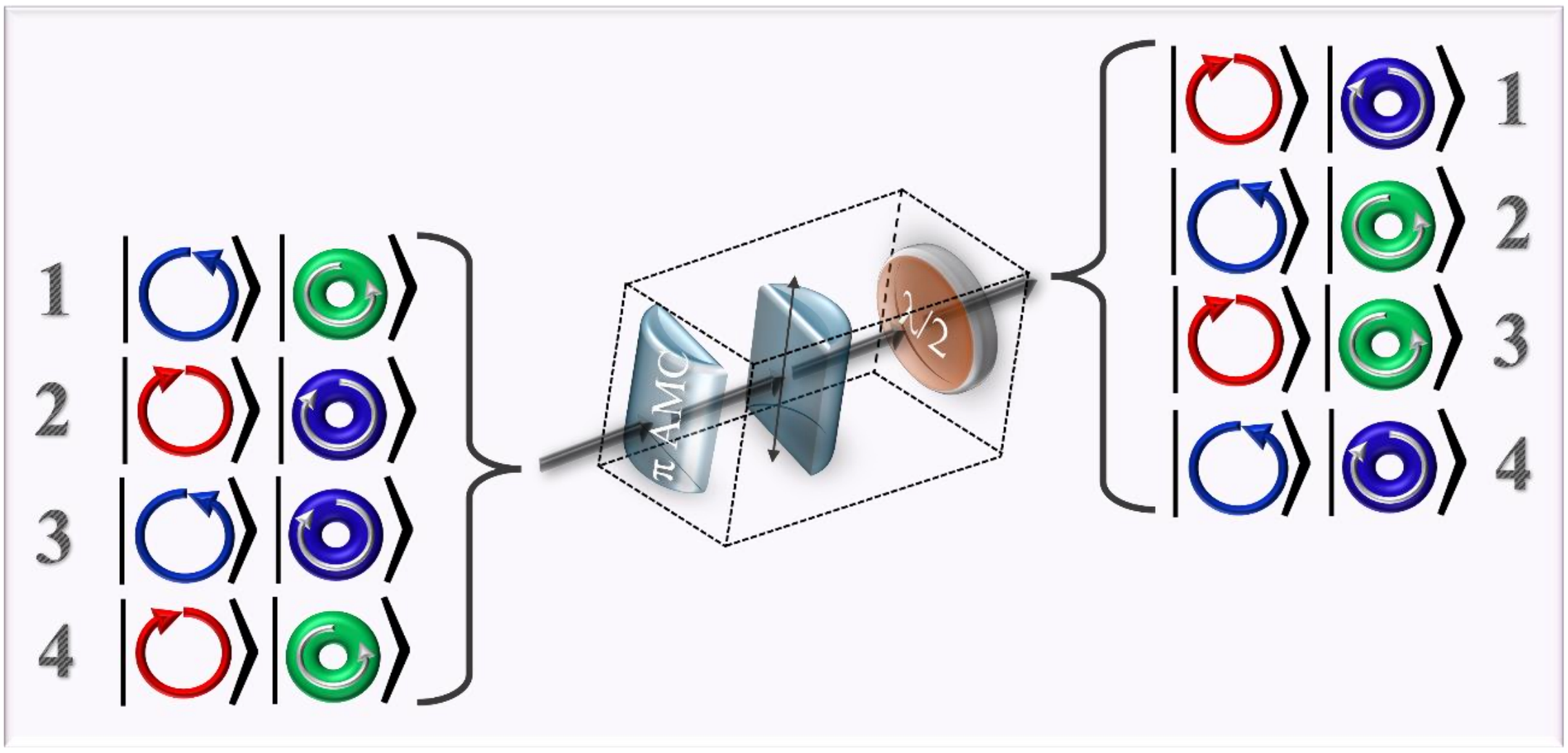

Fig. 32. Spatial and polarization mode converter simultaneously transformed the spin angular momentum and orbital angular momentum.

### *4.3. Experimental measurement of geometric phase in vector modes*

The vector mode provided by Eq. 102 can be experimentally generated by using different techniques based on diffractive optical elements like SLM, DMD, SPP, q-plate, WPSI, etc., with and without an interferometer [111-114]. The longitudinal position of the vector mode can be continuously tuned by controlling the relative amplitude of superposed LG modes, and the latitudinal position is tuned via varying the relative phase between the two superposed modes. The position of the vector mode on the parameter sphere can be transported from one position to another by SOC and produce a geometric phase. G. Milione et al. experimentally measured the geometric phase in the vector mode formed by LG modes of $\ell=\pm3$ [7]. To experimentally visualize the geometric phase, they have used two SOCs. In that first one was fixed while rotating the second one at a relative angle of $\Delta\phi$ [Fig. 33($a$)]. As a result, a closed optical circuit ($ABCDA$) on higher-order PS formed with its opening angle is given by $2\Delta\phi$ [Fig. 33($b$)]. Next to the SOCs, a single polarizer is used to project two orthogonal modes into a single polarization state so that we can get the interference between them. The structure and azimuthal position of the petal structure in terms of its intensity distribution are proportional to $1+\cos^2(2\ell\varphi+2\gamma_G+2\alpha)$ with $\alpha$ as the rotational angle of the polarizer. The petal structure formed after the polarizer rotated at the same angle as SOC2 and is theoretically given by $\Delta\phi = m\pi / 2\ (\ell + s)$ [Fig. 33($c$)]. From the rotation of the petal structure, we can get the geometric phase, $\gamma_G = 2(\ell + s)\Delta\phi$ .

Further, the geometric phase on higher-order PS was extended to the dark Poincaré beams generated by the superposition of LG modes with different non-zero topological charges, and experimentally demonstrated [116,117].

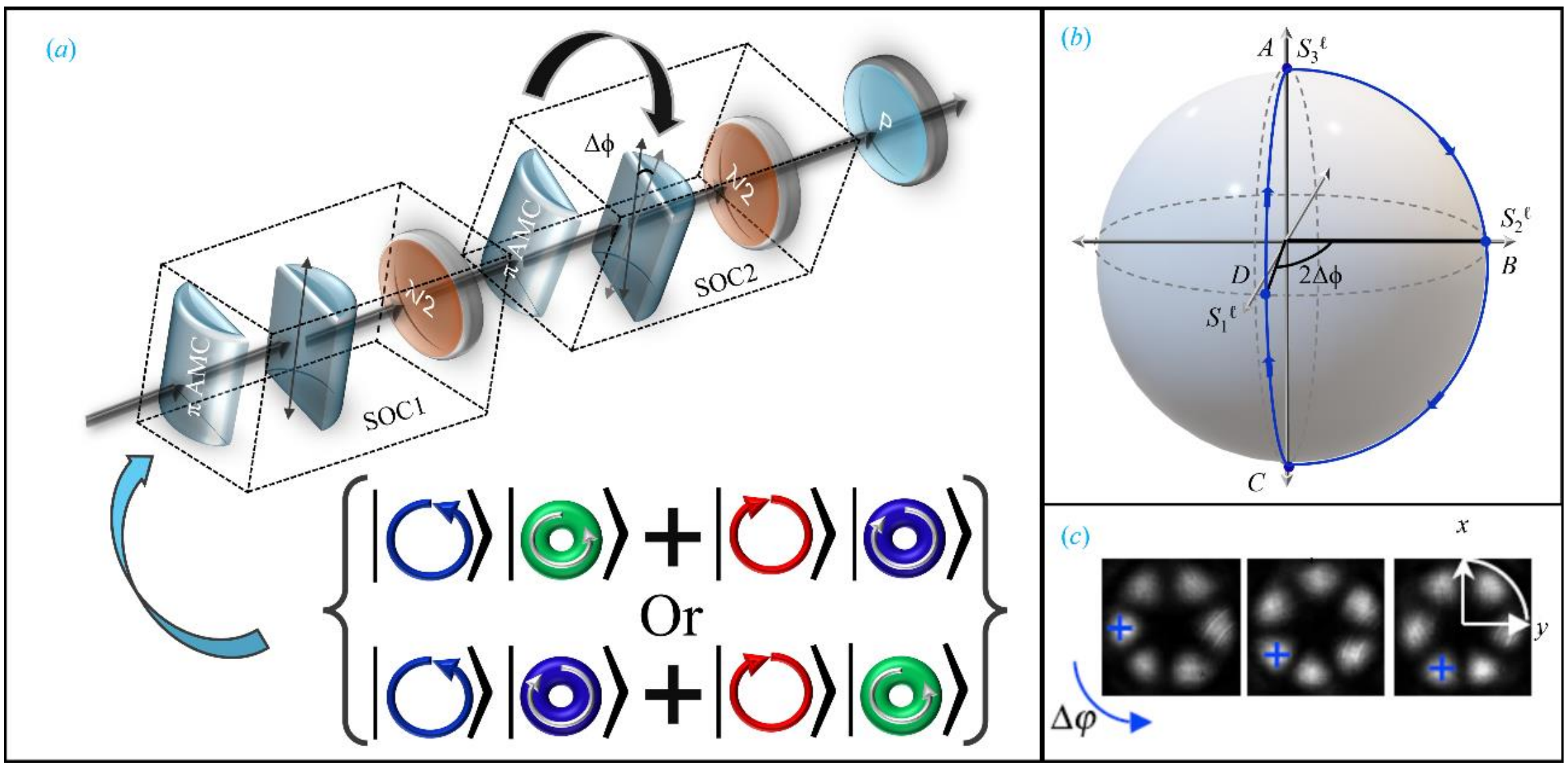

Fig. 33. (*a*) Synthesis of vector beams and their state transformation on the higher-order Poincaré sphere by using composite optical gadgets called spin-orbit converters (SOCs) created by the combination of an astigmatic mode converter (AMC) and a half-wave plate (λ/2). An interferogram created by projecting orthogonally polarized modes onto a single polarization by a linear polarizer (P). (*b*) Experimentally obtained optical circuit *ABCDA* on the modal Poincaré sphere by mode transformation of the vector mode created by LG modes of ℓ=±3 present in the orthogonal circular polarization states. (*c*) Experimentally observed rotation of the petal structure created after the polarizer by rotating SOC2 with reference to SOC1 [7].

## 5. Geometric phase in Electromagnetic field

The energy, momentum, spin, etc. of the electromagnetic field can be represented on a parameter sphere called the electromagnetic symmetry sphere (ESS) or electromagnetic sphere (ES) [118] (In this section, vectors are represented with bold letters.). The state vector is a bispinor, which has two components: electric field and magnetic field, and its generalized mathematical form is given by [118-120]

$$\boldsymbol{\Psi}(r) = \frac{1}{2}\begin{bmatrix}\sqrt{\varepsilon}\boldsymbol{E}\\ \sqrt{\mu}\boldsymbol{H}\end{bmatrix}. \tag{106}$$

Like a polarization spinor, the bispinor also has a choice of basis, which can be used for the construction of ESS. As shown in Fig. 34(*a*), we can have three bases of electric and magnetic field (EM), parallel and anti-parallel (PA), right and left-handed (RL), in the respective mathematical forms of

$$\boldsymbol{\Psi}_{EM}(r) = \begin{bmatrix}\boldsymbol{F}_e\\ \boldsymbol{F}_m\end{bmatrix} = \frac{1}{2}\begin{bmatrix}\sqrt{\varepsilon}\boldsymbol{E}\\ \sqrt{\mu}\boldsymbol{H}\end{bmatrix}, \tag{107a}$$

$$\boldsymbol{\Psi}_{PA}(r) = \begin{bmatrix}\boldsymbol{F}_p\\ \boldsymbol{F}_a\end{bmatrix} = \frac{1}{2\sqrt{2}}\begin{bmatrix}\sqrt{\varepsilon}\boldsymbol{E} + \sqrt{\mu}\boldsymbol{H}\\ \sqrt{\varepsilon}\boldsymbol{E} - \sqrt{\mu}\boldsymbol{H}\end{bmatrix}, \tag{107b}$$

and

$$\boldsymbol{\Psi}_{RL}(r) = \begin{bmatrix}\boldsymbol{F}_R\\ \boldsymbol{F}_L\end{bmatrix} = \frac{1}{2\sqrt{2}}\begin{bmatrix}\sqrt{\varepsilon}\boldsymbol{E} + i\sqrt{\mu}\boldsymbol{H}\\ \sqrt{\varepsilon}\boldsymbol{E} - i\sqrt{\mu}\boldsymbol{H}\end{bmatrix}. \tag{107c}$$

These three bases are related to the fundamental electromagnetic symmetries: parity inversion, time reversal, and discrete duality transformation. The electromagnetic energy density $W$ is given by the norm of the bispinor, i.e., $\|\boldsymbol{\Psi}\|^2$. Moreover, this energy density can be written as a sum of the squared norm of each of the two $\Psi$ components in any of the three bases, i.e., $W = W_e + W_m = W_p + W_a = W_R + W_L$ with $W_i = |\boldsymbol{F}_i|^2$. The ESS can be constructed with energy density parameters: $W_0$ (= $W$), $W_1$ (= $W_e$ - $W_m$), $W_2$ (= $W_a - W_p$), and $W_3$ (= $W_R - W_L$), which are equivalent to the Stokes vectors ($S_0$, $S_1$, $S_2$, $S_3$) in the PS [Fig. 34($b$)].

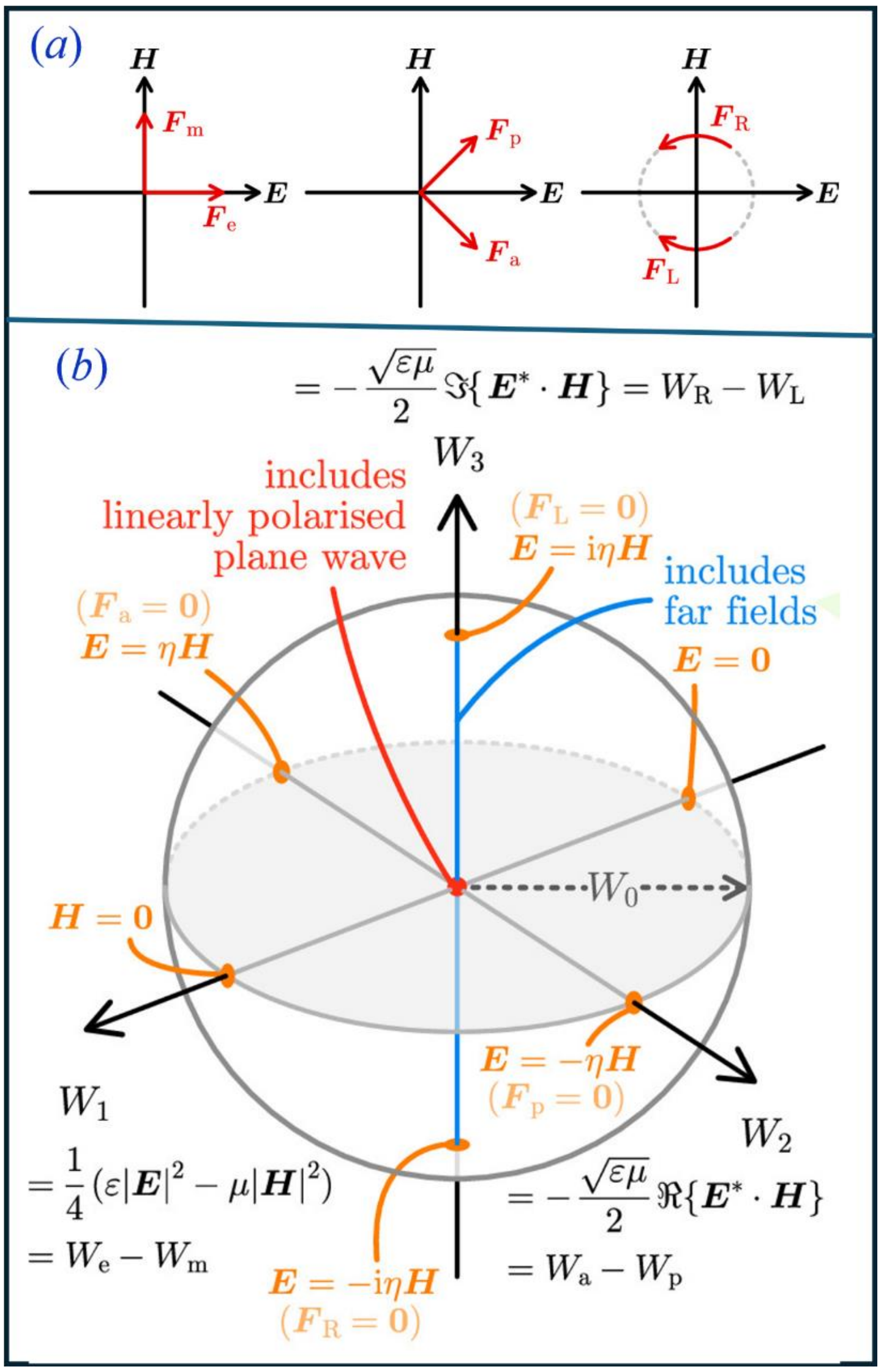

Fig. 34. ($a$) Components of the electromagnetic bispinor represented in the $\mathbb{C}^2$ subspace, using the electric and magnetic field components in three bases: electric-magnetic (EM), parallel-antiparallel (PA) and right-left handed (RL) bases, drawn with respect to $\boldsymbol{E}$ and $\boldsymbol{H}$ vectors. ($b$) The energy symmetry sphere (ESS) or electromagnetic sphere (ES) with axes are given by corresponding energy densities $W_i$. The field quantities of the axes are given in orange color. In ESS, the far-fields lie along the $W_3$ axis while a linearly polarised plane wave (preserving all symmetries) resides at the sphere centre [118].

Recently, A. J. Vernon and K. Y. Bliokh [12] used ES and derived the mathematical expression for electromagnetic geometric phase as

$$\gamma_G = i\int_C \boldsymbol{\Psi}^\dagger . \nabla\boldsymbol{\Psi} . dr + N\pi. \tag{108}$$

Here, $N = 0$ or 1 is a $\mathbb{Z}_2$ topological index which encodes the even or odd number of half-turns made by principal semi-axes, the 'six-dimensional EM polarization ellipse' of $\boldsymbol{\Psi}$ over C [121]. This Eq. is valid for both paraxial and non-paraxial conditions.

$$\boldsymbol{\Psi} = \begin{bmatrix} \cos\frac{\theta}{2} e^{i\chi_e}\hat{e} \\ \sin\frac{\theta}{2} e^{i\chi_m}\hat{h} \end{bmatrix}. \tag{109}$$

Here, the angular weight factor comes from the angle between electric and magnetic field components $\theta = 2\arctan(\sqrt{\mu}\boldsymbol{H}/\sqrt{\varepsilon}\boldsymbol{E})$ and complex unit polarization vectors: $\hat{e} = \exp[-i(\alpha + \chi_e)]\boldsymbol{E}/|\boldsymbol{E}|$ and $\hat{h} = \exp[-i(\alpha + \chi_m)]\boldsymbol{H}/|\boldsymbol{H}|$. The phases $\chi_e$ and $\chi_m$ added to the overall phase $\alpha$ such that $\mathrm{Arg}(\hat{e}.\hat{e}) = \mathrm{Arg}(\hat{h}.\hat{h}) = 0$. By substituting the state vector in Eq. 108, the geometric phase obtained is in the form of

$$\gamma_G = \gamma_{G\mathrm{I}} + \gamma_{G\mathrm{II}} + N\pi. \tag{110}$$

The first term is given by

$$\gamma_{G\mathrm{I}} = i\int_C \left[\cos^2\frac{\theta}{2}\hat{e}^*.\nabla\hat{e} + \sin^2\frac{\theta}{2}\hat{h}^*.\nabla\hat{h}\right].dr. \tag{111}$$

This expression provides the geometric phase, which comes as a result of variations in the electric and magnetic polarization ellipse along the curve *C*. This phase accounts for both the PB phase in 2D paraxial waves, where $\theta = \pi/2$ and $\gamma_{GI} + N\pi = \gamma_{PB}$, and the spin-redirection phase for 3D evolutions of polarization. Another condition stated that having fixed $\mathrm{Arg}(\hat{e}.\hat{e}) = \mathrm{Arg}(\hat{h}.\hat{h}) = 0$ means that $\gamma_{\mathrm{GI}} = 0$ if the electric and magnetic polarization states are uniform over C. The second term is the electromagnetic geometric phase, and its mathematical expression is given by

$$\gamma_{G\mathrm{II}} = -\int_C \left[\cos^2\frac{\theta}{2}d\chi_e + \sin^2\frac{\theta}{2}d\chi_e\right]. \tag{112}$$

This phase originates from the incorporation of the $\mathbb{C}^2$ electro-magnetic space in $\boldsymbol{\Psi}$ and arises exclusively in non-paraxial light, i.e., $\gamma_{G\mathrm{II}} = 0$ for paraxial waves. The electromagnetic geometric phase is given by

$$\gamma_{G\mathrm{EM}} = \gamma_{G\mathrm{II}} + N\pi = -\frac{1}{2}\int_C \left[(1 - \cos\theta)d\phi = -\frac{1}{2}\Omega\right]. \tag{113}$$

This analysis successfully applies and investigates the electromagnetic geometric phase in two physical phenomena: the standing wave and the focused structured beam [12].

## 6. Geometric phase in nonlinear optics

A. Karnieli et al. theoretically modelled and experimentally demonstrated the adiabatic geometric phase in nonlinear frequency conversion, wherein the coupling between the signal and idler frequencies constitutes the intrinsic two-level dynamics of the system [122,123]. In this context, the parameter-space surface is elongated along the z direction with idler frequency at north pole and signal frequency at south pole. As shown in Fig. 35(a), the geometric phase nonlinear crystal is fabricated in such a way that two sections have the opposite phase after selective etching of the poled surface. The field vector $\vec{B}$ rotates around the unit vector $\hat{n}$ forms a closed circuit [Fig. 35(b)]. The experimental configuration used by the authors is depicted in Fig. 35(c). The pulsed laser source at 1064.5 nm wavelength was used as a pump, and a continuous wave laser source of 1550 nm wavelength was selected for an idler. These two laser modes have Gaussian transverse profiles and are collinearly combined by a dichroic mirror (DM) and then pumped to an adiabatic KTP crystal for conversion to the 631 nm pulsed signal. The idler mode was filtered by the first filter (F1) from the signal and pump pulses. An additional adiabatic converter was used before the geometric phase crystal to efficiently generate a signal pulse arriving synchronously with the pump pulse. The synchronized pump and signal pulses were further focused using the lens (L1) into the geometric phase crystal. Next, the pump is filtered by the second filter (F2), and the output is imaged by a CCD camera using lens L2. The imaged consecutive planes from the crystal exit facet are given in under without pumping [Fig. 35(d1)] and with pumping [Fig. 35(d2)]. In the absence of a pump, the signal has a Gaussian transverse profile, and it becomes $HG_{0,1}$ mode in the presence of a pump. Here, the pump mode incident on the crystal allows for adiabatic circular trajectories with opposite orientations. The π geometric phase step converts the spatial mode to $HG_{0,1}$ while retaining the original input frequency of the signal wave.

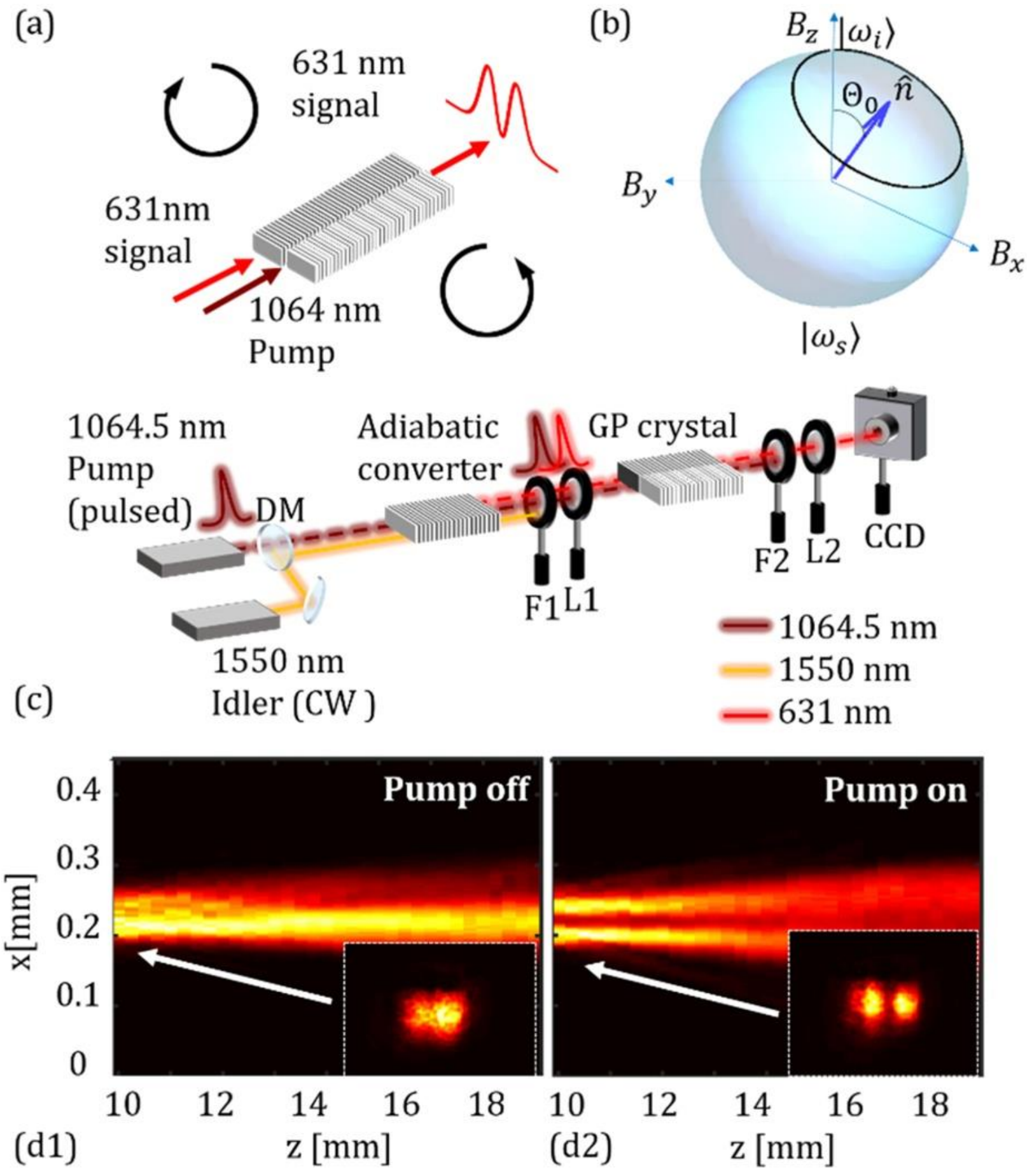

Fig. 35. Geometric phase in nonlinear wave mixing through a circular rotation scheme. (a) The geometric phase crystal is designed with two paths, each of which follows the same circular trajectory, with one path in a clockwise and other one is counter-clockwise direction. The signal has an $HG_{0,1}$ shape in its transverse profile in the presence of a pump. (b) The field vector $\vec{B}$ follows the black line around the unit vector $\hat{n}$, which forms an angle $\Theta_0$ with the $z$ axis. (c) The schematic diagram of the experimental setup used for the realization of the geometric phase. The pulsed pump and continuous wave idler are collinearly combined using a dichroic mirror (DM) in an adiabatic KTP crystal for conversion of the pulsed signal. The idler is filtered using the filter (F1), and the synchronized pump and signal pulses are focused using the lens (L1) into the geometric phase crystal. The pump is filtered by F2, and the output is imaged to a CCD camera using L2. The images provided in (d1) and (d2) correspond to the measured propagation after the crystal, when the pump is off and on, respectively [122].

## 7. Applications

The geometric phase approach can be used to control the phase and polarization in optical circuits on the PS using variable retarder operators, and we can develop a universal compensator. The linear increase of geometric phase with wave plates can be easily used to compensate for the random drifts observed in the interferometer sensors [124,125]. Another interesting phase-related application of geometric phase is utilizing it as an achromatic phase shifter. The achromatic phase shift can be produced by utilizing suitable birefringent materials and can be systematically controlled by rotating them to a suitable angle [126,127]. A white-light geometric phase interferometer was used for surface profiling [128]. The Pancharatnam phase can be used as a tool to control the phase of an atom interferometer [129]. The unbounded shift in the geometric phase by wave plates can produce a frequency shift. By controlling the rotation angle of wave plates, we can produce a controlled amount of frequency shift [130,131]. Another interesting application is the development of a geometric phase lens, which focuses the selective polarization [132,133]. Experimentally demonstrate an ultrasensitive displacement measurement based on the PB Phase of a liquid crystal optical element under vector beam illumination. In this case, both displacement magnitude and direction can be easily determined [134]. The geometric phase is successfully utilized in the SLMs, where we can spatially control the modulations in the phase and amplitude of light waves, and we can produce spatially structured light waves [135,136]. A. G. Fox utilized QHQ phase shifters in radio physics for developing phasing in an array of radar antennas [137]. Another interesting application of the Pancharatnam phase is creating phase jumps in SU (2) phase devices. This concept was well developed by R. Bhandari [138]. The geometric phase acquired while polarization state transferred between LCP and RCP spatially controlled in the beam cross-section and produced order-tunable LG modes with variable azimuthal index and zero radial index through with and without spin-orbit conversion [139]. An anisotropic medium with its optic axis lying orthogonal to the propagation direction of light is spatially modulated and the refractive index remains constant everywhere. A spin-controlled cumulative phase distortion is imposed on the beam to balance diffraction for a specific polarization attribute geometric phase. The continuously modulated geometric phase along the propagation acts as a wave guide for the light [140]. Propagation-invariant vectorial Bessel beams with tunable OAM were experimentally generated by utilizing space-variant subwavelength dielectric gratings, which primarily work based on PB phase [141]. A polarization-dependent focusing lens was successfully developed by a quantized geometrical blazed phase of polarization diffraction grating [142]. Recently, PB phase-based $q$-plate diffractive elements were used for order and shape-tunable optical skyrmions and Poincaré modes in paraxial laser beams [143]. A checkerboard-encoded design that enables simultaneous amplitude and phase modulation through PB phase engineering was used for experimentally realizing polarization-rotating beams with continuously varying polarization angles along the propagation axis, eliminating traditional dynamic phase requirements [144].

## 8. Conclusion

The geometric phase was first predicted and experimentally observed by Pancharatnam in the context of polarization, and later independently discovered by Berry in quantum systems undergoing time evolution. In optics, this phase is commonly referred to as the PB phase. The geometric phase arises solely from the geometry of the parameter space and is not necessarily restricted by the adiabatic condition. It reflects the sensitivity of the light's state along both its transverse and longitudinal evolution. The geometric phase is path-dependent and unbounded, increasing with the path length on the PS, whereas the dynamical phase is bounded modulo $2\pi$. Geodesic paths on the PS correspond to a purely geometric phase, while non-geodesic paths result in a combination of geometric and dynamical phases. For example, wave plates oriented at angles other than $\pi/4$ with respect to the incident polarization generate small circular paths (non-geodesic arcs) on the PS. In experiments, the dynamical phase can be effectively eliminated using compensating plates within the interferometer. Notably, in a Sagnac interferometer, no dynamical phase compensation is required, as both interfering beams perceive identical dynamical phases.

In its early development, the PB phase of light was fundamentally established and experimentally verified in polarization through simple interference experiments using optical retarders. Subsequent studies demonstrated its applications in light–matter interactions. The geometric phase of polarization can be generated in two ways. The first involves introducing a cyclic change in the direction of propagation of a beam (i.e., a change in the propagation vector) without altering its polarization state. The second method is achieved through a cyclic change in the polarization state itself. Analogous to polarization, the geometric phase is also observed in spatial scalar and vector modes of laser beams. More recently, it has been extended to electromagnetic fields and quantified on the electromagnetic symmetry sphere, or electromagnetic Poincaré sphere. Thus, there exist three primary parameter spheres: polarization PS, modal PS, higher-order PS, and the electromagnetic PS that are equivalent in their functional representation. The equivalence among the first three spheres and the corresponding evolution of the geometric phase is summarized in Table 1.

Table 1: Geometric Phase of light in terms of polarization, phase, and combination of polarization and phase.

| Geometric phase parameter | Polarization | Phase | Polarization + Phase |
|---|---|---|---|
| Angular momentum | SAM | OAM | TAM = SAM+OAM |
| Generalized mode | Elliptical polarization | Hermite-Laguerre-Gaussian mode | Non-uniformly distributed elliptical polarization and composed phase |
| Poincaré sphere | Polarization/SAM PS | Modal/OAM PS | Higher-order/Hybrid PS |
| States at poles | Circular polarization states (RCP and LCP) | Orthogonal LG modes | Orthogonal LG modes in circular polarization states |
| Quantum numbers | SAM, $s = \pm 1$ | OAM, $\ell = \pm 1, \pm 2, \pm 3, \ldots, \pm\infty$ | TAM, $J = \pm s \pm \ell$ |
| States at equator | Linear polarizations | HG modes | Non-uniformly distributed linear polarization and petal phase |
| Geometric phase, $\gamma_G$ | $\gamma_G = -s\ \Omega/2$ | $\gamma_G = -\ \ell\ \Omega/2$ | $\gamma_G = -\ J\ \Omega/2$ |
| Optical gadgets | Wave plates and optical active materials | Astigmatic mode converters and dove prisms | Spin-orbit converters |
| Basis vectors of linear optical gadgets | Linear polarization states | HG modes | HG modes + linear polarization states |

In experimental studies, several methods have been employed for the quantitative measurement of the geometric phase, most of which rely on interference. The fundamental principle in these measurements is the analysis of the final state of light relative to its initial state on the corresponding parameter sphere. Beyond its fundamental significance, the geometric phase has found numerous practical applications, including nonlinear optics, the development of geometric phase-based optical elements, fabrication of phase-compensation devices, creation of antenna arrays, surface profiling, and more.